\documentclass[10pt,twocolumn,showpacs,longbibliography,amsmath,amssymb,osajnl,floatfix,superscriptaddress]{revtex4-1}

\usepackage{graphicx}
\usepackage{dcolumn}
\usepackage{bm}
\usepackage{color}
\usepackage{txfonts}
\usepackage{microtype}
\usepackage{url}
\usepackage{epstopdf}
\usepackage{indentfirst}
\usepackage{textcomp}
\usepackage{mathrsfs}
\usepackage{braket}
\usepackage[caption=false,font=normalsize,labelfont=sf,textfont=sf]{subfig}
\newcommand{\setParDis}{\setlength {\parskip} {0.35cm}}
\newcommand{\setParDef}{\setlength {\parskip} {0.3pt}}

\usepackage[title,titletoc]{appendix}
\usepackage{nameref}
\usepackage{hyperref}
\hypersetup{hidelinks}
\begin{document}

\title{Generation of High-Density High-Polarization Positrons via Single-Shot Strong Laser-Foil Interaction}
\author{Kun Xue}
\affiliation{Key Laboratory for Nonequilibrium Synthesis and Modulation of Condensed Matter (MOE), Shaanxi Province Key Laboratory of Quantum Information and Quantum Optoelectronic Devices, School of Physics, Xi'an Jiaotong University, Xi'an 710049, China}
\author{Ting Sun}
\affiliation{Key Laboratory for Nonequilibrium Synthesis and Modulation of Condensed Matter (MOE), Shaanxi Province Key Laboratory of Quantum Information and Quantum Optoelectronic Devices, School of Physics, Xi'an Jiaotong University, Xi'an 710049, China}

\author{Ke-Jia Wei}
\affiliation{Key Laboratory for Nonequilibrium Synthesis and Modulation of Condensed Matter (MOE), Shaanxi Province Key Laboratory of Quantum Information and Quantum Optoelectronic Devices, School of Physics, Xi'an Jiaotong University, Xi'an 710049, China}

\author{Zhong-Peng Li}
\affiliation{Key Laboratory for Nonequilibrium Synthesis and Modulation of Condensed Matter (MOE), Shaanxi Province Key Laboratory of Quantum Information and Quantum Optoelectronic Devices, School of Physics, Xi'an Jiaotong University, Xi'an 710049, China}

\author{Qian Zhao}
\affiliation{Key Laboratory for Nonequilibrium Synthesis and Modulation of Condensed Matter (MOE), Shaanxi Province Key Laboratory of Quantum Information and Quantum Optoelectronic Devices, School of Physics, Xi'an Jiaotong University, Xi'an 710049, China}

\author{Feng Wan}	
\email{wanfeng@xjtu.edu.cn}
\affiliation{Key Laboratory for Nonequilibrium Synthesis and Modulation of Condensed Matter (MOE), Shaanxi Province Key Laboratory of Quantum Information and Quantum Optoelectronic Devices, School of Physics, Xi'an Jiaotong University, Xi'an 710049, China}

\author{Chong Lv}	
\email{lvchong@ciae.ac.cn}
\affiliation{Department of Nuclear Physics, China Institute of Atomic Energy, P.O. Box 275(7), Beijing 102413, China}	

\author{Yong-Tao Zhao}
\affiliation{Key Laboratory for Nonequilibrium Synthesis and Modulation of Condensed Matter (MOE), Shaanxi Province Key Laboratory of Quantum Information and Quantum Optoelectronic Devices, School of Physics, Xi'an Jiaotong University, Xi'an 710049, China}

\author{Zhong-Feng Xu}
\affiliation{Key Laboratory for Nonequilibrium Synthesis and Modulation of Condensed Matter (MOE), Shaanxi Province Key Laboratory of Quantum Information and Quantum Optoelectronic Devices, School of Physics, Xi'an Jiaotong University, Xi'an 710049, China}
\author{Jian-Xing Li}
\email{jianxing@xjtu.edu.cn}
\affiliation{Key Laboratory for Nonequilibrium Synthesis and Modulation of Condensed Matter (MOE), Shaanxi Province Key Laboratory of Quantum Information and Quantum Optoelectronic Devices, School of Physics, Xi'an Jiaotong University, Xi'an 710049, China}
\affiliation{Department of Nuclear Physics, China Institute of Atomic Energy, P.O. Box 275(7), Beijing 102413, China}	

\date{\today}

\begin{abstract}
	We put forward a novel method for producing ultrarelativistic high-density high-polarization positrons through a single-shot interaction of a strong laser with a tilted solid foil.
	In our method, the driving laser ionizes the target, and the emitted electrons are accelerated and subsequently generate abundant $\gamma$ photons via the nonlinear Compton scattering, dominated by the laser. These $\gamma$ photons then generate polarized positrons via the nonlinear Breit-Wheeler process, dominated by a strong self-generated quasistatic magnetic field $\mathbf{B}^{\rm S}$.
	We find that placing the foil at an appropriate angle can result in a directional orientation of $\mathbf{B}^{\rm S}$, thereby polarizing positrons.
	Manipulating the laser polarization direction can control the angle between the $\gamma$ photon polarization and $\mathbf{B}^{\rm S}$, significantly enhancing the positron polarization degree. 
	Our spin-resolved quantum electrodynamics particle-in-cell simulations demonstrate that employing a laser with a peak intensity of about $10^{23}$ W/cm$^2$ can obtain dense ($\gtrsim$ 10$^{18}$ cm$^{-3}$) polarized positrons with an average polarization degree of about 70\% and a yield of above 0.1 nC per shot. Moreover, our method is feasible using currently available or upcoming laser facilities and robust with respect to the laser and target parameters.
	Such high-density high-polarization positrons hold great significance in laboratory astrophysics, high-energy physics and new physics beyond the standard model.	
\end{abstract}

\maketitle

\setParDef
Ultrarelativistic spin-polarized positrons find wide use in laboratory astrophysics, high-energy physics, and new physics beyond the standard model~\cite{Leader2001,Zutic2004,Danielson2015,Ablikim2018,Remington2006,Moortgat2008}, such as simulating extreme cosmic environments in laboratories~\cite{Remington2006,novak2009,ruffini2010}, precisely measuring the effective weak mixing angle~\cite{BLONDEL1988145,DJOUADI20081}, and searching for supersymmetry particles and gravitons~\cite{boos2003,Bartl_2004,Bornhauser2012,Rizzo_2003}. These applications usually demand high density and high polarization of positrons. For example, relevant experiments in the International Linear Collider (ILC)~\cite{flottmann1993}, the Circular Electron Positron Collider (CEPC)~\cite{duan2019concepts} and the Jefferson Lab Electron Ion Collider (JLEIC)~\cite{lin2018polarized} require dense ($\sim$nanocoulombs per shot) highly polarized (30\% $\sim$ 60\%) positrons. 
There are mainly two methods used in experiments to obtain ultrarelativistic polarized positrons.
One is spontaneous polarization of positrons via the Sokolov-Ternov effect in storage rings, but it typically takes several minutes to hours due to the relatively weak magnetic field ($\sim$ Tesla)~\cite{Mane2005}. 
Another is the Bethe-Heitler (BH) electron-positron pair production~\cite{Heitler1954} in the interaction of circularly polarized $\gamma$ photons with high-$Z$ targets~\cite{Omori2006,Alexander2008,Abbott2016}, but the average polarization degree of positrons is only 30\%$\sim$40\% and the positron yield is limited to $10^{-6}$ nC per shot due to the low luminosity of $\gamma$ photon beams~\cite{dumas2009,dietrich2019status}.

The rapid development of modern ultraintense ultrashort laser facilities, with a record intensity of above $10^{23}$ ${\rm W}/{\rm cm}^2$~\cite{Kawanaka2016,Edwin2018,Danson2019,Yoon2021}, has led to the proposals of generating polarized positrons through laser-electron beam collisions in the strong-field quantum electrodynamics (QED) regime~\cite{ERBER1966,Ritus1985,baier1998,Yousef2006,sun2022,Fedotov2023,Piazza2012,Gonoskov2022,Wan2019,Chen2019,Xue2022,Dai2022,Li2020positron}.
This collision scheme mainly involves emitting $\gamma$ photons via the nonlinear Compton scattering (NCS) and producing positrons via the nonlinear Breit-Wheeler (NBW) process in the laser field~\cite{xie2017,Vranic2018,Zhao_2019,Dinu2020,Li2020Polarized,Xue2020,Torgrimsson2021}.
Positrons with a polarization degree of 30\%$\sim$40\% can be obtained by colliding an unpolarized GeV electron beam with an asymmetric laser pulse, such as an elliptically polarized~\cite{Wan2019} or bichromatic laser pulse~\cite{Chen2019}.  
However, the positron polarization degree is typically limited, since the parent photon polarization $\mathbf{P}_\gamma$ and the magnetic field $\mathbf{B}'$ in the positron rest frame are always nearly perpendicular (i.e., $\mathbf{P}_{\gamma}$ $\perp$ $\mathbf{B}'$). 
Positron polarization and yield are significantly influenced by the angle $\Theta$ between $\mathbf{P}_\gamma$ and $\mathbf{B}'$, reaching a maximum when $\mathbf{P}_{\gamma}$ $\parallel$ $\mathbf{B}'$  and minimum when $\mathbf{P}_{\gamma}$ $\perp$ $\mathbf{B}'$~\cite{ivanov2005,king2013,Seipt2020,Wan_2020,Xue2022,Dai2022}.
Alternatively, positrons with a polarization degree of 40\%$\sim$65\% can be obtained by a fully longitudinally  polarized GeV electron beam colliding with a circularly polarized laser pulse~\cite{Li2020positron}, but its applicability is limited by the flux and polarization of electron beams~\cite{adderley2010}. 
In the above laser-electron beam collision scheme, to achieve high density of positrons, dense GeV electron beams are envisioned to be obtained via plasma wakefield acceleration~\cite{bontoiu2023,pukhov2002,cho2018,Gschwendtner2019}, typically resulting in a yield of $10^{-4}$ nC per shot and a maximum density of $\sim10^{14}$ ${\rm cm}^{-3}$~\cite{Chen2019}. 
Moreover, achieving precise spatiotemporal synchronization between the laser and electron beams is also challenging.

By comparison, laser irradiation on solid targets can avoid the spatiotemporal synchronization issue, and is predicted to produce dense positrons in the  strong-field QED regime~\cite{Ridgers2012,Gu2016,Gu2018,Ji2014,Kostyukov2016,Zhu2016,Liu2016,Li2017,Liu2017,Zhang2021,fillipovic2022}. For instance, a 10 PW laser irradiating an aluminum foil can achieve a maximum positron density of $\sim10^{20}$ cm$^{-3}$~\cite{Ridgers2012}. 
Meanwhile, polarized positrons can be generated by a laser with a peak intensity exceeding $10^{24}$ ${\rm W}/{\rm cm}^2$ normally irradiating a foil target~\cite{Song2022}. In this method, positrons are polarized by an asymmetrical laser field in the skin layer of overdense plasma. However, the polarization degree is limited to about 30\% due to the intrinsic geometric relationship $\mathbf{P}_{\gamma}$ $\perp$ $\mathbf{B}'$, similar to Refs.~\cite{Wan2019,Chen2019}.
Therefore, it is still a great challenge to obtain dense polarized positrons with a high-polarization degree ($\gtrsim60\%$).

\begin{figure}	[htpb]	 
	\setlength{\belowcaptionskip}{-0.2cm}
	\centering
	\includegraphics[width=1.0\linewidth]{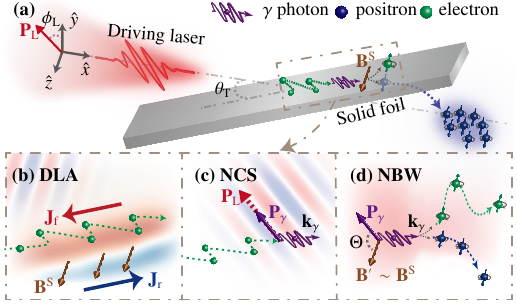}
	\caption{ (a) Interaction scenario. The laser propagates along $+\hat{{x}}$ and is polarized along $\mathbf{P}_{\rm L}$, with the laser polarization angle $\phi_{\rm L}$ between $\mathbf{P}_{\rm L}$ and $+\hat{{y}}$; the foil has a tilt angle $\theta_{\rm T}$ with respect to the $\hat{x}$-$\hat{z}$ plane.  
	(b) Direct laser acceleration (DLA). The brown, red, and blue arrows represent the directional quasistatic magnetic field $\mathbf{B}^{\rm S}$, the surface electric current $\mathbf{J}_{\rm f}$, and the return current $\mathbf{J}_{\rm r}$, respectively; $\mathbf{J}_{\rm f}$ and $\mathbf{J}_{\rm f}$ are both parallel to the front surface of the target; $\mathbf{B}^{\rm S}$ is oriented along $+\hat{{z}}$; the green-dotted line indicates the trajectory of the surface fast electrons; the heat maps in the top left and bottom right represent the incident laser field and the currents, respectively.
	(c) NCS process. The red and purple arrows represent the polarization of the laser and $\gamma$ photons, respectively; $\mathbf{k}_\gamma$ is the $\gamma$ photon wave vector; the heat map represents the reflected laser field, which propagates from left to right.
	(d) NBW process. The blue and green arrows represent the polarization directions of positrons and electrons, respectively, and the blue- and green-dotted lines represent their respective trajectories; the approximation of the magnetic field $\mathbf{B}'$ in the positron rest frame being equal to $\mathbf{B}^{\rm S}$ is employed; the heat map represents $\mathbf{B}^{\rm S}$.}
	\label{fig1}
\end{figure}

\begin{figure}[t]	
	\centering
	\setlength{\belowcaptionskip}{-0.2cm}
	\includegraphics[width=1.0\linewidth]{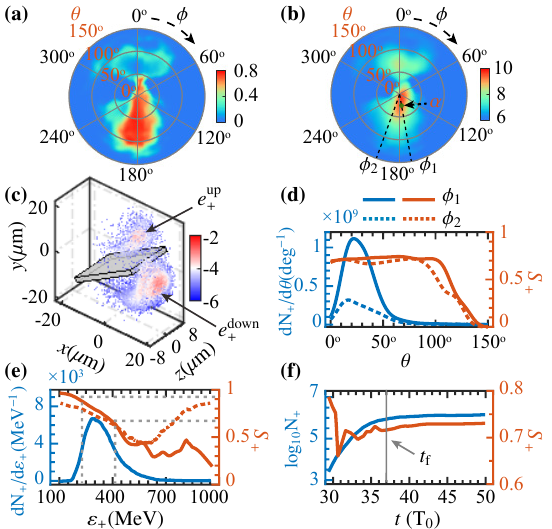}	
	\caption{(a) Angle-resolved positron polarization $S_+$ and (b) distribution $\log_{10}({\rm{d}}N_+/{\rm{d}} \Omega)$ with respect to the polar angle $\theta$ and the azimuth angle $\phi$, respectively. Here ${\rm d} \Omega$ = $\sin\theta{\rm d}\theta{\rm d}\phi$, $\theta$ = $0^{\circ}$ in $+\hat{{x}}$, and $\phi$ = $0^{\circ}$ in $+\hat{{y}}$. 
		(c) Positron density $\log_{10}(n_+/n_c)$. The gray box represents the initial position of the foil; {$e_+^{\rm down}$ represents positrons propagating downward through the target, while $e_+^{\rm up}$ represents positrons propagating upward along the target surface. }
		(d) Positron angle distribution ${\rm{d}}N_+/{\rm{d}}\theta$ and polarization $S_+$ vs $\theta$ at $\phi_{1}=176^{\circ}$ (solid lines) and $\phi_{2}=200^{\circ}$ (dotted lines). Here we collect positrons within a range of $\Delta\phi$ = 1$^\circ$.
		(e) Energy-resolved positron density ${\rm{d}}N_+/{\rm{d}}\varepsilon_+$ and polarization $S_+$ vs positron energy $\varepsilon_+$ at $\alpha$. The gray-dotted lines indicate the FWHM of positron energy; the red-dotted line represents the results of neglecting the radiative polarization effects of positrons. All above results are at simulation time $t$ = 50$T_0$. 
		(f) Positron yield $\log_{10}N_+$ and polarization $S_+$ vs $t$ at $\alpha$. $t_{\rm f}$ = $37T_0$ is the time of laser departure.}		
	\label{fig2}
\end{figure}

In this Letter, we investigate the generation of ultrarelativistic high-density high-polarization positrons via single-shot laser-foil interaction in the strong-field QED regime; see Fig.~\ref{fig1}(a). The driving laser ionizes the target and directly accelerates the emitted electrons~\cite{Sentoku1999,Gahn1999,Naumova2004,Nakamura2004,Nakamura2007,Chen2006, Li2006,tian2012,Thevenet2016}, and these electrons then emit abundant $\gamma$ photons via the NCS, which is dominated by the laser field. The emitted $\gamma$ photons subsequently generate polarized positrons via the NBW process, which is dominated by a strong self-generated quasistatic magnetic field $\mathbf{B}^{\rm S}$.
We find that placing the foil at an appropriate angle can result in a directional orientation of $\mathbf{B}^{\rm S}$ due to the electric currents along the foil surface [see Fig.~\ref{fig1}(b)], thereby polarizing positrons.
Manipulating the laser polarization direction can control the polarization of intermediate $\gamma$ photons $\mathbf{P}_\gamma$ [see Fig.~\ref{fig1}(c)], thereby controlling the angle $\Theta$ between $\mathbf{P}_\gamma$ and $\mathbf{B}^{\rm S}$, and significantly enhancing the polarization degree of positrons [see Fig.~\ref{fig1}(d)]. Under the influence of $\mathbf{B}^{\rm S}$, most positrons  move through the target, while newborn electrons propagate along the front surface of the target, ultimately mixing with unpolarized target electrons.
Our three-dimensional spin-resolved QED particle-in-cell (PIC) simulations show that using a laser with a peak intensity of about $10^{23}$ ${\rm W}/{\rm cm}^2$ can obtain dense ($\gtrsim 10^{18}$ cm$^{-3}$) transversely polarized  positrons with an average polarization degree of about 70\% and a yield of above 0.1 nC per shot; see Fig.~\ref{fig2}.  Our method is feasible using currently available or upcoming laser facilities, as it avoids the need for exact spatiotemporal synchronization. Moreover, our method is robust with respect to the laser and target parameters; see Appendix~\ref{impact}.

\setParDis

We utilize a Monte Carlo algorithm \cite{Wan2019,Xue2022,PIC_wan} in PIC code to investigate spin-resolved QED phenomena in laser-solid interactions.
Two primary QED processes, NCS and NBW, are implemented in the local constant field approximation~\cite{baier1998,Ritus1985,Fedotov2023}, which is valid for the invariant field parameter $a_0$ $\equiv$ $|e|E_0/mc\omega_0$ $\gg$ 1. 
These processes are characterized by two nonlinear QED parameters $\chi_e$ $\equiv$ $(|e|\hbar/m^3c^4)\sqrt{-(F_{\mu\nu}p^{\nu})^2}$ and $\chi_{\gamma}$ $\equiv$ $ (|e|\hbar/m^3c^4)\sqrt{-(F_{\mu\nu}k^{\nu})^2}$~\cite{Ritus1985,baier1998}, respectively. Here $e$ and $m$ are electron charge and mass, respectively, $E_0$ and $\omega_0$ are laser field amplitude and frequency, respectively, $c$ is the light speed in vacuum, $\hbar$ is the reduced Planck constant, $k^\nu$ and $p^\nu$ are the 4-momenta of $\gamma$ photon and electron (positron), respectively, and $F_{\mu\nu}$ is the field tensor.

Typical results are shown in Fig.~\ref{fig2}. The simulation box has dimensions of $x\times y\times z$ = 40 $\mu {\rm m}$ $\times$ 40 $\mu {\rm m}\times 30$ $\mu {\rm m}$, with the corresponding cells of 1000 $\times$ 1000 $\times$ 750. 
A linearly polarized laser pulse propagates along $+\hat{{x}}$, with a polarization angle $\phi_{\rm L}$ = $45^{\circ}$ [see Fig.~\ref{fig1}(a)], wavelength $\lambda_0$ = 1 $\mu {\rm m}$ and an envelope $a$ = $a_0$$\exp(-r^2/w_0^2)$$\exp[-(t-t_0)^2/\tau^2]$. 
Here $a_0$ = 500 [a smaller $a_0$ also works; see Fig.~\ref{figA1}(c) in Appendix~\ref{impact}] with a corresponding peak intensity of $3.4\times10^{23}$ ${\rm W}/{\rm cm}^2$, which can be achieved in upcoming 10 or 100 PW laser facilities~\cite{ELI,ECELS,ELI-beamlines,zou2015,Edwin2018,gales2018,Gan2021,du_shen_liang_wang_liu_li_2023}, $r$ = $\sqrt{y^2+z^2}$, $w_0$ = $5\lambda_0$ is the focal radius, $\tau$ = $6T_0$ is the pulse duration with a corresponding full width at half maximum (FWHM) $ \tau'$ $\approx$ $10T_0$, $T_0$ is the laser period, and $t_0$ = $12T_0$ is the time delay. ($t_0$ is used to shift the laser pulse out of the simulation box at time $t = 0$.)
A carbon foil with electron density $n_e$ = $550n_c$, length $L_{\rm T}$ = 30 $\mu {\rm m}$ and thickness $d_{\rm T}$ = 2 $\mu {\rm m}$ is placed in the box center at a tilt angle $\theta_{\rm{T}}$ = $30^{\circ}$, where $n_c$ = $m\omega_0^2/(4\pi e^2)$ $\approx$ 1.1 $\times$ $10^{21}$ ${\rm cm}^{-3}$ is the plasma critical density. Note that low-$Z$ target materials are required to effectively suppress the BH process~\cite{Heitler1954}.
Accounting for pulse leading edge or a prepulse~\cite{wagner2014}, a preplasma with density $n_{\rm pre}$ = $n_e$ $\exp(-L/L_{\rm pre})$ is used. Here $L$ is the distance between preplasma particles and the front surface of the target, and $L_{\rm pre}$ = 0.5 $\mu {\rm m}$ is the preplasma scale length. 
The numbers of macroparticles in each cell are 30 for electrons and 15 for fully ionized $\rm C^{6+}$.

\setParDef

\begin{figure}[t]	
	\centering	
	\setlength{\belowcaptionskip}{-0.2cm}	
	\includegraphics[width=1.0\linewidth]{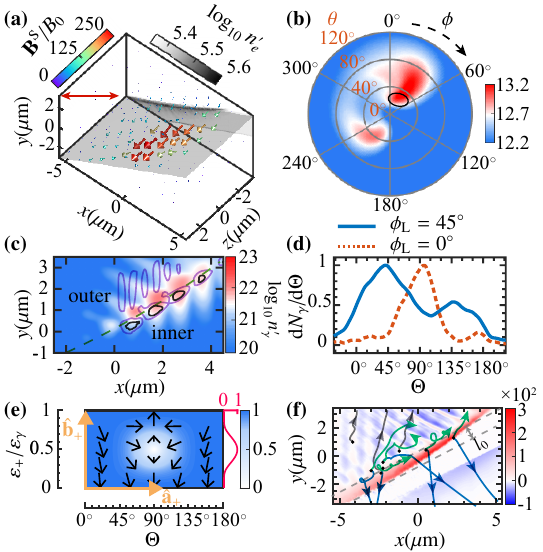}
	\caption{ (a) $\mathbf{B}^{\rm S}/B_0$ and projection of electron density $\log_{10}(n_e')$ in the $\hat{x}$-$\hat{y}$ plane. The red bidirectional arrow represents the laser polarization direction. 
		(b) Angle distribution of $\gamma$ photons $\log_{10}({\rm d}N_\gamma/{\rm d}\Omega)$ with respect to $\theta$ and $\phi$. Positron production mainly occurs within the black circle region ($10^\circ<\theta<30^\circ$, $0^\circ<\phi<60^\circ$)~\cite{SM}.  
		(c) Projection of photon density $\log_{10}(n_\gamma')$ for photon energy $\varepsilon_{\gamma}$ $>$ 300 MeV in the $\hat{x}$-$\hat{y}$ plane. The green-dashed line represents the front surface of the target; the black and purple lines represent the contour lines of the average nonlinear QED parameter $\overline{\chi}_{\gamma}$ = 0.35 and $\overline{\chi}_\gamma$ = 0.15, respectively.
		(d) Polarization-resolved distribution of $\gamma$ photons d$N_\gamma$/d$\Theta$ vs $\Theta$. $\mathbf{B}'$ $\approx$ $\varepsilon_{+}[\mathbf{B}^{\rm S}$ - $\hat{\mathbf{v}}_+(\hat{\mathbf{v}}_+$ $\cdot$ $\mathbf{B}^{\rm S})]$ and $\mathbf{B}^{\rm S}$ $\parallel$ $+\hat{\mathbf{z}}$ are employed; the $\gamma$ photons originate from the peak cone angles ($19^\circ<$ $\theta$ $<21^\circ$, $-1^\circ<$ $\phi$ $<1^\circ$) and ($24^\circ<$ $\theta$ $<26^\circ$, $25^\circ<$ $\phi$ $<27^\circ$) for the cases of $\phi_{\rm L}$ = $0^\circ$ (red dotted) and $\phi_{\rm L}$ = $45^\circ$ (blue solid), respectively.
		(e) $|\overline{\mathbf{S}}_+|$ with respect to $\Theta$ and $\varepsilon_{+}/\varepsilon_{\gamma}$. The black arrows represent $\overline{\mathbf{S}}_+$; the red line indicates the pair creation probability d$W$/d($\varepsilon_{+}$/$\varepsilon_{\gamma}$) for $\Theta$ = 90$^\circ$, normalized by its maximum; $\chi_{\gamma}$ = 0.4 is employed.
		(f) Projection of ${B}^{\rm S}_z/B_0$ in the $\hat{x}$-$\hat{y}$ plane. The blue, green and gray lines with arrows represent trajectories of the downward positrons $e_+^{\rm down}$, the newborn electrons and the upward positrons $e_+^{\rm up}$, respectively; $l_0$ is the spatial dimension of $\mathbf{B}^{\rm S}$. All of above simulation results are at $t=35T_0$.	}
	\label{fig3}
\end{figure}

The yield of positrons generated via the NBW process is roughly $8.0\times10^8$ ($\sim$ 0.13 nC), which is about 2 orders of magnitude larger than that of the BH process ($\sim$ 1.7 $\times$ $10^{6}$)(see Supplemental Material~\cite{SM}), and thus the contribution of the BH process is negligible.
Most positrons ($7.5\times10^8$ $e^+$$\sim$ 0.12 nC) travel downward the target [see $e_+^{\rm down}$ in Fig.~\ref{fig2}(c)], with an average polarization degree of 67.3\%, while a small amount propagate upward in the front of the target [see $e_+^{\rm up}$ in Fig.~\ref{fig2}(c)], with a slight polarization degree ($\sim$ 5.0\%); see Figs.~\ref{fig2}(a) and (b). 
Their distinct polarization properties arise from their different birth regions; see the reasons in Fig.~\ref{fig3}.  
As shown in Fig.~\ref{fig2}(c), the maximum positron density is above $10^{-3}n_c$ $\approx$ 10$^{18}$ ${\rm cm}^{-3}$, which is 4 orders of magnitude higher than these typically achieved in laser-electron beam collision schemes~\cite{Li2020positron,Wan2019,Chen2019}. 
At the peak angle of $\phi=176^\circ$, the positron polarization degree remains above 70\% for $0^\circ$ $\textless$ $\theta$ $\textless$ $100^\circ$; see Fig.~\ref{fig2}(d). 
Similar phenomena are observed at $\phi=200^\circ$.
The high polarization of positrons over a wide range of angles can be beneficial for detection.

Considering the capture and transfer of positrons for subsequent applications, we will focus on positrons within the peak cone angle $\alpha$ = ($25^\circ$ $\textless$ $\theta$ $\textless$ $27^\circ$, $175^\circ$ $\textless$ $\phi$ $\textless$ $177^\circ$); see Fig.~\ref{fig2}(b).
These positrons have an average polarization degree of about 72.3\%.
Their polarization degree $S_+$ decreases from 88.1\% to 62.6\% within the FWHM energy range of 225 MeV $<\varepsilon_{+}<$ 416 MeV, and $S_+$ $\approx$ 79.4\% at the energy peak of $\varepsilon_{+}$ $\approx$ 300 MeV; see Fig.~\ref{fig2}(e). 
These positrons have a yield of $1\times10^{6}$ ($\sim0.2$ pC), a flux of about $2\times10^{9}$ sr$^{-1}$ and an angle divergence of 35 $\times$ 35 mrad$^2$, with transverse and longitudinal sizes of 3 $\mu {\rm m}$ $\times$ 1.5 $\mu {\rm m}$ and 2 $\mu {\rm m}$ (at $t=50T_0$), respectively; see the comparisons with alternative sources in Appendix~\ref{impact}.
In principle, these positrons are suitable for injection into subsequent acceleration, such as radio-frequency accelerator and plasma wakefield acceleration~\cite{Alejo2019,Corde2015,Gonsalves2019}, owing to their sufficient charge and low divergence. Particularly, the small spatial size is desirable for injecting positrons in plasma wakefield acceleration due to the limited acceleration range of a few microns~\cite{Corde2015,Gonsalves2019}.
The brilliances are 0.65 $\times10^{19}$, 0.55 $\times10^{20}$, 0.42 $\times10^{20}$ and 0.17 $\times10^{20}$ $e_+/({\rm s}\cdot{\rm mm}^{2}\cdot{\rm mrad}^{2}\cdot0.1\% {\rm BW})$ for $\varepsilon_{+}=$ 200, 300, 400 and 500 MeV, respectively.
As shown in Fig.~\ref{fig2}(f), positrons are mainly produced during laser irradiation (27$T_0<t< 37T_0$), and their yield and polarization tend to stabilize after the laser departs. 
Moreover, the radiative polarization effects can enhance the polarization of the low-energy positrons as they pass through the directional magnetic field $\mathbf{B}^{\rm S}$; see the results of artificially neglecting the radiative polarization effects in Fig.~\ref{fig2}(e) and the corresponding reasons in Fig.~\ref{fig3}(e).
Additionally, the newborn electrons exhibit a high initial polarization, which gradually diminishes to a polarization degree of approximately 10\% due to the depolarization effects derived from the spin precession and the radiation during their propagation along the target surface ~\cite{SM,Thomas2020}.

\setParDis
The mechanisms for the production and polarization of positrons are shown in Fig.~\ref{fig3}. 
Upon laser irradiation, electrons are accelerated through $\mathbf{J} \times \mathbf{B}$ heating ~\cite{Brunel1987,Kruer1985} to form a fast electron current $\mathbf{J}_{\rm f}$ parallel to the front surface of the target; and a return current $\mathbf{J}_{\rm r}$ of cold electrons also appears inside the target to maintain charge balance, with a direction opposite to $\mathbf{J}_{\rm f}$~\cite{Chen2006,Li2006,zhang2007fast}; see Fig.~\ref{fig1}(b). Therefore, a quasistatic magnetic field $\mathbf{B}^{\rm S}$ is generated and oriented along $+\hat{{z}}$, with a peak strength ${B}^{\rm S}_{\rm max}\approx 250B_0 \approx 2.5 \times$ $10^6$ T, which is of the same order of magnitude as the magnetic field of the laser; see Fig.~\ref{fig3}(a). Here $B_0$ = $m\omega_0/|e|$ $\approx1\times10^4$ T. The high-density electron layer on the front surface of the target can divide the space into an outer region and an inner region; see Fig.~\ref{fig3}(c). 
In the inner region, the laser field rapidly decays within the skin depth and $\mathbf{B}^{\rm S}$ dominates the pair creation. 
When considering only $\mathbf{B}^{\rm S}$, one has $\chi_{\gamma}$ $\sim$ 0.36, resulting in abundant positron production. Here $\mathbf{B}^{\rm S}$ $\sim$ (0, 0, $B_z^{\rm S}$) $\approx$ (0, 0, 200$B_0$), $\hat{\mathbf{k}}_\gamma$ along ($\theta$, $\phi$) = (25$^\circ$, $26^\circ$), and $\varepsilon_{\gamma}$ = 400 MeV are employed (see Supplemental Material~\cite{SM}). In contrast, in the outer region, the laser field is much stronger than $\mathbf{B}^{\rm S}$, and the angle $\theta_{0}$ between the wave vectors of the laser and $\gamma$ photons is small.  Therefore, $\chi_{\gamma}$ $\propto$ $a_0(1-\cos\theta_{0})$ is typically small, leading to fewer pair creations than that of the inner region. Our simulations also support the difference in positron yields between the two regions, indicating that positron production mainly occurs for $\gamma$ photons that incident at a small angle with respect to the front surface of the target; see Fig.~\ref{fig3}(b).
\setParDef

In addition to dominating the positron production, $\mathbf{B}^{\rm s}$ also leads to the positron polarization. 
The average positron polarization $\overline{\mathbf{S}}_+$ can be written as~\cite{Xue2022}
\begin{eqnarray}
	\overline{\mathbf{S}}_+ =\frac{-(\frac{\varepsilon_{\gamma}}{\varepsilon_{+}}-\xi_{3}\frac{\varepsilon_{\gamma}}{\varepsilon_{-}}){K}_{\frac{1}{3}}(\rho)\hat{{\mathbf b}}_++\xi_{1}\frac{\varepsilon_{\gamma}}{\varepsilon_{-}}{ K}_{\frac{1}{3}}(\rho)\hat{{\mathbf a}}_+}{{\rm Int}K_{\frac{1}{3}}(\rho)+\frac{\varepsilon_{+}^2+\varepsilon_{-}^2}{\varepsilon_{+}\varepsilon_{-}}{ K}_{\frac{2}{3}}(\rho)-\xi_{3}{ K}_{\frac{2}{3}}(\rho)}
	\label{aveS}
\end{eqnarray}
where $\varepsilon_{-} = \varepsilon_{\gamma} - \varepsilon_{+}$ is the energy of newborn electrons, $\rho$ = 2$\varepsilon_{\gamma}^2/\left(3\chi_{\gamma}\varepsilon_{-}\varepsilon_{+}\right)$, ${\rm Int}K_{\frac{1}{3}}(\rho)$ $\equiv$ $ \int_{\rho}^{\infty} {\rm d}z {K}_{\frac{1}{3}}(z)$,  ${K}_n$ is the $n$-order modified Bessel function of the second kind, $\hat{{\mathbf b}}_+$ $=$ $\hat{\mathbf{v}}_+\times\hat{{\mathbf a}}_+/|\hat{\mathbf{v}}_+\times\hat{{\mathbf a}}_+|$ $\approx$ -$\mathbf{B}'/|\mathbf{B}'|$ with $\mathbf{B}'$ $\approx$ $\varepsilon_{+}[\mathbf{B}$ - $\hat{\mathbf{v}}_+\times\mathbf{E}$ - $\hat{\mathbf{v}}_+(\hat{\mathbf{v}}_+\cdot \mathbf{B})]$, $\hat{\mathbf{v}}_+$ and $\hat{{\mathbf a}}_+$ are the unit vectors along positron velocity and acceleration, respectively, $\mathbf{E}$ and $\mathbf{B}$ are the electric and magnetic fields, respectively. The photon polarization is characterized by the Stokes parameters ($\xi_{1}$, $\xi_{2}$, $\xi_{3}$) $\approx$ ($\sin2\Theta$, 0, -$\cos2\Theta$)~\cite{mcmaster1961} defined with respect to the axes $\hat{\mathbf{e} }_1$ $\equiv$ $[\mathbf{E}$ - $\hat{\mathbf{k}}_\gamma(\hat{\mathbf{k}}_\gamma\cdot\mathbf{E})$ + $\hat{\mathbf{k}}_\gamma\times\mathbf{B}]/|\mathbf{E}$ - $\hat{\mathbf{k}}_\gamma(\hat{\mathbf{k}}_\gamma\cdot\mathbf{E})$ + $\hat{\mathbf{k}}_\gamma\times\mathbf{B}|$ and $\hat{\mathbf{e} }_2$ $\equiv$ $\hat{\mathbf{k}}_\gamma\times\hat{\mathbf{e}}_1$ $\parallel$ $\mathbf{B}'$, $\hat{\mathbf{k}}_\gamma$ = $\mathbf{k}_\gamma/|\mathbf{k}_\gamma|$, and $\hat{\mathbf{v}}_+$ $\approx$ $\hat{\mathbf{k}}_\gamma$ is employed (the emission angle $\sim mc^2/\varepsilon_{\gamma}$ $\ll$ 1~\cite{Heitler1954}). As shown in Fig.~\ref{fig3}(e), $\overline{\mathbf{S}}_+$ always tends to align along $-\hat{{\mathbf b}}_+$, except in the vicinity of $\Theta = 90^\circ$, where the the reversal of $\overline{\mathbf{S}}_+$ around $\varepsilon_{+}/\varepsilon_{\gamma}$ = 0.5 results in a low average polarization degree. 
In the outer region, the temporal symmetry of the laser leads to roughly equal positron yields in each half cycle of the laser, but the positron polarization direction reverses with the reversal of the laser magnetic field direction, resulting in zero net polarization. In contrast, in the inner region, the directionality of $\mathbf{B}'$ $\sim$ $\mathbf{B}^{\rm s}$ can lead to net polarization.
Furthermore, due to the Lorentz force exerted by $\mathbf{B}^{\rm S}$, the positrons generated in the inner region mainly move through the target, i.e., the downward positrons $e_+^{\rm down}$; see the blue lines with arrows in Fig.~\ref{fig3}(f). Meanwhile, dominated by the laser field, the positrons produced in the outer region mainly propagate in the front of the target, i.e., the upward positrons $e_+^{\rm up}$; see the gray lines with arrows in Fig.~\ref{fig3}(f).

As previously discussed, the average polarization degree of positrons $\overline{S}_+$ is low when $\Theta$ = 90$^\circ$, whereas a high $\overline{S}_+$ can be obtained when $\Theta$ = 45$^\circ$; see Fig.~\ref{fig3}(e). 
Note that $\Theta$ can be manipulated by the laser polarization angle $\phi_{\rm L}$.
For most $\gamma$ photons, $\phi_{\rm L}$ + $\Theta$ $\approx$ $90^\circ$. 
As shown in Fig.~\ref{fig3}(d), for the cases of $\phi_{\rm L}$ = $0^\circ$ ($p$-polarized laser incident) and $\phi_{\rm L}$ = $45^\circ$, the $\gamma$ photons are mainly distributed around $\Theta$ = 90$^\circ$ and 45$^\circ$, respectively, leading to $\overline{S}_+$ $\approx$ 37\% and 70\%, respectively~\cite{SM}.
However, as $\phi_{\rm L}$ further increases, the polarization degree of $\gamma$ photons will decrease and the manipulation of $\Theta$ will be weaker. 

Furthermore, as positrons pass through the directional magnetic field $\mathbf{B}^{\rm S}$, the polarization effects of radiation cause positron spins to align preferentially with $\mathbf{B}^{\rm S}$. 
For positrons with $\varepsilon_{+}$/$\varepsilon_{\gamma}$ $<$ 0.5, $\overline{\mathbf{S}}_+$ at the positron creation moment is almost along $-\hat{{\mathbf b}}_+$ $\sim$ $\mathbf{B}^{\rm S}$ [see Fig.~\ref{fig3}(e)], leading to an increase in polarization upon radiation. 
On the contrary, for positrons with $\varepsilon_{+}$/$\varepsilon_{\gamma}$ $>$ 0.5, $\overline{\mathbf{S}}_+$ at the positron creation moment may not align with $\mathbf{B}^{\rm S}$ (such as $\Theta$ = 90$^\circ$ and 45$^\circ$, where $\overline{\mathbf{S}}_+$ points toward $\hat{{\mathbf b}}_+$ $\sim$ $-\mathbf{B}^{\rm S}$ and $+\hat{{\mathbf a}}$, respectively), leading to a decrease in polarization upon radiation. As a result, the average polarization degree $\overline{S}_+$ increases from 65.8\% to 67.3\%. Additionally, the depolarization caused by the spin precession of the downward positrons is relatively weak due to their rapid escape from the strong laser field under the influence of $\mathbf{B}^{\rm S}$~\cite{SM}.
The $\gamma$ photon attenuation (such as incoherent scatter, BH process, etc.) and the positron annihilation and collisions with other particles (such as Bhabha scattering, positron-nucleus scattering, etc.) during their propagation are estimated to be negligible~\cite{baier1998,berestetskii1982quantum,SM}.

\setParDis
For experimental feasibility, the impact of laser and target parameters on the yield $N_+$ and average polarization degree $\overline{S}_+$ of the downward positrons $e_+^{\rm down}$ has been investigated in Appendix~\ref{impact}, which shows that our method is robust with respect to the foil tilt angle $\theta_{\rm{T}}$, the laser polarization angle $\phi_{\rm L}$, the laser peak intensity $a_0$ and the preplasma scale length $L_{\rm pre}$. Moreover, the target configuration can also influence $N_+$ and $\overline{S}_+$. For example, using a conical target with the same laser pulse can remarkably increase the positron yield up to above 10 nC,  but at the cost of reducing positron polarization~\cite{SM}.
Additionally, the experimental implementation of a submicron preplasma usually requires a temporal contrast of about $10^{-13}\sim10^{-14}$ for ultraintense lasers with a peak intensity of about $10^{23}$ W/cm$^2$~\cite{SM,Lundh2007}, which would be generated in the near future via the utilization of plasma mirrors~\cite{Doumy2004,Levy2007}, plasma shutters~\cite{Reed2009}, frequency conversion~\cite{Mironov2009}, etc.

\setParDef

\setParDis
In conclusion, we put forward a novel method for generating ultrarelativistic high-density high-polarization positrons in the single-shot laser-foil interaction, simply by manipulating the laser polarization and the foil placement to generate a proper positron polarization field. The positrons generated by our method may be widely used in laboratory astrophysics, high-energy physics and new physics beyond the standard model, such as probing the spin-parity properties of hadrons~\cite{Aidala2013} and providing an alternative polarized positron source for polarized electron-positron colliders.
\setParDef

\setParDis
\textit{\textbf{Acknowledgement}} 
This work is supported by the National Natural Science Foundation of China (Grants  No. U2267204, No. 12022506, No. 12005305, and No. 12275209), the Foundation of Science and Technology on Plasma Physics Laboratory (No. JCKYS2021212008 and No. 6142A04220107), the Open Foundation of Key Laboratory of High Power Laser and Physics, Chinese Academy of Sciences (SGKF202101), the Shaanxi Fundamental Science Research Project for Mathematics and Physics (Grant No. 22JSY014), the Fundamental Research Funds for the Central Universities (No. xzy012023046), and the Foundation, China Institute of Atomic Energy (Grants No.
FY222506000201 and No. FC232412000201).\\
\setParDef

\begin{appendices}
\section{the experimental feasibility and positron qualities \label{impact}} 

\setlength {\parskip} {1pt}
\newcounter{Sfigure}
\setcounter{Sfigure}{1}
\renewcommand{\thefigure}{A\arabic{Sfigure}}

\begin{figure}	[htpt]	
	\setlength{\belowcaptionskip}{-0.2cm}
	\centering\includegraphics[width=1\linewidth]{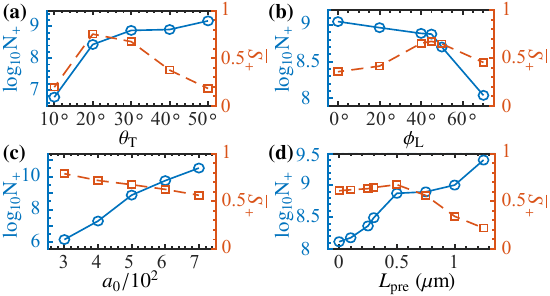}
	\caption{The yield $\log_{10}N_+$ and average polarization degree $\overline{S}_+$ of the downward positrons $e_+^{\rm down}$ vs (a) the foil tilt angle $\theta_{\rm{T}}$, (b) the laser polarization angle $\phi_{\rm L}$, (c) the laser peak intensity $a_0$ and (d) the preplasma scale length $L_{\rm pre}$, respectively. Other parameters are the same as those in Fig.\ref{fig2}.
	}
	\label{figA1}
	\addtocounter{Sfigure}{1}
\end{figure}

For experimental feasibility, the impact of laser and target parameters on the yield $N_+$ and average polarization degree $\overline{S}_+$ of the downward positrons $e_+^{\rm down}$ has been investigated in Fig.~\ref{figA1}. The target tilt angle $\theta_{\rm{T}}$ can affect the acceleration efficiency and angular distribution of fast electrons~\cite{zhang2007fast,Li2006}, and also affect the direction and magnitude of the quasistatic magnetic field $\mathbf{B}^{\rm S}$~\cite{Sentoku1999}, thereby further influencing $N_+$ and $\overline{S}_+$. 
If $\theta_{\rm{T}}$ is too small, the energy coupling efficiency between the laser and the target will be low, resulting in a weak quasistatic magnetic field $\mathbf{B}^{\rm S}$~\cite{Nakamura2007}. Consequently, both the polarization degree $\overline{S}_+$ and yield $N_+$ of positrons will be low; see Fig.~\ref{figA1}(a).
On the other hand, if $\theta_{\rm{T}}$ is too large, the propagation of fast electron current $\mathbf{J}_{\rm f}$ will gradually change from propagating along the front surface of the target to propagating into the target bulk, resulting in  oscillations of $\mathbf{B}^{\rm S}$ along $\hat{\mathbf{z}}$ and a reduction in its magnitude; see the magnitude and direction of $\mathbf{B}^{\rm S}$ at $\theta_{\rm{T}}=70^\circ$ in the Supplemental Material~\cite{SM}. Meanwhile, the propagation directions of the reflected laser and $\gamma$ photons change approximately from  parallel to  antiparallel, leading to the dominance of the reflected laser in pair production. Because of the symmetry of the laser field, the positrons generated in the laser field are almost unpolarized. Thus, a too large $\theta_{\rm{T}}$ will lead to a low  $\overline{S}_+$.
Our simulations show that for the given parameters, a high-polarization degree can be achieved when $20^\circ\lesssim$ $\theta_{\rm{T}}$ $\lesssim30^\circ$.
Additionally, when $\theta_{\rm{T}}$ increases slightly, $\mathbf{B}^{\rm S}$ dominates the positron creation and  $N_+$ $\propto$ $\chi_{\gamma}$ $\propto$ $\mathbf{B}^{\rm S}$ increases, while when $\theta_{\rm{T}}$ increases largely, the reflected laser dominates and $N_+$ $\propto$ $\chi_{\gamma}$ $\propto$ 1 $-$ $\cos\theta_{0}$ $\approx$ 1 $-$ $\cos(2\theta_{\rm T})$ increases. Consequently, as $\theta_{\rm{T}}$ increases, $N_+$ increases. 

Fig.~\ref{figA1}(b) shows the impact of the laser polarization angle $\phi_{\rm L}$ on $N_+$ and $\overline{S}_+$.
As $\phi_{\rm L}$ increases, the coupling efficiency between the laser and the target decreases, leading to a decrease in the magnitude of $\mathbf{B}^{\rm S}$~\cite{Sentoku1999,Nakamura2004}. Consequently, $N_+$ $\propto$ $\chi_{\gamma}$ $\propto$ $\mathbf{B}^{\rm S}$ decreases. Moreover, as  $\phi_{\rm L}$ increases, the direction of $\gamma$ photons $\mathbf{P}_\gamma$ gradually changes, leading to a reduction in the angle $\Theta$ between $\mathbf{P}_\gamma$ and $\mathbf{B}^{\rm S}$ from $90^\circ$; see Fig.~\ref{fig3}(d). As a result of this manipulation, $\overline{S}_+$ first increases.  However, as $\phi_{\rm L}$ further increases, the polarization degree of $\gamma$ photons will decrease, resulting in a weakened manipulation of $\Theta$, and thus $\overline{S}_+$ decreases.
Precise experimental control of $\theta_{\rm{T}}$ and $\phi_{\rm L}$ can be achieved by adjusting the foil placement via a two-dimensional motor.

Fig.~\ref{figA1}(c) shows the impact of the laser peak intensity $a_0$ on $N_+$ and $\overline{S}_+$. In the strong-field QED regime, as the laser peak intensity $a_0$ increases, the surface electric current $\mathbf{J}_{\rm f}$ also increases, resulting in strengthening $\mathbf{B}^{\rm S}$. Thus, $N_+$ $\propto$ $\chi_{\gamma}$ $\propto$ $\mathbf{B}^{\rm S}$ increases. 
Nevertheless, an excessively strong laser will severely damage the target~\cite{gamaly2002}, leading to a decrease in $\overline{S}_+$ due to the arrival of some upward positrons $e_+^{\rm up}$ at the back of the target.

Fig.~\ref{figA1}(d) shows the impact of the preplasma scale length $L_{\rm pre}$ on $N_+$ and $\overline{S}_+$. Increasing the preplasma scale length $L_{\rm pre}$ can amplify the surface electric current $\mathbf{J}_{\rm f}$, leading to an increase in the magnitude of $\mathbf{B}^{\rm S}$, thereby enhancing $N_+$ and $\overline{S}_+$. 
However, when $L_{\rm pre}$ increases largely, especially when $L_{\rm pre}>$ 0.5 $\mu$m for the given parameters, the resonance absorption will gradually dominate the electron acceleration~\cite{Chen2006,Nakamura2007}, resulting in the emergence of a strong current perpendicular to the target surface. Consequently, the directionality of $\mathbf{B}^{\rm S}$ deteriorates ~\cite{Li2006,ma2006}, leading to a decrease in $\overline{S}_+$.
Meanwhile, the spatial size of $\mathbf{B}^{\rm S}$, i.e., $l_0$ [see Fig. \ref{fig3}(f)], will increase, leading to electron energy $\varepsilon_e$ $\propto$ $l_0^2$ increases~\cite{Pukhov1999,Nakamura2007}, and thus $N_+$ $\propto$ $\chi_{\gamma}$ $\propto$ $\varepsilon_{\gamma}$ $\propto$ $\varepsilon_e$ increases.

Additionally, to facilitate a comparison with alternative sources of positrons driven by ultraintense lasers, we present a more comprehensive description on the qualities of positrons generated through our method. For subsequent applications, positrons within the peak cone angle $\alpha$ will be taken into account; see the selection of $\alpha$ in Fig.~\ref{fig2}(b). As shown in Fig.~\ref{fig2}(e), these positrons have an average polarization degree greater than 70\%, meeting the requirements for positron polarization in relevant experiments, such as ILC which acquires an average polarization degree of $\sim$60\%~\cite{Moortgat2008}. In contrast, positrons generated through the BH methods~\cite{Omori2006,Alexander2008,Abbott2016} and laser-electron beam collision methods~\cite{Dai2022,Xue2022,Wan2019,Chen2019,Li2020positron} typically have a polarization degree of 30\%$\sim$40\%. In addition, the peak flux of the positrons within $\alpha$ is about 1.5$\times$10$^{20}$ $e^+/$s, which is higher than the typical peak flux of about 10$^{17}\sim$10$^{19}$ $e^+/$s generated by the laser-electron beam collision methods~\cite{Wan2019,Chen2019,Li2020positron}, and significantly higher than the typical peak flux of 10$^{15}\sim$10$^{16}$ $e^+/$s produced by the BH methods~\cite{Omori2006,Alexander2008,Abbott2016}.

\end{appendices}

\bibliography{mybib}

\begin{thebibliography}{106}%
\makeatletter
\providecommand \@ifxundefined [1]{%
 \@ifx{#1\undefined}
}%
\providecommand \@ifnum [1]{%
 \ifnum #1\expandafter \@firstoftwo
 \else \expandafter \@secondoftwo
 \fi
}%
\providecommand \@ifx [1]{%
 \ifx #1\expandafter \@firstoftwo
 \else \expandafter \@secondoftwo
 \fi
}%
\providecommand \natexlab [1]{#1}%
\providecommand \enquote  [1]{``#1''}%
\providecommand \bibnamefont  [1]{#1}%
\providecommand \bibfnamefont [1]{#1}%
\providecommand \citenamefont [1]{#1}%
\providecommand \href@noop [0]{\@secondoftwo}%
\providecommand \href [0]{\begingroup \@sanitize@url \@href}%
\providecommand \@href[1]{\@@startlink{#1}\@@href}%
\providecommand \@@href[1]{\endgroup#1\@@endlink}%
\providecommand \@sanitize@url [0]{\catcode `\\12\catcode `\$12\catcode
  `\&12\catcode `\#12\catcode `\^12\catcode `\_12\catcode `\%12\relax}%
\providecommand \@@startlink[1]{}%
\providecommand \@@endlink[0]{}%
\providecommand \url  [0]{\begingroup\@sanitize@url \@url }%
\providecommand \@url [1]{\endgroup\@href {#1}{\urlprefix }}%
\providecommand \urlprefix  [0]{URL }%
\providecommand \Eprint [0]{\href }%
\providecommand \doibase [0]{http://dx.doi.org/}%
\providecommand \selectlanguage [0]{\@gobble}%
\providecommand \bibinfo  [0]{\@secondoftwo}%
\providecommand \bibfield  [0]{\@secondoftwo}%
\providecommand \translation [1]{[#1]}%
\providecommand \BibitemOpen [0]{}%
\providecommand \bibitemStop [0]{}%
\providecommand \bibitemNoStop [0]{.\EOS\space}%
\providecommand \EOS [0]{\spacefactor3000\relax}%
\providecommand \BibitemShut  [1]{\csname bibitem#1\endcsname}%
\let\auto@bib@innerbib\@empty
\bibitem [{\citenamefont {Leader}(2001)}]{Leader2001}%
  \BibitemOpen
  \bibfield  {author} {\bibinfo {author} {\bibfnamefont {Elliot}\ \bibnamefont
  {Leader}},\ }\href {\doibase 10.1017/CBO9780511524455} {\emph {\bibinfo
  {title} {Spin in Particle Physics}}},\ Cambridge Monographs on Particle
  Physics, Nuclear Physics and Cosmology\ (\bibinfo  {publisher} {Cambridge
  University Press},\ \bibinfo {address} {Cambridge, England},\ \bibinfo {year}
  {2001})\BibitemShut {NoStop}%
\bibitem [{\citenamefont {\ifmmode \check{Z}\else
  \v{Z}\fi{}uti\ifmmode~\acute{c}\else \'{c}\fi{}}\ \emph
  {et~al.}(2004)\citenamefont {\ifmmode \check{Z}\else
  \v{Z}\fi{}uti\ifmmode~\acute{c}\else \'{c}\fi{}}, \citenamefont {Fabian},\
  and\ \citenamefont {Das~Sarma}}]{Zutic2004}%
  \BibitemOpen
  \bibfield  {author} {\bibinfo {author} {\bibfnamefont {Igor}\ \bibnamefont
  {\ifmmode \check{Z}\else \v{Z}\fi{}uti\ifmmode~\acute{c}\else \'{c}\fi{}}},
  \bibinfo {author} {\bibfnamefont {Jaroslav}\ \bibnamefont {Fabian}}, \ and\
  \bibinfo {author} {\bibfnamefont {S.}~\bibnamefont {Das~Sarma}},\ }\bibfield
  {title} {\enquote {\bibinfo {title} {Spintronics: Fundamentals and
  applications},}\ }\href {\doibase 10.1103/RevModPhys.76.323} {\bibfield
  {journal} {\bibinfo  {journal} {Rev. Mod. Phys.}\ }\textbf {\bibinfo {volume}
  {76}},\ \bibinfo {pages} {323--410} (\bibinfo {year} {2004})}\BibitemShut
  {NoStop}%
\bibitem [{\citenamefont {Danielson}\ \emph {et~al.}(2015)\citenamefont
  {Danielson}, \citenamefont {Dubin}, \citenamefont {Greaves},\ and\
  \citenamefont {Surko}}]{Danielson2015}%
  \BibitemOpen
  \bibfield  {author} {\bibinfo {author} {\bibfnamefont {J.~R.}\ \bibnamefont
  {Danielson}}, \bibinfo {author} {\bibfnamefont {D.~H.~E.}\ \bibnamefont
  {Dubin}}, \bibinfo {author} {\bibfnamefont {R.~G.}\ \bibnamefont {Greaves}},
  \ and\ \bibinfo {author} {\bibfnamefont {C.~M.}\ \bibnamefont {Surko}},\
  }\bibfield  {title} {\enquote {\bibinfo {title} {Plasma and trap-based
  techniques for science with positrons},}\ }\href {\doibase
  10.1103/RevModPhys.87.247} {\bibfield  {journal} {\bibinfo  {journal} {Rev.
  Mod. Phys.}\ }\textbf {\bibinfo {volume} {87}},\ \bibinfo {pages} {247--306}
  (\bibinfo {year} {2015})}\BibitemShut {NoStop}%
\bibitem [{\citenamefont {Ablikim}\ \emph {et~al.}(2019)\citenamefont
  {Ablikim}, \citenamefont {Achasov}, \citenamefont {Ahmed},\ and\
  \citenamefont {\textit{et al.}}}]{Ablikim2018}%
  \BibitemOpen
  \bibfield  {author} {\bibinfo {author} {\bibfnamefont {M.}~\bibnamefont
  {Ablikim}}, \bibinfo {author} {\bibfnamefont {M.N.}\ \bibnamefont {Achasov}},
  \bibinfo {author} {\bibfnamefont {S.}~\bibnamefont {Ahmed}}, \ and\ \bibinfo
  {author} {\bibnamefont {\textit{et al.}}} (\bibinfo {collaboration} {The
  BESIII Collaboration}),\ }\bibfield  {title} {\enquote {\bibinfo {title}
  {{Polarization and Entanglement in Baryon-Antibaryon Pair Production in
  Electron-Positron Annihilation}},}\ }\href {\doibase
  10.1038/s41567-019-0494-8} {\bibfield  {journal} {\bibinfo  {journal} {Nat.
  Phys.}\ }\textbf {\bibinfo {volume} {15}},\ \bibinfo {pages} {631--634}
  (\bibinfo {year} {2019})}\BibitemShut {NoStop}%
\bibitem [{\citenamefont {Remington}\ \emph {et~al.}(2006)\citenamefont
  {Remington}, \citenamefont {Drake},\ and\ \citenamefont
  {Ryutov}}]{Remington2006}%
  \BibitemOpen
  \bibfield  {author} {\bibinfo {author} {\bibfnamefont {Bruce~A.}\
  \bibnamefont {Remington}}, \bibinfo {author} {\bibfnamefont {R.~Paul}\
  \bibnamefont {Drake}}, \ and\ \bibinfo {author} {\bibfnamefont {Dmitri~D.}\
  \bibnamefont {Ryutov}},\ }\bibfield  {title} {\enquote {\bibinfo {title}
  {Experimental astrophysics with high power lasers and $z$ pinches},}\ }\href
  {\doibase 10.1103/RevModPhys.78.755} {\bibfield  {journal} {\bibinfo
  {journal} {Rev. Mod. Phys.}\ }\textbf {\bibinfo {volume} {78}},\ \bibinfo
  {pages} {755--807} (\bibinfo {year} {2006})}\BibitemShut {NoStop}%
\bibitem [{\citenamefont {Moortgat-Pick}\ \emph {et~al.}(2008)\citenamefont
  {Moortgat-Pick}, \citenamefont {Abe}, \citenamefont {Alexander},
  \citenamefont {Ananthanarayan}, \citenamefont {Babich}, \citenamefont
  {Bharadwaj}, \citenamefont {Barber}, \citenamefont {Bartl}, \citenamefont
  {Brachmann}, \citenamefont {Chen},\ and\ \citenamefont {\textit{et
  al.}}}]{Moortgat2008}%
  \BibitemOpen
  \bibfield  {author} {\bibinfo {author} {\bibfnamefont {G.}~\bibnamefont
  {Moortgat-Pick}}, \bibinfo {author} {\bibfnamefont {T.}~\bibnamefont {Abe}},
  \bibinfo {author} {\bibfnamefont {G.}~\bibnamefont {Alexander}}, \bibinfo
  {author} {\bibfnamefont {B.}~\bibnamefont {Ananthanarayan}}, \bibinfo
  {author} {\bibfnamefont {A.~A.}\ \bibnamefont {Babich}}, \bibinfo {author}
  {\bibfnamefont {V.}~\bibnamefont {Bharadwaj}}, \bibinfo {author}
  {\bibfnamefont {D.}~\bibnamefont {Barber}}, \bibinfo {author} {\bibfnamefont
  {A.}~\bibnamefont {Bartl}}, \bibinfo {author} {\bibfnamefont
  {A.}~\bibnamefont {Brachmann}}, \bibinfo {author} {\bibfnamefont
  {Si}~\bibnamefont {Chen}}, \ and\ \bibinfo {author} {\bibnamefont {\textit{et
  al.}}},\ }\bibfield  {title} {\enquote {\bibinfo {title} {Polarized positrons
  and electrons at the linear collider},}\ }\href {\doibase
  https://doi.org/10.1016/j.physrep.2007.12.003} {\bibfield  {journal}
  {\bibinfo  {journal} {Phys. Rep.}\ }\textbf {\bibinfo {volume} {460}},\
  \bibinfo {pages} {131 -- 243} (\bibinfo {year} {2008})}\BibitemShut {NoStop}%
\bibitem [{\citenamefont {Novak}\ and\ \citenamefont
  {Kholodov}(2009)}]{novak2009}%
  \BibitemOpen
  \bibfield  {author} {\bibinfo {author} {\bibfnamefont {O.~P.}\ \bibnamefont
  {Novak}}\ and\ \bibinfo {author} {\bibfnamefont {R.~I.}\ \bibnamefont
  {Kholodov}},\ }\bibfield  {title} {\enquote {\bibinfo {title}
  {Spin-polarization effects in the processes of synchrotron radiation and
  electron-positron pair production by a photon in a magnetic field},}\ }\href
  {\doibase 10.1103/PhysRevD.80.025025} {\bibfield  {journal} {\bibinfo
  {journal} {Phys. Rev. D}\ }\textbf {\bibinfo {volume} {80}},\ \bibinfo
  {pages} {025025} (\bibinfo {year} {2009})}\BibitemShut {NoStop}%
\bibitem [{\citenamefont {Ruffini}\ \emph {et~al.}(2010)\citenamefont
  {Ruffini}, \citenamefont {Vereshchagin},\ and\ \citenamefont
  {Xue}}]{ruffini2010}%
  \BibitemOpen
  \bibfield  {author} {\bibinfo {author} {\bibfnamefont {Remo}\ \bibnamefont
  {Ruffini}}, \bibinfo {author} {\bibfnamefont {Gregory}\ \bibnamefont
  {Vereshchagin}}, \ and\ \bibinfo {author} {\bibfnamefont {She-Sheng}\
  \bibnamefont {Xue}},\ }\bibfield  {title} {\enquote {\bibinfo {title}
  {Electron–positron pairs in physics and astrophysics: From heavy nuclei to
  black holes},}\ }\href {\doibase
  https://doi.org/10.1016/j.physrep.2009.10.004} {\bibfield  {journal}
  {\bibinfo  {journal} {Phys. Rep.}\ }\textbf {\bibinfo {volume} {487}},\
  \bibinfo {pages} {1--140} (\bibinfo {year} {2010})}\BibitemShut {NoStop}%
\bibitem [{\citenamefont {Blondel}(1988)}]{BLONDEL1988145}%
  \BibitemOpen
  \bibfield  {author} {\bibinfo {author} {\bibfnamefont {Alain}\ \bibnamefont
  {Blondel}},\ }\bibfield  {title} {\enquote {\bibinfo {title} {{A scheme to
  measure the polarization asymmetry at the z pole in LEP}},}\ }\href {\doibase
  https://doi.org/10.1016/0370-2693(88)90869-6} {\bibfield  {journal} {\bibinfo
   {journal} {Phys. Lett. B}\ }\textbf {\bibinfo {volume} {202}},\ \bibinfo
  {pages} {145--148} (\bibinfo {year} {1988})}\BibitemShut {NoStop}%
\bibitem [{\citenamefont {Djouadi}(2008)}]{DJOUADI20081}%
  \BibitemOpen
  \bibfield  {author} {\bibinfo {author} {\bibfnamefont {Abdelhak}\
  \bibnamefont {Djouadi}},\ }\bibfield  {title} {\enquote {\bibinfo {title}
  {{The anatomy of electroweak symmetry breaking: Tome
  \uppercase\expandafter{\romannumeral1}: The Higgs boson in the Standard
  Model}},}\ }\href {\doibase https://doi.org/10.1016/j.physrep.2007.10.004}
  {\bibfield  {journal} {\bibinfo  {journal} {Phys. Rep.}\ }\textbf {\bibinfo
  {volume} {457}},\ \bibinfo {pages} {1--216} (\bibinfo {year}
  {2008})}\BibitemShut {NoStop}%
\bibitem [{\citenamefont {Boos}\ \emph {et~al.}(2003)\citenamefont {Boos},
  \citenamefont {Martyn}, \citenamefont {Moortgat-Pick}, \citenamefont
  {Sachwitz}, \citenamefont {Sherstnev},\ and\ \citenamefont
  {Zerwas}}]{boos2003}%
  \BibitemOpen
  \bibfield  {author} {\bibinfo {author} {\bibfnamefont {E.}~\bibnamefont
  {Boos}}, \bibinfo {author} {\bibfnamefont {H.~U.}\ \bibnamefont {Martyn}},
  \bibinfo {author} {\bibfnamefont {G.}~\bibnamefont {Moortgat-Pick}}, \bibinfo
  {author} {\bibfnamefont {M.}~\bibnamefont {Sachwitz}}, \bibinfo {author}
  {\bibfnamefont {A.}~\bibnamefont {Sherstnev}}, \ and\ \bibinfo {author}
  {\bibfnamefont {P.~M.}\ \bibnamefont {Zerwas}},\ }\bibfield  {title}
  {\enquote {\bibinfo {title} {Polarisation in sfermion decays: determining
  $\tan\beta$ and trilinear couplings},}\ }\href {\doibase
  10.1140/epjc/s2003-01288-y} {\bibfield  {journal} {\bibinfo  {journal} {Eur.
  Phys. J. C}\ }\textbf {\bibinfo {volume} {30}},\ \bibinfo {pages} {395--407}
  (\bibinfo {year} {2003})}\BibitemShut {NoStop}%
\bibitem [{\citenamefont {Bartl}\ \emph {et~al.}(2004)\citenamefont {Bartl},
  \citenamefont {Hesselbach}, \citenamefont {Hohenwarter-Sodek}, \citenamefont
  {Fraas},\ and\ \citenamefont {Moortgat-Pick}}]{Bartl_2004}%
  \BibitemOpen
  \bibfield  {author} {\bibinfo {author} {\bibfnamefont {Alfred}\ \bibnamefont
  {Bartl}}, \bibinfo {author} {\bibfnamefont {Stefan}\ \bibnamefont
  {Hesselbach}}, \bibinfo {author} {\bibfnamefont {Karl}\ \bibnamefont
  {Hohenwarter-Sodek}}, \bibinfo {author} {\bibfnamefont {Hans}\ \bibnamefont
  {Fraas}}, \ and\ \bibinfo {author} {\bibfnamefont {Gudrid}\ \bibnamefont
  {Moortgat-Pick}},\ }\bibfield  {title} {\enquote {\bibinfo {title} {{A T-odd
  asymmetry in neutralino production and decay}},}\ }\href {\doibase
  10.1088/1126-6708/2004/08/038} {\bibfield  {journal} {\bibinfo  {journal} {J.
  High Energy Phys.}\ }\textbf {\bibinfo {volume} {2004}},\ \bibinfo {pages}
  {038} (\bibinfo {year} {2004})}\BibitemShut {NoStop}%
\bibitem [{\citenamefont {Bornhauser}\ \emph {et~al.}(2012)\citenamefont
  {Bornhauser}, \citenamefont {Drees}, \citenamefont {Dreiner}, \citenamefont
  {Éboli}, \citenamefont {Kim},\ and\ \citenamefont
  {Kittel}}]{Bornhauser2012}%
  \BibitemOpen
  \bibfield  {author} {\bibinfo {author} {\bibfnamefont {S.}~\bibnamefont
  {Bornhauser}}, \bibinfo {author} {\bibfnamefont {M.}~\bibnamefont {Drees}},
  \bibinfo {author} {\bibfnamefont {H.}~\bibnamefont {Dreiner}}, \bibinfo
  {author} {\bibfnamefont {O.~J.~P.}\ \bibnamefont {Éboli}}, \bibinfo {author}
  {\bibfnamefont {J.~S.}\ \bibnamefont {Kim}}, \ and\ \bibinfo {author}
  {\bibfnamefont {O.}~\bibnamefont {Kittel}},\ }\bibfield  {title} {\enquote
  {\bibinfo {title} {{CP asymmetries in the supersymmetric trilepton signal at
  the LHC}},}\ }\href {\doibase 10.1140/epjc/s10052-012-1887-3} {\bibfield
  {journal} {\bibinfo  {journal} {Eur. Phys. J. C}\ }\textbf {\bibinfo {volume}
  {72}},\ \bibinfo {pages} {1887} (\bibinfo {year} {2012})}\BibitemShut
  {NoStop}%
\bibitem [{\citenamefont {Rizzo}(2003)}]{Rizzo_2003}%
  \BibitemOpen
  \bibfield  {author} {\bibinfo {author} {\bibfnamefont {Thomas~G.}\
  \bibnamefont {Rizzo}},\ }\bibfield  {title} {\enquote {\bibinfo {title}
  {Transverse polarization signatures of extra dimensions at linear
  colliders},}\ }\href {\doibase 10.1088/1126-6708/2003/02/008} {\bibfield
  {journal} {\bibinfo  {journal} {J. High Energy Phys.}\ }\textbf {\bibinfo
  {volume} {2003}},\ \bibinfo {pages} {008} (\bibinfo {year}
  {2003})}\BibitemShut {NoStop}%
\bibitem [{\citenamefont {Fl{\"o}ttmann}(1993)}]{flottmann1993}%
  \BibitemOpen
  \bibfield  {author} {\bibinfo {author} {\bibfnamefont {Klaus}\ \bibnamefont
  {Fl{\"o}ttmann}},\ }\href@noop {} {\emph {\bibinfo {title} {Investigations
  toward the development of polarized and unpolarized high intensity positron
  sources for linear colliders}}},\ Vol.~\bibinfo {volume} {93}\ (\bibinfo
  {publisher} {DESY},\ \bibinfo {address} {Berlin},\ \bibinfo {year}
  {1993})\BibitemShut {NoStop}%
\bibitem [{\citenamefont {Duan}\ \emph {et~al.}(2019)\citenamefont {Duan},
  \citenamefont {Gao}, \citenamefont {Li}, \citenamefont {Wang}, \citenamefont
  {Wang}, \citenamefont {Xia}, \citenamefont {Xu}, \citenamefont {Yu},
  \citenamefont {Zhang},\ and\ \citenamefont {\textit{et
  al.}}}]{duan2019concepts}%
  \BibitemOpen
  \bibfield  {author} {\bibinfo {author} {\bibfnamefont {Zhe}\ \bibnamefont
  {Duan}}, \bibinfo {author} {\bibfnamefont {Jie}\ \bibnamefont {Gao}},
  \bibinfo {author} {\bibfnamefont {XP}~\bibnamefont {Li}}, \bibinfo {author}
  {\bibfnamefont {Dou}\ \bibnamefont {Wang}}, \bibinfo {author} {\bibfnamefont
  {Yiwei}\ \bibnamefont {Wang}}, \bibinfo {author} {\bibfnamefont {Wenhao}\
  \bibnamefont {Xia}}, \bibinfo {author} {\bibfnamefont {Qingjin}\ \bibnamefont
  {Xu}}, \bibinfo {author} {\bibfnamefont {Chenghui}\ \bibnamefont {Yu}},
  \bibinfo {author} {\bibfnamefont {Yuan}\ \bibnamefont {Zhang}}, \ and\
  \bibinfo {author} {\bibnamefont {\textit{et al.}}},\ }\bibfield  {title}
  {\enquote {\bibinfo {title} {Concepts of longitudinally polarized electron
  and positron colliding beams in the circular electron positron collider},}\
  }in\ \href {\doibase 10.18429/JACoW-IPAC2019-MOPMP012} {\emph {\bibinfo
  {booktitle} {10th Int. Particle Accelerator Conf.(IPAC'19), Melbourne,
  Australia, 19-24 May 2019}}}\ (\bibinfo {organization} {JACOW Publishing,
  Geneva, Switzerland},\ \bibinfo {year} {2019})\ pp.\ \bibinfo {pages}
  {445--448}\BibitemShut {NoStop}%
\bibitem [{\citenamefont {Lin}\ \emph {et~al.}(2018)\citenamefont {Lin},
  \citenamefont {Grames}, \citenamefont {Guo}, \citenamefont {Morozov},\ and\
  \citenamefont {Zhang}}]{lin2018polarized}%
  \BibitemOpen
  \bibfield  {author} {\bibinfo {author} {\bibfnamefont {Fanglei}\ \bibnamefont
  {Lin}}, \bibinfo {author} {\bibfnamefont {Joe}\ \bibnamefont {Grames}},
  \bibinfo {author} {\bibfnamefont {Jiquan}\ \bibnamefont {Guo}}, \bibinfo
  {author} {\bibfnamefont {Vasiliy}\ \bibnamefont {Morozov}}, \ and\ \bibinfo
  {author} {\bibfnamefont {Yuhong}\ \bibnamefont {Zhang}},\ }\bibfield  {title}
  {\enquote {\bibinfo {title} {{Polarized positrons in Jefferson Lab electron
  ion collider (JLEIC)}},}\ }in\ \href
  {https://aip.scitation.org/doi/abs/10.1063/1.5040224} {\emph {\bibinfo
  {booktitle} {AIP Conf. Proc.}}},\ Vol.\ \bibinfo {volume} {1970}\ (\bibinfo
  {organization} {AIP Publishing LLC},\ \bibinfo {year} {2018})\ p.\ \bibinfo
  {pages} {050005}\BibitemShut {NoStop}%
\bibitem [{\citenamefont {Mane}\ \emph {et~al.}(2005)\citenamefont {Mane},
  \citenamefont {Shatunov},\ and\ \citenamefont {Yokoya}}]{Mane2005}%
  \BibitemOpen
  \bibfield  {author} {\bibinfo {author} {\bibfnamefont {S.~R.}\ \bibnamefont
  {Mane}}, \bibinfo {author} {\bibfnamefont {Yu~M.}\ \bibnamefont {Shatunov}},
  \ and\ \bibinfo {author} {\bibfnamefont {K.}~\bibnamefont {Yokoya}},\
  }\bibfield  {title} {\enquote {\bibinfo {title} {Spin-polarized charged
  particle beams in high-energy accelerators},}\ }\href {\doibase
  10.1088/0034-4885/68/9/R01} {\bibfield  {journal} {\bibinfo  {journal} {Rep.
  Prog. Phys.}\ }\textbf {\bibinfo {volume} {68}},\ \bibinfo {pages} {1997}
  (\bibinfo {year} {2005})}\BibitemShut {NoStop}%
\bibitem [{\citenamefont {Heitler}(1954)}]{Heitler1954}%
  \BibitemOpen
  \bibfield  {author} {\bibinfo {author} {\bibfnamefont {W.}~\bibnamefont
  {Heitler}},\ }\href@noop {} {\emph {\bibinfo {title} {The Quantum Theory of
  Radiation}}}\ (\bibinfo  {publisher} {Clarendon Press, Oxford},\ \bibinfo
  {year} {1954})\BibitemShut {NoStop}%
\bibitem [{\citenamefont {Omori}\ \emph {et~al.}(2006)\citenamefont {Omori},
  \citenamefont {Fukuda}, \citenamefont {Hirose}, \citenamefont {Kurihara},
  \citenamefont {Kuroda}, \citenamefont {Nomura}, \citenamefont {Ohashi},
  \citenamefont {Okugi}, \citenamefont {Sakaue}, \citenamefont {Saito},
  \citenamefont {Urakawa}, \citenamefont {Washio},\ and\ \citenamefont
  {Yamazaki}}]{Omori2006}%
  \BibitemOpen
  \bibfield  {author} {\bibinfo {author} {\bibfnamefont {T.}~\bibnamefont
  {Omori}}, \bibinfo {author} {\bibfnamefont {M.}~\bibnamefont {Fukuda}},
  \bibinfo {author} {\bibfnamefont {T.}~\bibnamefont {Hirose}}, \bibinfo
  {author} {\bibfnamefont {Y.}~\bibnamefont {Kurihara}}, \bibinfo {author}
  {\bibfnamefont {R.}~\bibnamefont {Kuroda}}, \bibinfo {author} {\bibfnamefont
  {M.}~\bibnamefont {Nomura}}, \bibinfo {author} {\bibfnamefont
  {A.}~\bibnamefont {Ohashi}}, \bibinfo {author} {\bibfnamefont
  {T.}~\bibnamefont {Okugi}}, \bibinfo {author} {\bibfnamefont
  {K.}~\bibnamefont {Sakaue}}, \bibinfo {author} {\bibfnamefont
  {T.}~\bibnamefont {Saito}}, \bibinfo {author} {\bibfnamefont
  {J.}~\bibnamefont {Urakawa}}, \bibinfo {author} {\bibfnamefont
  {M.}~\bibnamefont {Washio}}, \ and\ \bibinfo {author} {\bibfnamefont
  {I.}~\bibnamefont {Yamazaki}},\ }\bibfield  {title} {\enquote {\bibinfo
  {title} {{Efficient Propagation of Polarization from Laser Photons to
  Positrons through Compton Scattering and Electron-Positron Pair Creation}},}\
  }\href {\doibase 10.1103/PhysRevLett.96.114801} {\bibfield  {journal}
  {\bibinfo  {journal} {Phys. Rev. Lett.}\ }\textbf {\bibinfo {volume} {96}},\
  \bibinfo {pages} {114801} (\bibinfo {year} {2006})}\BibitemShut {NoStop}%
\bibitem [{\citenamefont {Alexander}\ \emph {et~al.}(2008)\citenamefont
  {Alexander}, \citenamefont {Barley}, \citenamefont {Batygin},\ and\
  \citenamefont {\textit{et al.}}}]{Alexander2008}%
  \BibitemOpen
  \bibfield  {author} {\bibinfo {author} {\bibfnamefont {G.}~\bibnamefont
  {Alexander}}, \bibinfo {author} {\bibfnamefont {J.}~\bibnamefont {Barley}},
  \bibinfo {author} {\bibfnamefont {Y.}~\bibnamefont {Batygin}}, \ and\
  \bibinfo {author} {\bibnamefont {\textit{et al.}}},\ }\bibfield  {title}
  {\enquote {\bibinfo {title} {Observation of polarized positrons from an
  undulator-based source},}\ }\href {\doibase 10.1103/PhysRevLett.100.210801}
  {\bibfield  {journal} {\bibinfo  {journal} {Phys. Rev. Lett.}\ }\textbf
  {\bibinfo {volume} {100}},\ \bibinfo {pages} {210801} (\bibinfo {year}
  {2008})}\BibitemShut {NoStop}%
\bibitem [{\citenamefont {Abbott}\ \emph {et~al.}(2016)\citenamefont {Abbott},
  \citenamefont {Adderley}, \citenamefont {Adeyemi},\ and\ \citenamefont {{\it
  et al.}}}]{Abbott2016}%
  \BibitemOpen
  \bibfield  {author} {\bibinfo {author} {\bibfnamefont {D.}~\bibnamefont
  {Abbott}}, \bibinfo {author} {\bibfnamefont {P.}~\bibnamefont {Adderley}},
  \bibinfo {author} {\bibfnamefont {A.}~\bibnamefont {Adeyemi}}, \ and\
  \bibinfo {author} {\bibnamefont {{\it et al.}}} (\bibinfo {collaboration}
  {PEPPo Collaboration}),\ }\bibfield  {title} {\enquote {\bibinfo {title}
  {{Production of Highly Polarized Positrons Using Polarized Electrons at MeV
  Energies}},}\ }\href {\doibase 10.1103/PhysRevLett.116.214801} {\bibfield
  {journal} {\bibinfo  {journal} {Phys. Rev. Lett.}\ }\textbf {\bibinfo
  {volume} {116}},\ \bibinfo {pages} {214801} (\bibinfo {year}
  {2016})}\BibitemShut {NoStop}%
\bibitem [{\citenamefont {Dumas}\ \emph {et~al.}(2009)\citenamefont {Dumas},
  \citenamefont {Grames},\ and\ \citenamefont {Voutier}}]{dumas2009}%
  \BibitemOpen
  \bibfield  {author} {\bibinfo {author} {\bibfnamefont {J.}~\bibnamefont
  {Dumas}}, \bibinfo {author} {\bibfnamefont {J.}~\bibnamefont {Grames}}, \
  and\ \bibinfo {author} {\bibfnamefont {E.}~\bibnamefont {Voutier}},\
  }\bibfield  {title} {\enquote {\bibinfo {title} {{Polarized Positrons at
  Jefferson Lab}},}\ }\href {\doibase 10.1063/1.3215617} {\bibfield  {journal}
  {\bibinfo  {journal} {AIP Conf. Proc.}\ }\textbf {\bibinfo {volume} {1149}},\
  \bibinfo {pages} {1184--1188} (\bibinfo {year} {2009})}\BibitemShut {NoStop}%
\bibitem [{\citenamefont {Dietrich}\ \emph {et~al.}(2019)\citenamefont
  {Dietrich}, \citenamefont {Moortgat-Pick}, \citenamefont {Riemann},
  \citenamefont {Sievers},\ and\ \citenamefont {Ushakov}}]{dietrich2019status}%
  \BibitemOpen
  \bibfield  {author} {\bibinfo {author} {\bibfnamefont {Felix}\ \bibnamefont
  {Dietrich}}, \bibinfo {author} {\bibfnamefont {Gudrid}\ \bibnamefont
  {Moortgat-Pick}}, \bibinfo {author} {\bibfnamefont {Sabine}\ \bibnamefont
  {Riemann}}, \bibinfo {author} {\bibfnamefont {Peter}\ \bibnamefont
  {Sievers}}, \ and\ \bibinfo {author} {\bibfnamefont {Andriy}\ \bibnamefont
  {Ushakov}},\ }\bibfield  {title} {\enquote {\bibinfo {title} {{Status of the
  undulator-based ILC positron source}},}\ }\href
  {https://doi.org/10.48550/arXiv.1902.07744} {\bibfield  {journal} {\bibinfo
  {journal} {arXiv:1902.07744}\ } (\bibinfo {year} {2019})}\BibitemShut
  {NoStop}%
\bibitem [{\citenamefont {Kawanaka}\ \emph {et~al.}(2016)\citenamefont
  {Kawanaka}, \citenamefont {Tsubakimoto}, \citenamefont {Yoshida},
  \citenamefont {Fujioka}, \citenamefont {Fujimoto}, \citenamefont {Tokita},
  \citenamefont {Jitsuno}, \citenamefont {Miyanaga},\ and\ \citenamefont
  {Team}}]{Kawanaka2016}%
  \BibitemOpen
  \bibfield  {author} {\bibinfo {author} {\bibfnamefont {J.}~\bibnamefont
  {Kawanaka}}, \bibinfo {author} {\bibfnamefont {K.}~\bibnamefont
  {Tsubakimoto}}, \bibinfo {author} {\bibfnamefont {H.}~\bibnamefont
  {Yoshida}}, \bibinfo {author} {\bibfnamefont {K.}~\bibnamefont {Fujioka}},
  \bibinfo {author} {\bibfnamefont {Y.}~\bibnamefont {Fujimoto}}, \bibinfo
  {author} {\bibfnamefont {S.}~\bibnamefont {Tokita}}, \bibinfo {author}
  {\bibfnamefont {T.}~\bibnamefont {Jitsuno}}, \bibinfo {author} {\bibfnamefont
  {N.}~\bibnamefont {Miyanaga}}, \ and\ \bibinfo {author} {\bibfnamefont
  {Gekko-EXA~Design}\ \bibnamefont {Team}},\ }\bibfield  {title} {\enquote
  {\bibinfo {title} {Conceptual design of sub-exa-watt system by using optical
  parametric chirped pulse amplification},}\ }\href {\doibase
  10.1088/1742-6596/688/1/012044} {\bibfield  {journal} {\bibinfo  {journal}
  {J. Phys.: Conf. Ser.}\ }\textbf {\bibinfo {volume} {688}},\ \bibinfo {pages}
  {012044} (\bibinfo {year} {2016})}\BibitemShut {NoStop}%
\bibitem [{\citenamefont {Cartlidge}(2018)}]{Edwin2018}%
  \BibitemOpen
  \bibfield  {author} {\bibinfo {author} {\bibfnamefont {Edwin}\ \bibnamefont
  {Cartlidge}},\ }\bibfield  {title} {\enquote {\bibinfo {title} {The light
  fantastic},}\ }\href {\doibase 10.1126/science.359.6374.382} {\bibfield
  {journal} {\bibinfo  {journal} {Science}\ }\textbf {\bibinfo {volume}
  {359}},\ \bibinfo {pages} {382--385} (\bibinfo {year} {2018})}\BibitemShut
  {NoStop}%
\bibitem [{\citenamefont {Danson}\ \emph {et~al.}(2019)\citenamefont {Danson},
  \citenamefont {Haefner}, \citenamefont {Bromage}, \citenamefont {Butcher},
  \citenamefont {Chanteloup}, \citenamefont {Chowdhury}, \citenamefont
  {Galvanauskas}, \citenamefont {Gizzi}, \citenamefont {Hein}, \citenamefont
  {Hillier},\ and\ \citenamefont {\textit{et al.}}}]{Danson2019}%
  \BibitemOpen
  \bibfield  {author} {\bibinfo {author} {\bibfnamefont {Colin~N.}\
  \bibnamefont {Danson}}, \bibinfo {author} {\bibfnamefont {Constantin}\
  \bibnamefont {Haefner}}, \bibinfo {author} {\bibfnamefont {Jake}\
  \bibnamefont {Bromage}}, \bibinfo {author} {\bibfnamefont {Thomas}\
  \bibnamefont {Butcher}}, \bibinfo {author} {\bibfnamefont
  {Jean-Christophe~F.}\ \bibnamefont {Chanteloup}}, \bibinfo {author}
  {\bibfnamefont {Enam~A.}\ \bibnamefont {Chowdhury}}, \bibinfo {author}
  {\bibfnamefont {Almantas}\ \bibnamefont {Galvanauskas}}, \bibinfo {author}
  {\bibfnamefont {Leonida~A.}\ \bibnamefont {Gizzi}}, \bibinfo {author}
  {\bibfnamefont {Joachim}\ \bibnamefont {Hein}}, \bibinfo {author}
  {\bibfnamefont {David~I.}\ \bibnamefont {Hillier}}, \ and\ \bibinfo {author}
  {\bibnamefont {\textit{et al.}}},\ }\bibfield  {title} {\enquote {\bibinfo
  {title} {Petawatt and exawatt class lasers worldwide},}\ }\href {\doibase
  10.1017/hpl.2019.36} {\bibfield  {journal} {\bibinfo  {journal} {High Power
  Laser Sci. Eng.}\ }\textbf {\bibinfo {volume} {7}},\ \bibinfo {pages} {e54}
  (\bibinfo {year} {2019})}\BibitemShut {NoStop}%
\bibitem [{\citenamefont {Yoon}\ \emph {et~al.}(2021)\citenamefont {Yoon},
  \citenamefont {Kim}, \citenamefont {Choi}, \citenamefont {Sung},
  \citenamefont {Lee}, \citenamefont {Lee},\ and\ \citenamefont
  {Nam}}]{Yoon2021}%
  \BibitemOpen
  \bibfield  {author} {\bibinfo {author} {\bibfnamefont {Jin~Woo}\ \bibnamefont
  {Yoon}}, \bibinfo {author} {\bibfnamefont {Yeong~Gyu}\ \bibnamefont {Kim}},
  \bibinfo {author} {\bibfnamefont {Il~Woo}\ \bibnamefont {Choi}}, \bibinfo
  {author} {\bibfnamefont {Jae~Hee}\ \bibnamefont {Sung}}, \bibinfo {author}
  {\bibfnamefont {Hwang~Woon}\ \bibnamefont {Lee}}, \bibinfo {author}
  {\bibfnamefont {Seong~Ku}\ \bibnamefont {Lee}}, \ and\ \bibinfo {author}
  {\bibfnamefont {Chang~Hee}\ \bibnamefont {Nam}},\ }\bibfield  {title}
  {\enquote {\bibinfo {title} {Realization of laser intensity over 10$^{23}$
  w/cm$^2$},}\ }\href {\doibase 10.1364/OPTICA.420520} {\bibfield  {journal}
  {\bibinfo  {journal} {Optica}\ }\textbf {\bibinfo {volume} {8}},\ \bibinfo
  {pages} {630--635} (\bibinfo {year} {2021})}\BibitemShut {NoStop}%
\bibitem [{\citenamefont {ERBER}(1966)}]{ERBER1966}%
  \BibitemOpen
  \bibfield  {author} {\bibinfo {author} {\bibfnamefont {THOMAS}\ \bibnamefont
  {ERBER}},\ }\bibfield  {title} {\enquote {\bibinfo {title} {High-energy
  electromagnetic conversion processes in intense magnetic fields},}\ }\href
  {\doibase 10.1103/RevModPhys.38.626} {\bibfield  {journal} {\bibinfo
  {journal} {Rev. Mod. Phys.}\ }\textbf {\bibinfo {volume} {38}},\ \bibinfo
  {pages} {626--659} (\bibinfo {year} {1966})}\BibitemShut {NoStop}%
\bibitem [{\citenamefont {Ritus}(1985)}]{Ritus1985}%
  \BibitemOpen
  \bibfield  {author} {\bibinfo {author} {\bibfnamefont {V.~I.}\ \bibnamefont
  {Ritus}},\ }\bibfield  {title} {\enquote {\bibinfo {title} {Quantum effects
  of the interaction of elementary particles with an intense electromagnetic
  field},}\ }\href {https://doi.org/10.1007/BF01120220} {\bibfield  {journal}
  {\bibinfo  {journal} {J. Sov. Laser Res.}\ }\textbf {\bibinfo {volume} {6}},\
  \bibinfo {pages} {497} (\bibinfo {year} {1985})}\BibitemShut {NoStop}%
\bibitem [{\citenamefont {Baier}\ \emph {et~al.}(1998)\citenamefont {Baier},
  \citenamefont {Katkov},\ and\ \citenamefont {Strakhovenko}}]{baier1998}%
  \BibitemOpen
  \bibfield  {author} {\bibinfo {author} {\bibfnamefont {V.~N.}\ \bibnamefont
  {Baier}}, \bibinfo {author} {\bibfnamefont {V.~M.}\ \bibnamefont {Katkov}}, \
  and\ \bibinfo {author} {\bibfnamefont {V.~M.}\ \bibnamefont {Strakhovenko}},\
  }\href@noop {} {\emph {\bibinfo {title} {Electromagnetic processes at high
  energies in oriented single crystals}}}\ (\bibinfo  {publisher} {World
  Scientific},\ \bibinfo {address} {Singapore},\ \bibinfo {year}
  {1998})\BibitemShut {NoStop}%
\bibitem [{\citenamefont {Salamin}\ \emph {et~al.}(2006)\citenamefont
  {Salamin}, \citenamefont {Hu}, \citenamefont {Hatsagortsyan},\ and\
  \citenamefont {Keitel}}]{Yousef2006}%
  \BibitemOpen
  \bibfield  {author} {\bibinfo {author} {\bibfnamefont {Yousef~I.}\
  \bibnamefont {Salamin}}, \bibinfo {author} {\bibfnamefont {S.X.}\
  \bibnamefont {Hu}}, \bibinfo {author} {\bibfnamefont {Karen~Z.}\ \bibnamefont
  {Hatsagortsyan}}, \ and\ \bibinfo {author} {\bibfnamefont {Christoph~H.}\
  \bibnamefont {Keitel}},\ }\bibfield  {title} {\enquote {\bibinfo {title}
  {Relativistic high-power laser–matter interactions},}\ }\href {\doibase
  https://doi.org/10.1016/j.physrep.2006.01.002} {\bibfield  {journal}
  {\bibinfo  {journal} {Phys. Rep.}\ }\textbf {\bibinfo {volume} {427}},\
  \bibinfo {pages} {41--155} (\bibinfo {year} {2006})}\BibitemShut {NoStop}%
\bibitem [{\citenamefont {Sun}\ \emph {et~al.}(2022)\citenamefont {Sun},
  \citenamefont {Zhao}, \citenamefont {Xue}, \citenamefont {Lu}, \citenamefont
  {Ji}, \citenamefont {Wan}, \citenamefont {Wang}, \citenamefont {Salamin},\
  and\ \citenamefont {Li}}]{sun2022}%
  \BibitemOpen
  \bibfield  {author} {\bibinfo {author} {\bibfnamefont {Ting}\ \bibnamefont
  {Sun}}, \bibinfo {author} {\bibfnamefont {Qian}\ \bibnamefont {Zhao}},
  \bibinfo {author} {\bibfnamefont {Kun}\ \bibnamefont {Xue}}, \bibinfo
  {author} {\bibfnamefont {Zhi-Wei}\ \bibnamefont {Lu}}, \bibinfo {author}
  {\bibfnamefont {Liang-Liang}\ \bibnamefont {Ji}}, \bibinfo {author}
  {\bibfnamefont {Feng}\ \bibnamefont {Wan}}, \bibinfo {author} {\bibfnamefont
  {Yu}~\bibnamefont {Wang}}, \bibinfo {author} {\bibfnamefont {Yousef~I.}\
  \bibnamefont {Salamin}}, \ and\ \bibinfo {author} {\bibfnamefont {Jian-Xing}\
  \bibnamefont {Li}},\ }\bibfield  {title} {\enquote {\bibinfo {title}
  {Production of polarized particle beams via ultraintense laser pulses},}\
  }\href {\doibase 10.1007/s41614-022-00099-9} {\bibfield  {journal} {\bibinfo
  {journal} {Rev. Mod. Plasma Phys.}\ }\textbf {\bibinfo {volume} {6}},\
  \bibinfo {pages} {38} (\bibinfo {year} {2022})}\BibitemShut {NoStop}%
\bibitem [{\citenamefont {Fedotov}\ \emph {et~al.}(2023)\citenamefont
  {Fedotov}, \citenamefont {Ilderton}, \citenamefont {Karbstein}, \citenamefont
  {King}, \citenamefont {Seipt}, \citenamefont {Taya},\ and\ \citenamefont
  {Torgrimsson}}]{Fedotov2023}%
  \BibitemOpen
  \bibfield  {author} {\bibinfo {author} {\bibfnamefont {A.}~\bibnamefont
  {Fedotov}}, \bibinfo {author} {\bibfnamefont {A.}~\bibnamefont {Ilderton}},
  \bibinfo {author} {\bibfnamefont {F.}~\bibnamefont {Karbstein}}, \bibinfo
  {author} {\bibfnamefont {B.}~\bibnamefont {King}}, \bibinfo {author}
  {\bibfnamefont {D.}~\bibnamefont {Seipt}}, \bibinfo {author} {\bibfnamefont
  {H.}~\bibnamefont {Taya}}, \ and\ \bibinfo {author} {\bibfnamefont
  {G.}~\bibnamefont {Torgrimsson}},\ }\bibfield  {title} {\enquote {\bibinfo
  {title} {{Advances in QED with intense background fields}},}\ }\href
  {\doibase https://doi.org/10.1016/j.physrep.2023.01.003} {\bibfield
  {journal} {\bibinfo  {journal} {Phys. Rep.}\ }\textbf {\bibinfo {volume}
  {1010}},\ \bibinfo {pages} {1--138} (\bibinfo {year} {2023})}\BibitemShut
  {NoStop}%
\bibitem [{\citenamefont {Di~Piazza}\ \emph {et~al.}(2012)\citenamefont
  {Di~Piazza}, \citenamefont {M{\"u}ller}, \citenamefont {Hatsagortsyan},\ and\
  \citenamefont {Keitel}}]{Piazza2012}%
  \BibitemOpen
  \bibfield  {author} {\bibinfo {author} {\bibfnamefont {A.}~\bibnamefont
  {Di~Piazza}}, \bibinfo {author} {\bibfnamefont {C.}~\bibnamefont
  {M{\"u}ller}}, \bibinfo {author} {\bibfnamefont {K.~Z.}\ \bibnamefont
  {Hatsagortsyan}}, \ and\ \bibinfo {author} {\bibfnamefont {Ch.~H.}\
  \bibnamefont {Keitel}},\ }\bibfield  {title} {\enquote {\bibinfo {title}
  {Extremely high-intensity laser interactions with fundamental quantum
  systems},}\ }\href {\doibase 10.1103/RevModPhys.84.1177} {\bibfield
  {journal} {\bibinfo  {journal} {Rev. Mod. Phys.}\ }\textbf {\bibinfo {volume}
  {84}},\ \bibinfo {pages} {1177--1228} (\bibinfo {year} {2012})}\BibitemShut
  {NoStop}%
\bibitem [{\citenamefont {Gonoskov}\ \emph {et~al.}(2022)\citenamefont
  {Gonoskov}, \citenamefont {Blackburn}, \citenamefont {Marklund},\ and\
  \citenamefont {Bulanov}}]{Gonoskov2022}%
  \BibitemOpen
  \bibfield  {author} {\bibinfo {author} {\bibfnamefont {A.}~\bibnamefont
  {Gonoskov}}, \bibinfo {author} {\bibfnamefont {T.~G.}\ \bibnamefont
  {Blackburn}}, \bibinfo {author} {\bibfnamefont {M.}~\bibnamefont {Marklund}},
  \ and\ \bibinfo {author} {\bibfnamefont {S.~S.}\ \bibnamefont {Bulanov}},\
  }\bibfield  {title} {\enquote {\bibinfo {title} {Charged particle motion and
  radiation in strong electromagnetic fields},}\ }\href {\doibase
  10.1103/RevModPhys.94.045001} {\bibfield  {journal} {\bibinfo  {journal}
  {Rev. Mod. Phys.}\ }\textbf {\bibinfo {volume} {94}},\ \bibinfo {pages}
  {045001} (\bibinfo {year} {2022})}\BibitemShut {NoStop}%
\bibitem [{\citenamefont {Wan}\ \emph {et~al.}(2020{\natexlab{a}})\citenamefont
  {Wan}, \citenamefont {Shaisultanov}, \citenamefont {Li}, \citenamefont
  {Hatsagortsyan}, \citenamefont {Keitel},\ and\ \citenamefont {Li}}]{Wan2019}%
  \BibitemOpen
  \bibfield  {author} {\bibinfo {author} {\bibfnamefont {Feng}\ \bibnamefont
  {Wan}}, \bibinfo {author} {\bibfnamefont {Rashid}\ \bibnamefont
  {Shaisultanov}}, \bibinfo {author} {\bibfnamefont {Yan-Fei}\ \bibnamefont
  {Li}}, \bibinfo {author} {\bibfnamefont {Karen~Z.}\ \bibnamefont
  {Hatsagortsyan}}, \bibinfo {author} {\bibfnamefont {Christoph~H.}\
  \bibnamefont {Keitel}}, \ and\ \bibinfo {author} {\bibfnamefont {Jian-Xing}\
  \bibnamefont {Li}},\ }\bibfield  {title} {\enquote {\bibinfo {title}
  {Ultrarelativistic polarized positron jets via collision of electron and
  ultraintense laser beams},}\ }\href
  {https://doi.org/10.1016/j.physletb.2019.135120} {\bibfield  {journal}
  {\bibinfo  {journal} {Phys. Lett. B}\ }\textbf {\bibinfo {volume} {800}},\
  \bibinfo {pages} {135120} (\bibinfo {year} {2020}{\natexlab{a}})}\BibitemShut
  {NoStop}%
\bibitem [{\citenamefont {Chen}\ \emph {et~al.}(2019)\citenamefont {Chen},
  \citenamefont {He}, \citenamefont {Shaisultanov}, \citenamefont
  {Hatsagortsyan},\ and\ \citenamefont {Keitel}}]{Chen2019}%
  \BibitemOpen
  \bibfield  {author} {\bibinfo {author} {\bibfnamefont {Yue-Yue}\ \bibnamefont
  {Chen}}, \bibinfo {author} {\bibfnamefont {Pei-Lun}\ \bibnamefont {He}},
  \bibinfo {author} {\bibfnamefont {Rashid}\ \bibnamefont {Shaisultanov}},
  \bibinfo {author} {\bibfnamefont {Karen~Z.}\ \bibnamefont {Hatsagortsyan}}, \
  and\ \bibinfo {author} {\bibfnamefont {Christoph~H.}\ \bibnamefont
  {Keitel}},\ }\bibfield  {title} {\enquote {\bibinfo {title} {Polarized
  positron beams via intense two-color laser pulses},}\ }\href {\doibase
  10.1103/PhysRevLett.123.174801} {\bibfield  {journal} {\bibinfo  {journal}
  {Phys. Rev. Lett.}\ }\textbf {\bibinfo {volume} {123}},\ \bibinfo {pages}
  {174801} (\bibinfo {year} {2019})}\BibitemShut {NoStop}%
\bibitem [{\citenamefont {Xue}\ \emph {et~al.}(2022)\citenamefont {Xue},
  \citenamefont {Guo}, \citenamefont {Wan}, \citenamefont {Shaisultanov},
  \citenamefont {Chen}, \citenamefont {Xu}, \citenamefont {Ren}, \citenamefont
  {Hatsagortsyan}, \citenamefont {Keitel},\ and\ \citenamefont {Li}}]{Xue2022}%
  \BibitemOpen
  \bibfield  {author} {\bibinfo {author} {\bibfnamefont {Kun}\ \bibnamefont
  {Xue}}, \bibinfo {author} {\bibfnamefont {Ren-Tong}\ \bibnamefont {Guo}},
  \bibinfo {author} {\bibfnamefont {Feng}\ \bibnamefont {Wan}}, \bibinfo
  {author} {\bibfnamefont {Rashid}\ \bibnamefont {Shaisultanov}}, \bibinfo
  {author} {\bibfnamefont {Yue-Yue}\ \bibnamefont {Chen}}, \bibinfo {author}
  {\bibfnamefont {Zhong-Feng}\ \bibnamefont {Xu}}, \bibinfo {author}
  {\bibfnamefont {Xue-Guang}\ \bibnamefont {Ren}}, \bibinfo {author}
  {\bibfnamefont {Karen~Z.}\ \bibnamefont {Hatsagortsyan}}, \bibinfo {author}
  {\bibfnamefont {Christoph~H.}\ \bibnamefont {Keitel}}, \ and\ \bibinfo
  {author} {\bibfnamefont {Jian-Xing}\ \bibnamefont {Li}},\ }\bibfield  {title}
  {\enquote {\bibinfo {title} {{Generation of arbitrarily polarized GeV lepton
  beams via nonlinear Breit-Wheeler process}},}\ }\href {\doibase
  https://doi.org/10.1016/j.fmre.2021.11.022} {\bibfield  {journal} {\bibinfo
  {journal} {Fundam. Res.}\ }\textbf {\bibinfo {volume} {2}},\ \bibinfo {pages}
  {539--545} (\bibinfo {year} {2022})}\BibitemShut {NoStop}%
\bibitem [{\citenamefont {Dai}\ \emph {et~al.}(2022)\citenamefont {Dai},
  \citenamefont {Shen}, \citenamefont {Li}, \citenamefont {Shaisultanov},
  \citenamefont {Hatsagortsyan}, \citenamefont {Keitel},\ and\ \citenamefont
  {Chen}}]{Dai2022}%
  \BibitemOpen
  \bibfield  {author} {\bibinfo {author} {\bibfnamefont {Ya-Nan}\ \bibnamefont
  {Dai}}, \bibinfo {author} {\bibfnamefont {Bai-Fei}\ \bibnamefont {Shen}},
  \bibinfo {author} {\bibfnamefont {Jian-Xing}\ \bibnamefont {Li}}, \bibinfo
  {author} {\bibfnamefont {Rashid}\ \bibnamefont {Shaisultanov}}, \bibinfo
  {author} {\bibfnamefont {Karen~Z.}\ \bibnamefont {Hatsagortsyan}}, \bibinfo
  {author} {\bibfnamefont {Christoph~H.}\ \bibnamefont {Keitel}}, \ and\
  \bibinfo {author} {\bibfnamefont {Yue-Yue}\ \bibnamefont {Chen}},\ }\bibfield
   {title} {\enquote {\bibinfo {title} {Photon polarization effects in
  polarized electron–positron pair production in a strong laser field},}\
  }\href {\doibase 10.1063/5.0063633} {\bibfield  {journal} {\bibinfo
  {journal} {Matter Radiat. Extremes}\ }\textbf {\bibinfo {volume} {7}},\
  \bibinfo {pages} {014401} (\bibinfo {year} {2022})}\BibitemShut {NoStop}%
\bibitem [{\citenamefont {Li}\ \emph {et~al.}(2020{\natexlab{a}})\citenamefont
  {Li}, \citenamefont {Chen}, \citenamefont {Wang},\ and\ \citenamefont
  {Hu}}]{Li2020positron}%
  \BibitemOpen
  \bibfield  {author} {\bibinfo {author} {\bibfnamefont {Yan-Fei}\ \bibnamefont
  {Li}}, \bibinfo {author} {\bibfnamefont {Yue-Yue}\ \bibnamefont {Chen}},
  \bibinfo {author} {\bibfnamefont {Wei-Min}\ \bibnamefont {Wang}}, \ and\
  \bibinfo {author} {\bibfnamefont {Hua-Si}\ \bibnamefont {Hu}},\ }\bibfield
  {title} {\enquote {\bibinfo {title} {Production of highly polarized positron
  beams via helicity transfer from polarized electrons in a strong laser
  field},}\ }\href {\doibase 10.1103/PhysRevLett.125.044802} {\bibfield
  {journal} {\bibinfo  {journal} {Phys. Rev. Lett.}\ }\textbf {\bibinfo
  {volume} {125}},\ \bibinfo {pages} {044802} (\bibinfo {year}
  {2020}{\natexlab{a}})}\BibitemShut {NoStop}%
\bibitem [{\citenamefont {Xie}\ \emph {et~al.}(2017)\citenamefont {Xie},
  \citenamefont {Li},\ and\ \citenamefont {Tang}}]{xie2017}%
  \BibitemOpen
  \bibfield  {author} {\bibinfo {author} {\bibfnamefont {Bai~Song}\
  \bibnamefont {Xie}}, \bibinfo {author} {\bibfnamefont {Zi~Liang}\
  \bibnamefont {Li}}, \ and\ \bibinfo {author} {\bibfnamefont {Suo}\
  \bibnamefont {Tang}},\ }\bibfield  {title} {\enquote {\bibinfo {title}
  {{Electron-positron pair production in ultrastrong laser fields}},}\ }\href
  {\doibase 10.1016/j.mre.2017.07.002} {\bibfield  {journal} {\bibinfo
  {journal} {Matter Radiat. Extremes}\ }\textbf {\bibinfo {volume} {2}},\
  \bibinfo {pages} {225--242} (\bibinfo {year} {2017})}\BibitemShut {NoStop}%
\bibitem [{\citenamefont {Vranic}\ \emph {et~al.}(2018)\citenamefont {Vranic},
  \citenamefont {Klimo}, \citenamefont {Korn},\ and\ \citenamefont
  {Weber}}]{Vranic2018}%
  \BibitemOpen
  \bibfield  {author} {\bibinfo {author} {\bibfnamefont {Marija}\ \bibnamefont
  {Vranic}}, \bibinfo {author} {\bibfnamefont {Ondrej}\ \bibnamefont {Klimo}},
  \bibinfo {author} {\bibfnamefont {Georg}\ \bibnamefont {Korn}}, \ and\
  \bibinfo {author} {\bibfnamefont {Stefan}\ \bibnamefont {Weber}},\ }\bibfield
   {title} {\enquote {\bibinfo {title} {{Multi-GeV electron-positron beam
  generation from laser-electron scattering}},}\ }\href {\doibase
  10.1038/s41598-018-23126-7} {\bibfield  {journal} {\bibinfo  {journal} {Sci.
  Rep.}\ }\textbf {\bibinfo {volume} {8}},\ \bibinfo {pages} {4702} (\bibinfo
  {year} {2018})}\BibitemShut {NoStop}%
\bibitem [{\citenamefont {Zhao}\ \emph {et~al.}(2019)\citenamefont {Zhao},
  \citenamefont {Liu}, \citenamefont {Li},\ and\ \citenamefont
  {Xia}}]{Zhao_2019}%
  \BibitemOpen
  \bibfield  {author} {\bibinfo {author} {\bibfnamefont {Yuan}\ \bibnamefont
  {Zhao}}, \bibinfo {author} {\bibfnamefont {Jianxun}\ \bibnamefont {Liu}},
  \bibinfo {author} {\bibfnamefont {Yangmei}\ \bibnamefont {Li}}, \ and\
  \bibinfo {author} {\bibfnamefont {Guoxing}\ \bibnamefont {Xia}},\ }\bibfield
  {title} {\enquote {\bibinfo {title} {{Ultra-bright $\gamma$-ray emission by
  using PW laser irradiating solid target obliquely}},}\ }\href {\doibase
  10.1088/1361-6587/ab132e} {\bibfield  {journal} {\bibinfo  {journal} {Plasma
  Phys. Control. Fusion}\ }\textbf {\bibinfo {volume} {61}},\ \bibinfo {pages}
  {065010} (\bibinfo {year} {2019})}\BibitemShut {NoStop}%
\bibitem [{\citenamefont {Dinu}\ and\ \citenamefont
  {Torgrimsson}(2020)}]{Dinu2020}%
  \BibitemOpen
  \bibfield  {author} {\bibinfo {author} {\bibfnamefont {Victor}\ \bibnamefont
  {Dinu}}\ and\ \bibinfo {author} {\bibfnamefont {Greger}\ \bibnamefont
  {Torgrimsson}},\ }\bibfield  {title} {\enquote {\bibinfo {title}
  {{Approximating higher-order nonlinear QED processes with first-order
  building blocks}},}\ }\href {\doibase 10.1103/PhysRevD.102.016018} {\bibfield
   {journal} {\bibinfo  {journal} {Phys. Rev. D}\ }\textbf {\bibinfo {volume}
  {102}},\ \bibinfo {pages} {016018} (\bibinfo {year} {2020})}\BibitemShut
  {NoStop}%
\bibitem [{\citenamefont {Li}\ \emph {et~al.}(2020{\natexlab{b}})\citenamefont
  {Li}, \citenamefont {Shaisultanov}, \citenamefont {Chen}, \citenamefont
  {Wan}, \citenamefont {Hatsagortsyan}, \citenamefont {Keitel},\ and\
  \citenamefont {Li}}]{Li2020Polarized}%
  \BibitemOpen
  \bibfield  {author} {\bibinfo {author} {\bibfnamefont {Yan-Fei}\ \bibnamefont
  {Li}}, \bibinfo {author} {\bibfnamefont {Rashid}\ \bibnamefont
  {Shaisultanov}}, \bibinfo {author} {\bibfnamefont {Yue-Yue}\ \bibnamefont
  {Chen}}, \bibinfo {author} {\bibfnamefont {Feng}\ \bibnamefont {Wan}},
  \bibinfo {author} {\bibfnamefont {Karen~Z.}\ \bibnamefont {Hatsagortsyan}},
  \bibinfo {author} {\bibfnamefont {Christoph~H.}\ \bibnamefont {Keitel}}, \
  and\ \bibinfo {author} {\bibfnamefont {Jian-Xing}\ \bibnamefont {Li}},\
  }\bibfield  {title} {\enquote {\bibinfo {title} {{Polarized Ultrashort
  Brilliant Multi-GeV $\ensuremath{\gamma}$ Rays via Single-Shot Laser-Electron
  Interaction}},}\ }\href {\doibase 10.1103/PhysRevLett.124.014801} {\bibfield
  {journal} {\bibinfo  {journal} {Phys. Rev. Lett.}\ }\textbf {\bibinfo
  {volume} {124}},\ \bibinfo {pages} {014801} (\bibinfo {year}
  {2020}{\natexlab{b}})}\BibitemShut {NoStop}%
\bibitem [{\citenamefont {Xue}\ \emph {et~al.}(2020)\citenamefont {Xue},
  \citenamefont {Dou}, \citenamefont {Wan}, \citenamefont {Yu}, \citenamefont
  {Wang}, \citenamefont {Ren}, \citenamefont {Zhao}, \citenamefont {Zhao},
  \citenamefont {Xu},\ and\ \citenamefont {Li}}]{Xue2020}%
  \BibitemOpen
  \bibfield  {author} {\bibinfo {author} {\bibfnamefont {Kun}\ \bibnamefont
  {Xue}}, \bibinfo {author} {\bibfnamefont {Zhen-Ke}\ \bibnamefont {Dou}},
  \bibinfo {author} {\bibfnamefont {Feng}\ \bibnamefont {Wan}}, \bibinfo
  {author} {\bibfnamefont {Tong-Pu}\ \bibnamefont {Yu}}, \bibinfo {author}
  {\bibfnamefont {Wei-Min}\ \bibnamefont {Wang}}, \bibinfo {author}
  {\bibfnamefont {Jie-Ru}\ \bibnamefont {Ren}}, \bibinfo {author}
  {\bibfnamefont {Qian}\ \bibnamefont {Zhao}}, \bibinfo {author} {\bibfnamefont
  {Yong-Tao}\ \bibnamefont {Zhao}}, \bibinfo {author} {\bibfnamefont
  {Zhong-Feng}\ \bibnamefont {Xu}}, \ and\ \bibinfo {author} {\bibfnamefont
  {Jian-Xing}\ \bibnamefont {Li}},\ }\bibfield  {title} {\enquote {\bibinfo
  {title} {Generation of highly-polarized high-energy brilliant $\gamma$-rays
  via laser-plasma interaction},}\ }\href {\doibase 10.1063/5.0007734}
  {\bibfield  {journal} {\bibinfo  {journal} {Matter Radiat. Extremes}\
  }\textbf {\bibinfo {volume} {5}},\ \bibinfo {pages} {054402} (\bibinfo {year}
  {2020})}\BibitemShut {NoStop}%
\bibitem [{\citenamefont {Torgrimsson}(2021)}]{Torgrimsson2021}%
  \BibitemOpen
  \bibfield  {author} {\bibinfo {author} {\bibfnamefont {Greger}\ \bibnamefont
  {Torgrimsson}},\ }\bibfield  {title} {\enquote {\bibinfo {title} {{ Loops and
  polarization in strong-field QED}},}\ }\href {\doibase
  10.1088/1367-2630/abf274} {\bibfield  {journal} {\bibinfo  {journal} {New J.
  Phys.}\ }\textbf {\bibinfo {volume} {23}},\ \bibinfo {pages} {065001}
  (\bibinfo {year} {2021})}\BibitemShut {NoStop}%
\bibitem [{\citenamefont {Ivanov}\ \emph {et~al.}(2005)\citenamefont {Ivanov},
  \citenamefont {Kotkin},\ and\ \citenamefont {Serbo}}]{ivanov2005}%
  \BibitemOpen
  \bibfield  {author} {\bibinfo {author} {\bibfnamefont {D.~Y.}\ \bibnamefont
  {Ivanov}}, \bibinfo {author} {\bibfnamefont {G.~L.}\ \bibnamefont {Kotkin}},
  \ and\ \bibinfo {author} {\bibfnamefont {V.~G.}\ \bibnamefont {Serbo}},\
  }\bibfield  {title} {\enquote {\bibinfo {title} {Complete description of
  polarization effects in e$^+$e$^-$ pair production by a photon in the field
  of a strong laser wave},}\ }\href {\doibase 10.1140/epjc/s2005-02125-1}
  {\bibfield  {journal} {\bibinfo  {journal} {Eur. Phys. J. C}\ }\textbf
  {\bibinfo {volume} {40}},\ \bibinfo {pages} {27--40} (\bibinfo {year}
  {2005})}\BibitemShut {NoStop}%
\bibitem [{\citenamefont {King}\ \emph {et~al.}(2013)\citenamefont {King},
  \citenamefont {Elkina},\ and\ \citenamefont {Ruhl}}]{king2013}%
  \BibitemOpen
  \bibfield  {author} {\bibinfo {author} {\bibfnamefont {B.}~\bibnamefont
  {King}}, \bibinfo {author} {\bibfnamefont {N.}~\bibnamefont {Elkina}}, \ and\
  \bibinfo {author} {\bibfnamefont {H.}~\bibnamefont {Ruhl}},\ }\bibfield
  {title} {\enquote {\bibinfo {title} {Photon polarization in electron-seeded
  pair-creation cascades},}\ }\href {\doibase 10.1103/PhysRevA.87.042117}
  {\bibfield  {journal} {\bibinfo  {journal} {Phys. Rev. A}\ }\textbf {\bibinfo
  {volume} {87}},\ \bibinfo {pages} {042117} (\bibinfo {year}
  {2013})}\BibitemShut {NoStop}%
\bibitem [{\citenamefont {Seipt}\ and\ \citenamefont {King}(2020)}]{Seipt2020}%
  \BibitemOpen
  \bibfield  {author} {\bibinfo {author} {\bibfnamefont {D.}~\bibnamefont
  {Seipt}}\ and\ \bibinfo {author} {\bibfnamefont {B.}~\bibnamefont {King}},\
  }\bibfield  {title} {\enquote {\bibinfo {title} {{Spin- and
  polarization-dependent locally-constant-field-approximation rates for
  nonlinear Compton and Breit-Wheeler processes}},}\ }\href {\doibase
  10.1103/PhysRevA.102.052805} {\bibfield  {journal} {\bibinfo  {journal}
  {Phys. Rev. A}\ }\textbf {\bibinfo {volume} {102}},\ \bibinfo {pages}
  {052805} (\bibinfo {year} {2020})}\BibitemShut {NoStop}%
\bibitem [{\citenamefont {Wan}\ \emph {et~al.}(2020{\natexlab{b}})\citenamefont
  {Wan}, \citenamefont {Wang}, \citenamefont {Guo}, \citenamefont {Chen},
  \citenamefont {Shaisultanov}, \citenamefont {Xu}, \citenamefont
  {Hatsagortsyan}, \citenamefont {Keitel},\ and\ \citenamefont
  {Li}}]{Wan_2020}%
  \BibitemOpen
  \bibfield  {author} {\bibinfo {author} {\bibfnamefont {Feng}\ \bibnamefont
  {Wan}}, \bibinfo {author} {\bibfnamefont {Yu}~\bibnamefont {Wang}}, \bibinfo
  {author} {\bibfnamefont {Ren-Tong}\ \bibnamefont {Guo}}, \bibinfo {author}
  {\bibfnamefont {Yue-Yue}\ \bibnamefont {Chen}}, \bibinfo {author}
  {\bibfnamefont {Rashid}\ \bibnamefont {Shaisultanov}}, \bibinfo {author}
  {\bibfnamefont {Zhong-Feng}\ \bibnamefont {Xu}}, \bibinfo {author}
  {\bibfnamefont {Karen~Z.}\ \bibnamefont {Hatsagortsyan}}, \bibinfo {author}
  {\bibfnamefont {Christoph~H.}\ \bibnamefont {Keitel}}, \ and\ \bibinfo
  {author} {\bibfnamefont {Jian-Xing}\ \bibnamefont {Li}},\ }\bibfield  {title}
  {\enquote {\bibinfo {title} {{High-energy $\gamma$-photon polarization in
  nonlinear Breit-Wheeler pair production and $\ensuremath{\gamma}$
  polarimetry}},}\ }\href {\doibase 10.1103/PhysRevResearch.2.032049}
  {\bibfield  {journal} {\bibinfo  {journal} {Phys. Rev. Res.}\ }\textbf
  {\bibinfo {volume} {2}},\ \bibinfo {pages} {032049} (\bibinfo {year}
  {2020}{\natexlab{b}})}\BibitemShut {NoStop}%
\bibitem [{\citenamefont {Adderley}\ \emph {et~al.}(2010)\citenamefont
  {Adderley}, \citenamefont {Clark}, \citenamefont {Grames}, \citenamefont
  {Hansknecht}, \citenamefont {Surles-Law}, \citenamefont {Machie},
  \citenamefont {Poelker}, \citenamefont {Stutzman},\ and\ \citenamefont
  {Suleiman}}]{adderley2010}%
  \BibitemOpen
  \bibfield  {author} {\bibinfo {author} {\bibfnamefont {P.~A.}\ \bibnamefont
  {Adderley}}, \bibinfo {author} {\bibfnamefont {J.}~\bibnamefont {Clark}},
  \bibinfo {author} {\bibfnamefont {Joe}\ \bibnamefont {Grames}}, \bibinfo
  {author} {\bibfnamefont {John}\ \bibnamefont {Hansknecht}}, \bibinfo {author}
  {\bibfnamefont {K.}~\bibnamefont {Surles-Law}}, \bibinfo {author}
  {\bibfnamefont {D.}~\bibnamefont {Machie}}, \bibinfo {author} {\bibfnamefont
  {M.}~\bibnamefont {Poelker}}, \bibinfo {author} {\bibfnamefont {M.~L.}\
  \bibnamefont {Stutzman}}, \ and\ \bibinfo {author} {\bibfnamefont {Rafiu}\
  \bibnamefont {Suleiman}},\ }\bibfield  {title} {\enquote {\bibinfo {title}
  {Load-locked dc high voltage gaas photogun with an inverted-geometry ceramic
  insulator},}\ }\href {\doibase 10.1103/PhysRevSTAB.13.010101} {\bibfield
  {journal} {\bibinfo  {journal} {Phys. Rev. ST Accel. Beams}\ }\textbf
  {\bibinfo {volume} {13}},\ \bibinfo {pages} {010101} (\bibinfo {year}
  {2010})}\BibitemShut {NoStop}%
\bibitem [{\citenamefont {Bonţoiu}\ \emph {et~al.}(2023)\citenamefont
  {Bonţoiu}, \citenamefont {Apsimon}, \citenamefont {Kukstas}, \citenamefont
  {Rodin}, \citenamefont {Yadav}, \citenamefont {Welsch}, \citenamefont
  {Resta-López}, \citenamefont {Bonatto},\ and\ \citenamefont
  {Xia}}]{bontoiu2023}%
  \BibitemOpen
  \bibfield  {author} {\bibinfo {author} {\bibfnamefont {Cristian}\
  \bibnamefont {Bonţoiu}}, \bibinfo {author} {\bibfnamefont {Öznur}\
  \bibnamefont {Apsimon}}, \bibinfo {author} {\bibfnamefont {Egidijus}\
  \bibnamefont {Kukstas}}, \bibinfo {author} {\bibfnamefont {Volodymyr}\
  \bibnamefont {Rodin}}, \bibinfo {author} {\bibfnamefont {Monika}\
  \bibnamefont {Yadav}}, \bibinfo {author} {\bibfnamefont {Carsten}\
  \bibnamefont {Welsch}}, \bibinfo {author} {\bibfnamefont {Javier}\
  \bibnamefont {Resta-López}}, \bibinfo {author} {\bibfnamefont {Alexandre}\
  \bibnamefont {Bonatto}}, \ and\ \bibinfo {author} {\bibfnamefont {Guoxing}\
  \bibnamefont {Xia}},\ }\bibfield  {title} {\enquote {\bibinfo {title} {{TeV/m
  catapult acceleration of electrons in graphene layers}},}\ }\href {\doibase
  10.1038/s41598-023-28617-w} {\bibfield  {journal} {\bibinfo  {journal} {Sci.
  Rep.}\ }\textbf {\bibinfo {volume} {13}},\ \bibinfo {pages} {1330} (\bibinfo
  {year} {2023})}\BibitemShut {NoStop}%
\bibitem [{\citenamefont {Pukhov}\ and\ \citenamefont {Meyer-ter
  Vehn}(2002)}]{pukhov2002}%
  \BibitemOpen
  \bibfield  {author} {\bibinfo {author} {\bibfnamefont {A.}~\bibnamefont
  {Pukhov}}\ and\ \bibinfo {author} {\bibfnamefont {J.}~\bibnamefont {Meyer-ter
  Vehn}},\ }\bibfield  {title} {\enquote {\bibinfo {title} {Laser wake field
  acceleration: the highly non-linear broken-wave regime},}\ }\href {\doibase
  10.1007/s003400200795} {\bibfield  {journal} {\bibinfo  {journal} {Appl.
  Phys. B}\ }\textbf {\bibinfo {volume} {74}},\ \bibinfo {pages} {355--361}
  (\bibinfo {year} {2002})}\BibitemShut {NoStop}%
\bibitem [{\citenamefont {Cho}\ \emph {et~al.}(2018)\citenamefont {Cho},
  \citenamefont {Pathak}, \citenamefont {Kim},\ and\ \citenamefont
  {Nam}}]{cho2018}%
  \BibitemOpen
  \bibfield  {author} {\bibinfo {author} {\bibfnamefont {Myung~Hoon}\
  \bibnamefont {Cho}}, \bibinfo {author} {\bibfnamefont {Vishwa~Bandhu}\
  \bibnamefont {Pathak}}, \bibinfo {author} {\bibfnamefont {Hyung~Taek}\
  \bibnamefont {Kim}}, \ and\ \bibinfo {author} {\bibfnamefont {Chang~Hee}\
  \bibnamefont {Nam}},\ }\bibfield  {title} {\enquote {\bibinfo {title}
  {Controlled electron injection facilitated by nanoparticles for laser
  wakefield acceleration},}\ }\href {\doibase 10.1038/s41598-018-34998-0}
  {\bibfield  {journal} {\bibinfo  {journal} {Sci. Rep.}\ }\textbf {\bibinfo
  {volume} {8}},\ \bibinfo {pages} {16924} (\bibinfo {year}
  {2018})}\BibitemShut {NoStop}%
\bibitem [{\citenamefont {Gschwendtner}\ and\ \citenamefont
  {Muggli}(2019)}]{Gschwendtner2019}%
  \BibitemOpen
  \bibfield  {author} {\bibinfo {author} {\bibfnamefont {Edda}\ \bibnamefont
  {Gschwendtner}}\ and\ \bibinfo {author} {\bibfnamefont {Patric}\ \bibnamefont
  {Muggli}},\ }\bibfield  {title} {\enquote {\bibinfo {title} {Plasma wakefield
  accelerators},}\ }\href {\doibase 10.1038/s42254-019-0049-z} {\bibfield
  {journal} {\bibinfo  {journal} {Nat. Rev. Phys}\ }\textbf {\bibinfo {volume}
  {1}},\ \bibinfo {pages} {246--248} (\bibinfo {year} {2019})}\BibitemShut
  {NoStop}%
\bibitem [{\citenamefont {Ridgers}\ \emph {et~al.}(2012)\citenamefont
  {Ridgers}, \citenamefont {Brady}, \citenamefont {Duclous}, \citenamefont
  {Kirk}, \citenamefont {Bennett}, \citenamefont {Arber}, \citenamefont
  {Robinson},\ and\ \citenamefont {Bell}}]{Ridgers2012}%
  \BibitemOpen
  \bibfield  {author} {\bibinfo {author} {\bibfnamefont {C.~P.}\ \bibnamefont
  {Ridgers}}, \bibinfo {author} {\bibfnamefont {Christopher~S.}\ \bibnamefont
  {Brady}}, \bibinfo {author} {\bibfnamefont {R.}~\bibnamefont {Duclous}},
  \bibinfo {author} {\bibfnamefont {J.~G.}\ \bibnamefont {Kirk}}, \bibinfo
  {author} {\bibfnamefont {K.}~\bibnamefont {Bennett}}, \bibinfo {author}
  {\bibfnamefont {T.~D.}\ \bibnamefont {Arber}}, \bibinfo {author}
  {\bibfnamefont {A.~P.~L.}\ \bibnamefont {Robinson}}, \ and\ \bibinfo {author}
  {\bibfnamefont {A.~R.}\ \bibnamefont {Bell}},\ }\bibfield  {title} {\enquote
  {\bibinfo {title} {Dense electron-positron plasmas and ultraintense
  $\ensuremath{\gamma}$ rays from laser-irradiated solids},}\ }\href {\doibase
  10.1103/PhysRevLett.108.165006} {\bibfield  {journal} {\bibinfo  {journal}
  {Phys. Rev. Lett.}\ }\textbf {\bibinfo {volume} {108}},\ \bibinfo {pages}
  {165006} (\bibinfo {year} {2012})}\BibitemShut {NoStop}%
\bibitem [{\citenamefont {Gu}\ \emph {et~al.}(2016)\citenamefont {Gu},
  \citenamefont {Klimo}, \citenamefont {Weber},\ and\ \citenamefont
  {Korn}}]{Gu2016}%
  \BibitemOpen
  \bibfield  {author} {\bibinfo {author} {\bibfnamefont {Y.~J.}\ \bibnamefont
  {Gu}}, \bibinfo {author} {\bibfnamefont {O.}~\bibnamefont {Klimo}}, \bibinfo
  {author} {\bibfnamefont {S.}~\bibnamefont {Weber}}, \ and\ \bibinfo {author}
  {\bibfnamefont {G.}~\bibnamefont {Korn}},\ }\bibfield  {title} {\enquote
  {\bibinfo {title} {High density ultrashort relativistic positron beam
  generation by laser-plasma interaction},}\ }\href {\doibase
  10.1088/1367-2630/18/11/113023} {\bibfield  {journal} {\bibinfo  {journal}
  {New J. Phys.}\ }\textbf {\bibinfo {volume} {18}},\ \bibinfo {pages} {113023}
  (\bibinfo {year} {2016})}\BibitemShut {NoStop}%
\bibitem [{\citenamefont {Gu}\ \emph {et~al.}(2018)\citenamefont {Gu},
  \citenamefont {Klimo}, \citenamefont {Bulanov},\ and\ \citenamefont
  {Weber}}]{Gu2018}%
  \BibitemOpen
  \bibfield  {author} {\bibinfo {author} {\bibfnamefont {Yan-Jun}\ \bibnamefont
  {Gu}}, \bibinfo {author} {\bibfnamefont {Ondrej}\ \bibnamefont {Klimo}},
  \bibinfo {author} {\bibfnamefont {Sergei~V}\ \bibnamefont {Bulanov}}, \ and\
  \bibinfo {author} {\bibfnamefont {Stefan}\ \bibnamefont {Weber}},\ }\bibfield
   {title} {\enquote {\bibinfo {title} {Brilliant gamma-ray beam and
  electron–positron pair production by enhanced attosecond pulses},}\ }\href
  {\doibase 10.1038/s42005-018-0095-3} {\bibfield  {journal} {\bibinfo
  {journal} {Commun. Phys.}\ }\textbf {\bibinfo {volume} {1}},\ \bibinfo
  {pages} {93} (\bibinfo {year} {2018})}\BibitemShut {NoStop}%
\bibitem [{\citenamefont {Ji}\ \emph {et~al.}(2014)\citenamefont {Ji},
  \citenamefont {Pukhov}, \citenamefont {Kostyukov}, \citenamefont {Shen},\
  and\ \citenamefont {Akli}}]{Ji2014}%
  \BibitemOpen
  \bibfield  {author} {\bibinfo {author} {\bibfnamefont {L.~L.}\ \bibnamefont
  {Ji}}, \bibinfo {author} {\bibfnamefont {A.}~\bibnamefont {Pukhov}}, \bibinfo
  {author} {\bibfnamefont {I.~Yu.}\ \bibnamefont {Kostyukov}}, \bibinfo
  {author} {\bibfnamefont {B.~F.}\ \bibnamefont {Shen}}, \ and\ \bibinfo
  {author} {\bibfnamefont {K.}~\bibnamefont {Akli}},\ }\bibfield  {title}
  {\enquote {\bibinfo {title} {Radiation-reaction trapping of electrons in
  extreme laser fields},}\ }\href {\doibase 10.1103/PhysRevLett.112.145003}
  {\bibfield  {journal} {\bibinfo  {journal} {Phys. Rev. Lett.}\ }\textbf
  {\bibinfo {volume} {112}},\ \bibinfo {pages} {145003} (\bibinfo {year}
  {2014})}\BibitemShut {NoStop}%
\bibitem [{\citenamefont {Kostyukov}\ and\ \citenamefont
  {Nerush}(2016)}]{Kostyukov2016}%
  \BibitemOpen
  \bibfield  {author} {\bibinfo {author} {\bibfnamefont {I.~Yu.}\ \bibnamefont
  {Kostyukov}}\ and\ \bibinfo {author} {\bibfnamefont {E.~N.}\ \bibnamefont
  {Nerush}},\ }\bibfield  {title} {\enquote {\bibinfo {title} {Production and
  dynamics of positrons in ultrahigh intensity laser-foil interactions},}\
  }\href {\doibase 10.1063/1.4962567} {\bibfield  {journal} {\bibinfo
  {journal} {Phys. Plasmas}\ }\textbf {\bibinfo {volume} {23}},\ \bibinfo
  {pages} {093119} (\bibinfo {year} {2016})}\BibitemShut {NoStop}%
\bibitem [{\citenamefont {Zhu}\ \emph {et~al.}(2016)\citenamefont {Zhu},
  \citenamefont {Yu}, \citenamefont {Sheng}, \citenamefont {Yin}, \citenamefont
  {Turcu},\ and\ \citenamefont {Pukhov}}]{Zhu2016}%
  \BibitemOpen
  \bibfield  {author} {\bibinfo {author} {\bibfnamefont {Xing-Long}\
  \bibnamefont {Zhu}}, \bibinfo {author} {\bibfnamefont {Tong-Pu}\ \bibnamefont
  {Yu}}, \bibinfo {author} {\bibfnamefont {Zheng-Ming}\ \bibnamefont {Sheng}},
  \bibinfo {author} {\bibfnamefont {Yan}\ \bibnamefont {Yin}}, \bibinfo
  {author} {\bibfnamefont {Ion Cristian~Edmond}\ \bibnamefont {Turcu}}, \ and\
  \bibinfo {author} {\bibfnamefont {Alexander}\ \bibnamefont {Pukhov}},\
  }\bibfield  {title} {\enquote {\bibinfo {title} {{Dense GeV
  electron–positron pairs generated by lasers in near-critical-density
  plasmas}},}\ }\href {\doibase 10.1038/ncomms13686} {\bibfield  {journal}
  {\bibinfo  {journal} {Nat. Commun.}\ }\textbf {\bibinfo {volume} {7}},\
  \bibinfo {pages} {13686} (\bibinfo {year} {2016})}\BibitemShut {NoStop}%
\bibitem [{\citenamefont {Liu}\ \emph {et~al.}(2016)\citenamefont {Liu},
  \citenamefont {Yu}, \citenamefont {Yin}, \citenamefont {Zhu},\ and\
  \citenamefont {Shao}}]{Liu2016}%
  \BibitemOpen
  \bibfield  {author} {\bibinfo {author} {\bibfnamefont {Jin-Jin}\ \bibnamefont
  {Liu}}, \bibinfo {author} {\bibfnamefont {Tong-Pu}\ \bibnamefont {Yu}},
  \bibinfo {author} {\bibfnamefont {Yan}\ \bibnamefont {Yin}}, \bibinfo
  {author} {\bibfnamefont {Xing-Long}\ \bibnamefont {Zhu}}, \ and\ \bibinfo
  {author} {\bibfnamefont {Fu-Qiu}\ \bibnamefont {Shao}},\ }\bibfield  {title}
  {\enquote {\bibinfo {title} {All-optical bright $\gamma$-ray and dense
  positron source by laser driven plasmas-filled cone},}\ }\href {\doibase
  10.1364/OE.24.015978} {\bibfield  {journal} {\bibinfo  {journal} {Opt.
  Express}\ }\textbf {\bibinfo {volume} {24}},\ \bibinfo {pages} {15978--15986}
  (\bibinfo {year} {2016})}\BibitemShut {NoStop}%
\bibitem [{\citenamefont {Li}\ \emph {et~al.}(2017)\citenamefont {Li},
  \citenamefont {Yu}, \citenamefont {Hu}, \citenamefont {Yin}, \citenamefont
  {Zou}, \citenamefont {Liu}, \citenamefont {Wang}, \citenamefont {Hu},\ and\
  \citenamefont {Shao}}]{Li2017}%
  \BibitemOpen
  \bibfield  {author} {\bibinfo {author} {\bibfnamefont {Han-Zhen}\
  \bibnamefont {Li}}, \bibinfo {author} {\bibfnamefont {Tong-Pu}\ \bibnamefont
  {Yu}}, \bibinfo {author} {\bibfnamefont {Li-Xiang}\ \bibnamefont {Hu}},
  \bibinfo {author} {\bibfnamefont {Yan}\ \bibnamefont {Yin}}, \bibinfo
  {author} {\bibfnamefont {De-Bin}\ \bibnamefont {Zou}}, \bibinfo {author}
  {\bibfnamefont {Jian-Xun}\ \bibnamefont {Liu}}, \bibinfo {author}
  {\bibfnamefont {Wei-Quan}\ \bibnamefont {Wang}}, \bibinfo {author}
  {\bibfnamefont {Shun}\ \bibnamefont {Hu}}, \ and\ \bibinfo {author}
  {\bibfnamefont {Fu-Qiu}\ \bibnamefont {Shao}},\ }\bibfield  {title} {\enquote
  {\bibinfo {title} {Ultra-bright $\gamma$-ray flashes and dense attosecond
  positron bunches from two counter-propagating laser pulses irradiating a
  micro-wire target},}\ }\href {\doibase 10.1364/OE.25.021583} {\bibfield
  {journal} {\bibinfo  {journal} {Opt. Express}\ }\textbf {\bibinfo {volume}
  {25}},\ \bibinfo {pages} {21583--21593} (\bibinfo {year} {2017})}\BibitemShut
  {NoStop}%
\bibitem [{\citenamefont {Liu}\ \emph {et~al.}(2017)\citenamefont {Liu},
  \citenamefont {Luo}, \citenamefont {Yuan}, \citenamefont {Yu}, \citenamefont
  {Chen},\ and\ \citenamefont {Sheng}}]{Liu2017}%
  \BibitemOpen
  \bibfield  {author} {\bibinfo {author} {\bibfnamefont {W.~Y.}\ \bibnamefont
  {Liu}}, \bibinfo {author} {\bibfnamefont {W.}~\bibnamefont {Luo}}, \bibinfo
  {author} {\bibfnamefont {T.}~\bibnamefont {Yuan}}, \bibinfo {author}
  {\bibfnamefont {J.~Y.}\ \bibnamefont {Yu}}, \bibinfo {author} {\bibfnamefont
  {M.}~\bibnamefont {Chen}}, \ and\ \bibinfo {author} {\bibfnamefont {Z.~M.}\
  \bibnamefont {Sheng}},\ }\bibfield  {title} {\enquote {\bibinfo {title}
  {Enhanced pair plasma generation in the relativistic transparency regime},}\
  }\href {\doibase 10.1063/1.5001457} {\bibfield  {journal} {\bibinfo
  {journal} {Phys. Plasmas}\ }\textbf {\bibinfo {volume} {24}},\ \bibinfo
  {pages} {103130} (\bibinfo {year} {2017})}\BibitemShut {NoStop}%
\bibitem [{\citenamefont {Zhang}\ \emph {et~al.}(2021)\citenamefont {Zhang},
  \citenamefont {Wu}, \citenamefont {Huang}, \citenamefont {Lan}, \citenamefont
  {Liu}, \citenamefont {Wu}, \citenamefont {Yang}, \citenamefont {Zhao},
  \citenamefont {Zhu},\ and\ \citenamefont {Luo}}]{Zhang2021}%
  \BibitemOpen
  \bibfield  {author} {\bibinfo {author} {\bibfnamefont {Liang-qi}\
  \bibnamefont {Zhang}}, \bibinfo {author} {\bibfnamefont {Shao-dong}\
  \bibnamefont {Wu}}, \bibinfo {author} {\bibfnamefont {Hai-rong}\ \bibnamefont
  {Huang}}, \bibinfo {author} {\bibfnamefont {Hao-yang}\ \bibnamefont {Lan}},
  \bibinfo {author} {\bibfnamefont {Wei-yuan}\ \bibnamefont {Liu}}, \bibinfo
  {author} {\bibfnamefont {Yu-chi}\ \bibnamefont {Wu}}, \bibinfo {author}
  {\bibfnamefont {Yue}\ \bibnamefont {Yang}}, \bibinfo {author} {\bibfnamefont
  {Zong-qing}\ \bibnamefont {Zhao}}, \bibinfo {author} {\bibfnamefont
  {Zhi-chao}\ \bibnamefont {Zhu}}, \ and\ \bibinfo {author} {\bibfnamefont
  {Wen}\ \bibnamefont {Luo}},\ }\bibfield  {title} {\enquote {\bibinfo {title}
  {Brilliant attosecond $\gamma$-ray emission and high-yield positron
  production from intense laser-irradiated nano-micro array},}\ }\href
  {\doibase 10.1063/5.0030909} {\bibfield  {journal} {\bibinfo  {journal}
  {Phys. Plasmas}\ }\textbf {\bibinfo {volume} {28}},\ \bibinfo {pages}
  {023110} (\bibinfo {year} {2021})}\BibitemShut {NoStop}%
\bibitem [{\citenamefont {Filipovic}\ and\ \citenamefont
  {Pukhov}(2022)}]{fillipovic2022}%
  \BibitemOpen
  \bibfield  {author} {\bibinfo {author} {\bibfnamefont {Marko}\ \bibnamefont
  {Filipovic}}\ and\ \bibinfo {author} {\bibfnamefont {Alexander}\ \bibnamefont
  {Pukhov}},\ }\bibfield  {title} {\enquote {\bibinfo {title} {{QED effects at
  grazing incidence on solid-state targets}},}\ }\href {\doibase
  10.1140/epjd/s10053-022-00494-4} {\bibfield  {journal} {\bibinfo  {journal}
  {Eur. Phys. J. D}\ }\textbf {\bibinfo {volume} {76}},\ \bibinfo {pages} {187}
  (\bibinfo {year} {2022})}\BibitemShut {NoStop}%
\bibitem [{\citenamefont {Song}\ \emph {et~al.}(2022)\citenamefont {Song},
  \citenamefont {Wang},\ and\ \citenamefont {Li}}]{Song2022}%
  \BibitemOpen
  \bibfield  {author} {\bibinfo {author} {\bibfnamefont {Huai-Hang}\
  \bibnamefont {Song}}, \bibinfo {author} {\bibfnamefont {Wei-Min}\
  \bibnamefont {Wang}}, \ and\ \bibinfo {author} {\bibfnamefont {Yu-Tong}\
  \bibnamefont {Li}},\ }\bibfield  {title} {\enquote {\bibinfo {title} {{Dense
  Polarized Positrons from Laser-Irradiated Foil Targets in the QED Regime}},}\
  }\href {\doibase 10.1103/PhysRevLett.129.035001} {\bibfield  {journal}
  {\bibinfo  {journal} {Phys. Rev. Lett.}\ }\textbf {\bibinfo {volume} {129}},\
  \bibinfo {pages} {035001} (\bibinfo {year} {2022})}\BibitemShut {NoStop}%
\bibitem [{\citenamefont {Sentoku}\ \emph {et~al.}(1999)\citenamefont
  {Sentoku}, \citenamefont {Ruhl}, \citenamefont {Mima}, \citenamefont
  {Kodama}, \citenamefont {Tanaka},\ and\ \citenamefont
  {Kishimoto}}]{Sentoku1999}%
  \BibitemOpen
  \bibfield  {author} {\bibinfo {author} {\bibfnamefont {Y.}~\bibnamefont
  {Sentoku}}, \bibinfo {author} {\bibfnamefont {H.}~\bibnamefont {Ruhl}},
  \bibinfo {author} {\bibfnamefont {K.}~\bibnamefont {Mima}}, \bibinfo {author}
  {\bibfnamefont {R.}~\bibnamefont {Kodama}}, \bibinfo {author} {\bibfnamefont
  {K.~A.}\ \bibnamefont {Tanaka}}, \ and\ \bibinfo {author} {\bibfnamefont
  {Y.}~\bibnamefont {Kishimoto}},\ }\bibfield  {title} {\enquote {\bibinfo
  {title} {Plasma jet formation and magnetic-field generation in the intense
  laser plasma under oblique incidence},}\ }\href {\doibase 10.1063/1.873243}
  {\bibfield  {journal} {\bibinfo  {journal} {Phys. Plasmas}\ }\textbf
  {\bibinfo {volume} {6}},\ \bibinfo {pages} {2855--2861} (\bibinfo {year}
  {1999})}\BibitemShut {NoStop}%
\bibitem [{\citenamefont {Gahn}\ \emph {et~al.}(1999)\citenamefont {Gahn},
  \citenamefont {Tsakiris}, \citenamefont {Pukhov}, \citenamefont {Meyer-ter
  Vehn}, \citenamefont {Pretzler}, \citenamefont {Thirolf}, \citenamefont
  {Habs},\ and\ \citenamefont {Witte}}]{Gahn1999}%
  \BibitemOpen
  \bibfield  {author} {\bibinfo {author} {\bibfnamefont {C.}~\bibnamefont
  {Gahn}}, \bibinfo {author} {\bibfnamefont {G.~D.}\ \bibnamefont {Tsakiris}},
  \bibinfo {author} {\bibfnamefont {A.}~\bibnamefont {Pukhov}}, \bibinfo
  {author} {\bibfnamefont {J.}~\bibnamefont {Meyer-ter Vehn}}, \bibinfo
  {author} {\bibfnamefont {G.}~\bibnamefont {Pretzler}}, \bibinfo {author}
  {\bibfnamefont {P.}~\bibnamefont {Thirolf}}, \bibinfo {author} {\bibfnamefont
  {D.}~\bibnamefont {Habs}}, \ and\ \bibinfo {author} {\bibfnamefont {K.~J.}\
  \bibnamefont {Witte}},\ }\bibfield  {title} {\enquote {\bibinfo {title}
  {{Multi-MeV Electron Beam Generation by Direct Laser Acceleration in
  High-Density Plasma Channels}},}\ }\href {\doibase
  10.1103/PhysRevLett.83.4772} {\bibfield  {journal} {\bibinfo  {journal}
  {Phys. Rev. Lett.}\ }\textbf {\bibinfo {volume} {83}},\ \bibinfo {pages}
  {4772--4775} (\bibinfo {year} {1999})}\BibitemShut {NoStop}%
\bibitem [{\citenamefont {Naumova}\ \emph {et~al.}(2004)\citenamefont
  {Naumova}, \citenamefont {Sokolov}, \citenamefont {Nees}, \citenamefont
  {Maksimchuk}, \citenamefont {Yanovsky},\ and\ \citenamefont
  {Mourou}}]{Naumova2004}%
  \BibitemOpen
  \bibfield  {author} {\bibinfo {author} {\bibfnamefont {N.}~\bibnamefont
  {Naumova}}, \bibinfo {author} {\bibfnamefont {I.}~\bibnamefont {Sokolov}},
  \bibinfo {author} {\bibfnamefont {J.}~\bibnamefont {Nees}}, \bibinfo {author}
  {\bibfnamefont {A.}~\bibnamefont {Maksimchuk}}, \bibinfo {author}
  {\bibfnamefont {V.}~\bibnamefont {Yanovsky}}, \ and\ \bibinfo {author}
  {\bibfnamefont {G.}~\bibnamefont {Mourou}},\ }\bibfield  {title} {\enquote
  {\bibinfo {title} {{Attosecond Electron Bunches}},}\ }\href {\doibase
  10.1103/PhysRevLett.93.195003} {\bibfield  {journal} {\bibinfo  {journal}
  {Phys. Rev. Lett.}\ }\textbf {\bibinfo {volume} {93}},\ \bibinfo {pages}
  {195003} (\bibinfo {year} {2004})}\BibitemShut {NoStop}%
\bibitem [{\citenamefont {Nakamura}\ \emph {et~al.}(2004)\citenamefont
  {Nakamura}, \citenamefont {Kato}, \citenamefont {Nagatomo},\ and\
  \citenamefont {Mima}}]{Nakamura2004}%
  \BibitemOpen
  \bibfield  {author} {\bibinfo {author} {\bibfnamefont {Tatsufumi}\
  \bibnamefont {Nakamura}}, \bibinfo {author} {\bibfnamefont {Susumu}\
  \bibnamefont {Kato}}, \bibinfo {author} {\bibfnamefont {Hideo}\ \bibnamefont
  {Nagatomo}}, \ and\ \bibinfo {author} {\bibfnamefont {Kunioki}\ \bibnamefont
  {Mima}},\ }\bibfield  {title} {\enquote {\bibinfo {title}
  {Surface-magnetic-field and fast-electron current-layer formation by
  ultraintense laser irradiation},}\ }\href
  {https://link.aps.org/doi/10.1103/PhysRevLett.93.265002} {\bibfield
  {journal} {\bibinfo  {journal} {Phys. Rev. Lett.}\ }\textbf {\bibinfo
  {volume} {93}} (\bibinfo {year} {2004})}\BibitemShut {NoStop}%
\bibitem [{\citenamefont {Nakamura}\ \emph {et~al.}(2007)\citenamefont
  {Nakamura}, \citenamefont {Mima}, \citenamefont {Sakagami},\ and\
  \citenamefont {Johzaki}}]{Nakamura2007}%
  \BibitemOpen
  \bibfield  {author} {\bibinfo {author} {\bibfnamefont {Tatsufumi}\
  \bibnamefont {Nakamura}}, \bibinfo {author} {\bibfnamefont {Kunioki}\
  \bibnamefont {Mima}}, \bibinfo {author} {\bibfnamefont {Hitoshi}\
  \bibnamefont {Sakagami}}, \ and\ \bibinfo {author} {\bibfnamefont {Tomoyuki}\
  \bibnamefont {Johzaki}},\ }\bibfield  {title} {\enquote {\bibinfo {title}
  {Electron surface acceleration on a solid capillary target inner wall
  irradiated with ultraintense laser pulses},}\ }\href
  {https://doi.org/10.1063/1.2731383} {\bibfield  {journal} {\bibinfo
  {journal} {Phys. Plasma}\ }\textbf {\bibinfo {volume} {14}},\ \bibinfo
  {pages} {053112} (\bibinfo {year} {2007})}\BibitemShut {NoStop}%
\bibitem [{\citenamefont {Chen}\ \emph {et~al.}(2006)\citenamefont {Chen},
  \citenamefont {Shenga}, \citenamefont {Zheng}, \citenamefont {Ma},
  \citenamefont {Bari}, \citenamefont {Li},\ and\ \citenamefont
  {Zhang}}]{Chen2006}%
  \BibitemOpen
  \bibfield  {author} {\bibinfo {author} {\bibfnamefont {Min}\ \bibnamefont
  {Chen}}, \bibinfo {author} {\bibfnamefont {Zheng-Ming}\ \bibnamefont
  {Shenga}}, \bibinfo {author} {\bibfnamefont {Jun}\ \bibnamefont {Zheng}},
  \bibinfo {author} {\bibfnamefont {Yan-Yun}\ \bibnamefont {Ma}}, \bibinfo
  {author} {\bibfnamefont {Muhammad~Abbas}\ \bibnamefont {Bari}}, \bibinfo
  {author} {\bibfnamefont {Yu-Tong}\ \bibnamefont {Li}}, \ and\ \bibinfo
  {author} {\bibfnamefont {Jie}\ \bibnamefont {Zhang}},\ }\bibfield  {title}
  {\enquote {\bibinfo {title} {Surface electron acceleration in relativistic
  laser-solid interactions},}\ }\href {\doibase 10.1364/OE.14.003093}
  {\bibfield  {journal} {\bibinfo  {journal} {Opt. Express}\ }\textbf {\bibinfo
  {volume} {14}},\ \bibinfo {pages} {3093--3098} (\bibinfo {year}
  {2006})}\BibitemShut {NoStop}%
\bibitem [{\citenamefont {Li}\ \emph {et~al.}(2006)\citenamefont {Li},
  \citenamefont {Yuan}, \citenamefont {Xu}, \citenamefont {Zheng},
  \citenamefont {Sheng}, \citenamefont {Chen}, \citenamefont {Ma},
  \citenamefont {Liang}, \citenamefont {Yu}, \citenamefont {Zhang},
  \citenamefont {Liu}, \citenamefont {Wang}, \citenamefont {Wei}, \citenamefont
  {Zhao}, \citenamefont {Jin},\ and\ \citenamefont {Zhang}}]{Li2006}%
  \BibitemOpen
  \bibfield  {author} {\bibinfo {author} {\bibfnamefont {Y.~T.}\ \bibnamefont
  {Li}}, \bibinfo {author} {\bibfnamefont {X.~H.}\ \bibnamefont {Yuan}},
  \bibinfo {author} {\bibfnamefont {M.~H.}\ \bibnamefont {Xu}}, \bibinfo
  {author} {\bibfnamefont {Z.~Y.}\ \bibnamefont {Zheng}}, \bibinfo {author}
  {\bibfnamefont {Z.~M.}\ \bibnamefont {Sheng}}, \bibinfo {author}
  {\bibfnamefont {M.}~\bibnamefont {Chen}}, \bibinfo {author} {\bibfnamefont
  {Y.~Y.}\ \bibnamefont {Ma}}, \bibinfo {author} {\bibfnamefont {W.~X.}\
  \bibnamefont {Liang}}, \bibinfo {author} {\bibfnamefont {Q.~Z.}\ \bibnamefont
  {Yu}}, \bibinfo {author} {\bibfnamefont {Y.}~\bibnamefont {Zhang}}, \bibinfo
  {author} {\bibfnamefont {F.}~\bibnamefont {Liu}}, \bibinfo {author}
  {\bibfnamefont {Z.~H.}\ \bibnamefont {Wang}}, \bibinfo {author}
  {\bibfnamefont {Z.~Y.}\ \bibnamefont {Wei}}, \bibinfo {author} {\bibfnamefont
  {W.}~\bibnamefont {Zhao}}, \bibinfo {author} {\bibfnamefont {Z.}~\bibnamefont
  {Jin}}, \ and\ \bibinfo {author} {\bibfnamefont {J.}~\bibnamefont {Zhang}},\
  }\bibfield  {title} {\enquote {\bibinfo {title} {Observation of a fast
  electron beam emitted along the surface of a target irradiated by intense
  femtosecond laser pulses},}\ }\href {\doibase 10.1103/PhysRevLett.96.165003}
  {\bibfield  {journal} {\bibinfo  {journal} {Phys. Rev. Lett.}\ }\textbf
  {\bibinfo {volume} {96}},\ \bibinfo {pages} {165003} (\bibinfo {year}
  {2006})}\BibitemShut {NoStop}%
\bibitem [{\citenamefont {Tian}\ \emph {et~al.}(2012)\citenamefont {Tian},
  \citenamefont {Liu}, \citenamefont {Wang}, \citenamefont {Wang},
  \citenamefont {Deng}, \citenamefont {Xia}, \citenamefont {Li}, \citenamefont
  {Cao}, \citenamefont {Lu}, \citenamefont {Zhang}, \citenamefont {Xu},
  \citenamefont {Leng}, \citenamefont {Li},\ and\ \citenamefont
  {Xu}}]{tian2012}%
  \BibitemOpen
  \bibfield  {author} {\bibinfo {author} {\bibfnamefont {Ye}~\bibnamefont
  {Tian}}, \bibinfo {author} {\bibfnamefont {Jiansheng}\ \bibnamefont {Liu}},
  \bibinfo {author} {\bibfnamefont {Wentao}\ \bibnamefont {Wang}}, \bibinfo
  {author} {\bibfnamefont {Cheng}\ \bibnamefont {Wang}}, \bibinfo {author}
  {\bibfnamefont {Aihua}\ \bibnamefont {Deng}}, \bibinfo {author}
  {\bibfnamefont {Changquan}\ \bibnamefont {Xia}}, \bibinfo {author}
  {\bibfnamefont {Wentao}\ \bibnamefont {Li}}, \bibinfo {author} {\bibfnamefont
  {Lihua}\ \bibnamefont {Cao}}, \bibinfo {author} {\bibfnamefont {Haiyang}\
  \bibnamefont {Lu}}, \bibinfo {author} {\bibfnamefont {Hui}\ \bibnamefont
  {Zhang}}, \bibinfo {author} {\bibfnamefont {Yi}~\bibnamefont {Xu}}, \bibinfo
  {author} {\bibfnamefont {Yuxin}\ \bibnamefont {Leng}}, \bibinfo {author}
  {\bibfnamefont {Ruxin}\ \bibnamefont {Li}}, \ and\ \bibinfo {author}
  {\bibfnamefont {Zhizhan}\ \bibnamefont {Xu}},\ }\bibfield  {title} {\enquote
  {\bibinfo {title} {Electron emission at locked phases from the laser-driven
  surface plasma wave},}\ }\href {\doibase 10.1103/PhysRevLett.109.115002}
  {\bibfield  {journal} {\bibinfo  {journal} {Phys. Rev. Lett.}\ }\textbf
  {\bibinfo {volume} {109}},\ \bibinfo {pages} {115002} (\bibinfo {year}
  {2012})}\BibitemShut {NoStop}%
\bibitem [{\citenamefont {Th{\'e}venet}\ \emph {et~al.}(2016)\citenamefont
  {Th{\'e}venet}, \citenamefont {Leblanc}, \citenamefont {Kahaly},
  \citenamefont {Vincenti}, \citenamefont {Vernier}, \citenamefont
  {Qu{\'e}r{\'e}},\ and\ \citenamefont {Faure}}]{Thevenet2016}%
  \BibitemOpen
  \bibfield  {author} {\bibinfo {author} {\bibfnamefont {M.}~\bibnamefont
  {Th{\'e}venet}}, \bibinfo {author} {\bibfnamefont {A.}~\bibnamefont
  {Leblanc}}, \bibinfo {author} {\bibfnamefont {S.}~\bibnamefont {Kahaly}},
  \bibinfo {author} {\bibfnamefont {H.}~\bibnamefont {Vincenti}}, \bibinfo
  {author} {\bibfnamefont {A.}~\bibnamefont {Vernier}}, \bibinfo {author}
  {\bibfnamefont {F.}~\bibnamefont {Qu{\'e}r{\'e}}}, \ and\ \bibinfo {author}
  {\bibfnamefont {J{\'e}r{\^o}me}\ \bibnamefont {Faure}},\ }\bibfield  {title}
  {\enquote {\bibinfo {title} {Vacuum laser acceleration of relativistic
  electrons using plasma mirror injectors},}\ }\href {\doibase
  10.1038/nphys3597} {\bibfield  {journal} {\bibinfo  {journal} {Nat. Phys.}\
  }\textbf {\bibinfo {volume} {12}},\ \bibinfo {pages} {355--360} (\bibinfo
  {year} {2016})}\BibitemShut {NoStop}%
\bibitem [{\citenamefont {Wan}\ \emph {et~al.}(2023)\citenamefont {Wan},
  \citenamefont {Lv}, \citenamefont {Xue}, \citenamefont {Dou}, \citenamefont
  {Zhao}, \citenamefont {Ababekri}, \citenamefont {Wei}, \citenamefont {Li},
  \citenamefont {Zhao},\ and\ \citenamefont {Li}}]{PIC_wan}%
  \BibitemOpen
  \bibfield  {author} {\bibinfo {author} {\bibfnamefont {Feng}\ \bibnamefont
  {Wan}}, \bibinfo {author} {\bibfnamefont {Chong}\ \bibnamefont {Lv}},
  \bibinfo {author} {\bibfnamefont {Kun}\ \bibnamefont {Xue}}, \bibinfo
  {author} {\bibfnamefont {Zhen-Ke}\ \bibnamefont {Dou}}, \bibinfo {author}
  {\bibfnamefont {Qian}\ \bibnamefont {Zhao}}, \bibinfo {author} {\bibfnamefont
  {Mamutjan}\ \bibnamefont {Ababekri}}, \bibinfo {author} {\bibfnamefont
  {Wen-Qing}\ \bibnamefont {Wei}}, \bibinfo {author} {\bibfnamefont
  {Zhong-Peng}\ \bibnamefont {Li}}, \bibinfo {author} {\bibfnamefont
  {Yong-Tao}\ \bibnamefont {Zhao}}, \ and\ \bibinfo {author} {\bibfnamefont
  {Jian-Xing}\ \bibnamefont {Li}},\ }\bibfield  {title} {\enquote {\bibinfo
  {title} {Simulations of spin/polarization-resolved laser–plasma
  interactions in the nonlinear qed regime},}\ }\href {\doibase
  10.1063/5.0163929} {\bibfield  {journal} {\bibinfo  {journal} {Matter Radiat.
  Extremes}\ }\textbf {\bibinfo {volume} {8}} (\bibinfo {year} {2023}),\
  10.1063/5.0163929}\BibitemShut {NoStop}%
\bibitem [{\citenamefont {{Extreme Light Infrastructure (ELI)}}()}]{ELI}%
  \BibitemOpen
  \bibfield  {author} {\bibinfo {author} {\bibnamefont {{Extreme Light
  Infrastructure (ELI)}}},\ }\href@noop {} {}\bibinfo {howpublished}
  {\url{https://eli-laser.eu/}}\BibitemShut {NoStop}%
\bibitem [{\citenamefont {{Exawatt Center for Extreme Light Studies
  (XCELS)}}()}]{ECELS}%
  \BibitemOpen
  \bibfield  {author} {\bibinfo {author} {\bibnamefont {{Exawatt Center for
  Extreme Light Studies (XCELS)}}},\ }\href@noop {} {}\bibinfo {howpublished}
  {\url{https://xcels.iapras.ru/}}\BibitemShut {NoStop}%
\bibitem [{\citenamefont {Garrec}\ \emph {et~al.}(2014)\citenamefont {Garrec},
  \citenamefont {Sebban}, \citenamefont {Margarone}, \citenamefont {Precek},
  \citenamefont {Weber}, \citenamefont {Klimo}, \citenamefont {Korn},\ and\
  \citenamefont {Rus}}]{ELI-beamlines}%
  \BibitemOpen
  \bibfield  {author} {\bibinfo {author} {\bibfnamefont {Bruno~Le}\
  \bibnamefont {Garrec}}, \bibinfo {author} {\bibfnamefont {Stephane}\
  \bibnamefont {Sebban}}, \bibinfo {author} {\bibfnamefont {Daniele}\
  \bibnamefont {Margarone}}, \bibinfo {author} {\bibfnamefont {Martin}\
  \bibnamefont {Precek}}, \bibinfo {author} {\bibfnamefont {Stefan}\
  \bibnamefont {Weber}}, \bibinfo {author} {\bibfnamefont {Ondrej}\
  \bibnamefont {Klimo}}, \bibinfo {author} {\bibfnamefont {Georg}\ \bibnamefont
  {Korn}}, \ and\ \bibinfo {author} {\bibfnamefont {Bedrich}\ \bibnamefont
  {Rus}},\ }\bibfield  {title} {\enquote {\bibinfo {title} {{ELI-beamlines:
  extreme light infrastructure science and technology with ultra-intense
  lasers}},}\ }in\ \href {\doibase 10.1117/12.2039165} {\emph {\bibinfo
  {booktitle} {High Energy/Average Power Lasers and Intense Beam Applications
  VII}}},\ Vol.\ \bibinfo {volume} {8962},\ \bibinfo {editor} {edited by\
  \bibinfo {editor} {\bibfnamefont {Steven~J.}\ \bibnamefont {Davis}}, \bibinfo
  {editor} {\bibfnamefont {Michael~C.}\ \bibnamefont {Heaven}}, \ and\ \bibinfo
  {editor} {\bibfnamefont {J.~Thomas}\ \bibnamefont {Schriempf}}},\ \bibinfo
  {organization} {International Society for Optics and Photonics}\ (\bibinfo
  {publisher} {SPIE},\ \bibinfo {address} {Bellingham},\ \bibinfo {year}
  {2014})\ p.\ \bibinfo {pages} {89620I}\BibitemShut {NoStop}%
\bibitem [{\citenamefont {Zou}\ \emph {et~al.}(2015)\citenamefont {Zou},
  \citenamefont {Le~Blanc}, \citenamefont {Papadopoulos}, \citenamefont
  {Ch{\'e}riaux}, \citenamefont {Georges}, \citenamefont {Mennerat},
  \citenamefont {Druon}, \citenamefont {Lecherbourg}, \citenamefont
  {Pellegrina}, \citenamefont {Ramirez},\ and\ \citenamefont {\textit{et
  al.}}}]{zou2015}%
  \BibitemOpen
  \bibfield  {author} {\bibinfo {author} {\bibfnamefont {Ji-Ping}\ \bibnamefont
  {Zou}}, \bibinfo {author} {\bibfnamefont {Catherine}\ \bibnamefont
  {Le~Blanc}}, \bibinfo {author} {\bibfnamefont {Dimitrios~N}\ \bibnamefont
  {Papadopoulos}}, \bibinfo {author} {\bibfnamefont {Gilles}\ \bibnamefont
  {Ch{\'e}riaux}}, \bibinfo {author} {\bibfnamefont {Patrick}\ \bibnamefont
  {Georges}}, \bibinfo {author} {\bibfnamefont {G}~\bibnamefont {Mennerat}},
  \bibinfo {author} {\bibfnamefont {Fr{\'e}d{\'e}ric}\ \bibnamefont {Druon}},
  \bibinfo {author} {\bibfnamefont {Ludovic}\ \bibnamefont {Lecherbourg}},
  \bibinfo {author} {\bibfnamefont {Alain}\ \bibnamefont {Pellegrina}},
  \bibinfo {author} {\bibfnamefont {Patricia}\ \bibnamefont {Ramirez}}, \ and\
  \bibinfo {author} {\bibnamefont {\textit{et al.}}},\ }\bibfield  {title}
  {\enquote {\bibinfo {title} {{Design and current progress of the Apollon 10
  PW project}},}\ }\href {\doibase 10.1017/hpl.2014.41} {\bibfield  {journal}
  {\bibinfo  {journal} {High Power Laser Sci. Eng.}\ }\textbf {\bibinfo
  {volume} {3}},\ \bibinfo {pages} {e2} (\bibinfo {year} {2015})}\BibitemShut
  {NoStop}%
\bibitem [{\citenamefont {Gales}\ \emph {et~al.}(2018)\citenamefont {Gales},
  \citenamefont {Tanaka}, \citenamefont {Balabanski}, \citenamefont {Negoita},
  \citenamefont {Stutman}, \citenamefont {Tesileanu}, \citenamefont {Ur},
  \citenamefont {Ursescu}, \citenamefont {Andrei}, \citenamefont {Ataman},
  \citenamefont {Cernaianu}, \citenamefont {D’Alessi}, \citenamefont
  {Dancus}, \citenamefont {Diaconescu}, \citenamefont {Djourelov},
  \citenamefont {Filipescu}, \citenamefont {Ghenuche}, \citenamefont {Ghita},
  \citenamefont {Matei}, \citenamefont {Seto}, \citenamefont {Zeng},\ and\
  \citenamefont {Zamfir}}]{gales2018}%
  \BibitemOpen
  \bibfield  {author} {\bibinfo {author} {\bibfnamefont {S.}~\bibnamefont
  {Gales}}, \bibinfo {author} {\bibfnamefont {K.~A.}\ \bibnamefont {Tanaka}},
  \bibinfo {author} {\bibfnamefont {D.~L.}\ \bibnamefont {Balabanski}},
  \bibinfo {author} {\bibfnamefont {F.}~\bibnamefont {Negoita}}, \bibinfo
  {author} {\bibfnamefont {D.}~\bibnamefont {Stutman}}, \bibinfo {author}
  {\bibfnamefont {O.}~\bibnamefont {Tesileanu}}, \bibinfo {author}
  {\bibfnamefont {C.~A.}\ \bibnamefont {Ur}}, \bibinfo {author} {\bibfnamefont
  {D.}~\bibnamefont {Ursescu}}, \bibinfo {author} {\bibfnamefont
  {I.}~\bibnamefont {Andrei}}, \bibinfo {author} {\bibfnamefont
  {S.}~\bibnamefont {Ataman}}, \bibinfo {author} {\bibfnamefont {M.~O.}\
  \bibnamefont {Cernaianu}}, \bibinfo {author} {\bibfnamefont {L.}~\bibnamefont
  {D’Alessi}}, \bibinfo {author} {\bibfnamefont {I.}~\bibnamefont {Dancus}},
  \bibinfo {author} {\bibfnamefont {B.}~\bibnamefont {Diaconescu}}, \bibinfo
  {author} {\bibfnamefont {N.}~\bibnamefont {Djourelov}}, \bibinfo {author}
  {\bibfnamefont {D.}~\bibnamefont {Filipescu}}, \bibinfo {author}
  {\bibfnamefont {P.}~\bibnamefont {Ghenuche}}, \bibinfo {author}
  {\bibfnamefont {D.~G.}\ \bibnamefont {Ghita}}, \bibinfo {author}
  {\bibfnamefont {C.}~\bibnamefont {Matei}}, \bibinfo {author} {\bibfnamefont
  {K.}~\bibnamefont {Seto}}, \bibinfo {author} {\bibfnamefont {M.}~\bibnamefont
  {Zeng}}, \ and\ \bibinfo {author} {\bibfnamefont {N.~V.}\ \bibnamefont
  {Zamfir}},\ }\bibfield  {title} {\enquote {\bibinfo {title} {{The extreme
  light infrastructure—nuclear physics (ELI-NP) facility: new horizons in
  physics with 10 PW ultra-intense lasers and 20 MeV brilliant gamma beams}},}\
  }\href {\doibase 10.1088/1361-6633/aacfe8} {\bibfield  {journal} {\bibinfo
  {journal} {Rep. Prog. Phys.}\ }\textbf {\bibinfo {volume} {81}},\ \bibinfo
  {pages} {094301} (\bibinfo {year} {2018})}\BibitemShut {NoStop}%
\bibitem [{\citenamefont {Gan}\ \emph {et~al.}(2021)\citenamefont {Gan},
  \citenamefont {Yu}, \citenamefont {Wang}, \citenamefont {Liu}, \citenamefont
  {Xu}, \citenamefont {Li}, \citenamefont {Li}, \citenamefont {Yu},
  \citenamefont {Wang}, \citenamefont {Liu}, \citenamefont {Chen},
  \citenamefont {Peng}, \citenamefont {Xu}, \citenamefont {Yao}, \citenamefont
  {Zhang}, \citenamefont {Chen}, \citenamefont {Tang}, \citenamefont {Wang},
  \citenamefont {Yin}, \citenamefont {Liang}, \citenamefont {Leng},
  \citenamefont {Li},\ and\ \citenamefont {Xu}}]{Gan2021}%
  \BibitemOpen
  \bibfield  {author} {\bibinfo {author} {\bibfnamefont {Zebiao}\ \bibnamefont
  {Gan}}, \bibinfo {author} {\bibfnamefont {Lianghong}\ \bibnamefont {Yu}},
  \bibinfo {author} {\bibfnamefont {Cheng}\ \bibnamefont {Wang}}, \bibinfo
  {author} {\bibfnamefont {Yanqi}\ \bibnamefont {Liu}}, \bibinfo {author}
  {\bibfnamefont {Yi}~\bibnamefont {Xu}}, \bibinfo {author} {\bibfnamefont
  {Wenqi}\ \bibnamefont {Li}}, \bibinfo {author} {\bibfnamefont {Shuai}\
  \bibnamefont {Li}}, \bibinfo {author} {\bibfnamefont {Linpeng}\ \bibnamefont
  {Yu}}, \bibinfo {author} {\bibfnamefont {Xinliang}\ \bibnamefont {Wang}},
  \bibinfo {author} {\bibfnamefont {Xinyan}\ \bibnamefont {Liu}}, \bibinfo
  {author} {\bibfnamefont {Junchi}\ \bibnamefont {Chen}}, \bibinfo {author}
  {\bibfnamefont {Yujie}\ \bibnamefont {Peng}}, \bibinfo {author}
  {\bibfnamefont {Lu}~\bibnamefont {Xu}}, \bibinfo {author} {\bibfnamefont
  {Bo}~\bibnamefont {Yao}}, \bibinfo {author} {\bibfnamefont {Xiaobo}\
  \bibnamefont {Zhang}}, \bibinfo {author} {\bibfnamefont {Lingru}\
  \bibnamefont {Chen}}, \bibinfo {author} {\bibfnamefont {Yunhai}\ \bibnamefont
  {Tang}}, \bibinfo {author} {\bibfnamefont {Xiaobin}\ \bibnamefont {Wang}},
  \bibinfo {author} {\bibfnamefont {Dinjun}\ \bibnamefont {Yin}}, \bibinfo
  {author} {\bibfnamefont {Xiaoyan}\ \bibnamefont {Liang}}, \bibinfo {author}
  {\bibfnamefont {Yuxin}\ \bibnamefont {Leng}}, \bibinfo {author}
  {\bibfnamefont {Ruxin}\ \bibnamefont {Li}}, \ and\ \bibinfo {author}
  {\bibfnamefont {Zhizhan}\ \bibnamefont {Xu}},\ }\enquote {\bibinfo {title}
  {{The Shanghai Superintense Ultrafast Laser Facility (SULF) Project}},}\ in\
  \href {\doibase 10.1007/978-3-030-75089-3_10} {\emph {\bibinfo {booktitle}
  {Progress in Ultrafast Intense Laser Science XVI}}},\ \bibinfo {editor}
  {edited by\ \bibinfo {editor} {\bibfnamefont {Kaoru}\ \bibnamefont
  {Yamanouchi}}, \bibinfo {editor} {\bibfnamefont {Katsumi}\ \bibnamefont
  {Midorikawa}}, \ and\ \bibinfo {editor} {\bibfnamefont {Luis}\ \bibnamefont
  {Roso}}}\ (\bibinfo  {publisher} {Springer International Publishing},\
  \bibinfo {address} {Cham},\ \bibinfo {year} {2021})\ pp.\ \bibinfo {pages}
  {199--217}\BibitemShut {NoStop}%
\bibitem [{\citenamefont {Du}\ \emph {et~al.}(2023)\citenamefont {Du},
  \citenamefont {Shen}, \citenamefont {Liang}, \citenamefont {Wang},
  \citenamefont {Liu},\ and\ \citenamefont
  {Li}}]{du_shen_liang_wang_liu_li_2023}%
  \BibitemOpen
  \bibfield  {author} {\bibinfo {author} {\bibfnamefont {Shuman}\ \bibnamefont
  {Du}}, \bibinfo {author} {\bibfnamefont {Xiong}\ \bibnamefont {Shen}},
  \bibinfo {author} {\bibfnamefont {Wenhai}\ \bibnamefont {Liang}}, \bibinfo
  {author} {\bibfnamefont {Peng}\ \bibnamefont {Wang}}, \bibinfo {author}
  {\bibfnamefont {Jun}\ \bibnamefont {Liu}}, \ and\ \bibinfo {author}
  {\bibfnamefont {Ruxin}\ \bibnamefont {Li}},\ }\bibfield  {title} {\enquote
  {\bibinfo {title} {{A 100-PW compressor based on single-pass single-grating
  pair}},}\ }\href {\doibase 10.1017/hpl.2023.5} {\bibfield  {journal}
  {\bibinfo  {journal} {High Power Laser Sci. Eng.}\ }\textbf {\bibinfo
  {volume} {11}},\ \bibinfo {pages} {e4} (\bibinfo {year} {2023})}\BibitemShut
  {NoStop}%
\bibitem [{\citenamefont {Wagner}\ \emph {et~al.}(2014)\citenamefont {Wagner},
  \citenamefont {Bedacht}, \citenamefont {Ortner}, \citenamefont {Roth},
  \citenamefont {Tauschwitz}, \citenamefont {Zielbauer},\ and\ \citenamefont
  {Bagnoud}}]{wagner2014}%
  \BibitemOpen
  \bibfield  {author} {\bibinfo {author} {\bibfnamefont {Florian}\ \bibnamefont
  {Wagner}}, \bibinfo {author} {\bibfnamefont {Stefan}\ \bibnamefont
  {Bedacht}}, \bibinfo {author} {\bibfnamefont {Alex}\ \bibnamefont {Ortner}},
  \bibinfo {author} {\bibfnamefont {Markus}\ \bibnamefont {Roth}}, \bibinfo
  {author} {\bibfnamefont {Anna}\ \bibnamefont {Tauschwitz}}, \bibinfo {author}
  {\bibfnamefont {Bernhard}\ \bibnamefont {Zielbauer}}, \ and\ \bibinfo
  {author} {\bibfnamefont {Vincent}\ \bibnamefont {Bagnoud}},\ }\bibfield
  {title} {\enquote {\bibinfo {title} {Pre-plasma formation in experiments
  using petawatt lasers},}\ }\href {\doibase 10.1364/OE.22.029505} {\bibfield
  {journal} {\bibinfo  {journal} {Opt. Express}\ }\textbf {\bibinfo {volume}
  {22}},\ \bibinfo {pages} {29505--29514} (\bibinfo {year} {2014})}\BibitemShut
  {NoStop}%
\bibitem [{SM()}]{SM}%
  \BibitemOpen
  \href@noop {} {}\bibinfo {note} {See Supplemental Material for details on the
  applied theoretical model, the estimations of other physical processes that
  affect positron yield and polarization and the simulation results for other
  parameters.}\BibitemShut {Stop}%
\bibitem [{\citenamefont {Alejo}\ \emph {et~al.}(2019)\citenamefont {Alejo},
  \citenamefont {Walczak},\ and\ \citenamefont {Sarri}}]{Alejo2019}%
  \BibitemOpen
  \bibfield  {author} {\bibinfo {author} {\bibfnamefont {A.}~\bibnamefont
  {Alejo}}, \bibinfo {author} {\bibfnamefont {R.}~\bibnamefont {Walczak}}, \
  and\ \bibinfo {author} {\bibfnamefont {G.}~\bibnamefont {Sarri}},\ }\bibfield
   {title} {\enquote {\bibinfo {title} {Laser-driven high-quality positron
  sources as possible injectors for plasma-based accelerators},}\ }\href
  {https://doi.org/10.1038/s41598-019-41650-y} {\bibfield  {journal} {\bibinfo
  {journal} {Sci. Rep.}\ }\textbf {\bibinfo {volume} {9}} (\bibinfo {year}
  {2019})}\BibitemShut {NoStop}%
\bibitem [{\citenamefont {Corde}\ \emph {et~al.}(2015)\citenamefont {Corde},
  \citenamefont {Adli}, \citenamefont {Allen}, \citenamefont {An},
  \citenamefont {Clarke}, \citenamefont {Clayton}, \citenamefont {Delahaye},
  \citenamefont {Frederico}, \citenamefont {Gessner}, \citenamefont {Green},\
  and\ \citenamefont {\textit{et al.}}}]{Corde2015}%
  \BibitemOpen
  \bibfield  {author} {\bibinfo {author} {\bibfnamefont {S{\'e}bastien}\
  \bibnamefont {Corde}}, \bibinfo {author} {\bibfnamefont {E}~\bibnamefont
  {Adli}}, \bibinfo {author} {\bibfnamefont {JM}~\bibnamefont {Allen}},
  \bibinfo {author} {\bibfnamefont {W}~\bibnamefont {An}}, \bibinfo {author}
  {\bibfnamefont {CI}~\bibnamefont {Clarke}}, \bibinfo {author} {\bibfnamefont
  {CE}~\bibnamefont {Clayton}}, \bibinfo {author} {\bibfnamefont
  {JP}~\bibnamefont {Delahaye}}, \bibinfo {author} {\bibfnamefont
  {J}~\bibnamefont {Frederico}}, \bibinfo {author} {\bibfnamefont
  {S}~\bibnamefont {Gessner}}, \bibinfo {author} {\bibfnamefont
  {SZ}~\bibnamefont {Green}}, \ and\ \bibinfo {author} {\bibnamefont
  {\textit{et al.}}},\ }\bibfield  {title} {\enquote {\bibinfo {title}
  {Multi-gigaelectronvolt acceleration of positrons in a self-loaded plasma
  wakefield},}\ }\href {\doibase 10.1038/nature14890} {\bibfield  {journal}
  {\bibinfo  {journal} {Nature}\ }\textbf {\bibinfo {volume} {524}},\ \bibinfo
  {pages} {442--445} (\bibinfo {year} {2015})}\BibitemShut {NoStop}%
\bibitem [{\citenamefont {Gonsalves}\ \emph {et~al.}(2019)\citenamefont
  {Gonsalves}, \citenamefont {Nakamura}, \citenamefont {Daniels}, \citenamefont
  {Benedetti}, \citenamefont {Pieronek}, \citenamefont {de~Raadt},
  \citenamefont {Steinke}, \citenamefont {Bin}, \citenamefont {Bulanov},
  \citenamefont {van Tilborg}, \citenamefont {Geddes}, \citenamefont
  {Schroeder}, \citenamefont {T\'oth}, \citenamefont {Esarey}, \citenamefont
  {Swanson}, \citenamefont {Fan-Chiang}, \citenamefont {Bagdasarov},
  \citenamefont {Bobrova}, \citenamefont {Gasilov}, \citenamefont {Korn},
  \citenamefont {Sasorov},\ and\ \citenamefont {Leemans}}]{Gonsalves2019}%
  \BibitemOpen
  \bibfield  {author} {\bibinfo {author} {\bibfnamefont {A.~J.}\ \bibnamefont
  {Gonsalves}}, \bibinfo {author} {\bibfnamefont {K.}~\bibnamefont {Nakamura}},
  \bibinfo {author} {\bibfnamefont {J.}~\bibnamefont {Daniels}}, \bibinfo
  {author} {\bibfnamefont {C.}~\bibnamefont {Benedetti}}, \bibinfo {author}
  {\bibfnamefont {C.}~\bibnamefont {Pieronek}}, \bibinfo {author}
  {\bibfnamefont {T.~C.~H.}\ \bibnamefont {de~Raadt}}, \bibinfo {author}
  {\bibfnamefont {S.}~\bibnamefont {Steinke}}, \bibinfo {author} {\bibfnamefont
  {J.~H.}\ \bibnamefont {Bin}}, \bibinfo {author} {\bibfnamefont {S.~S.}\
  \bibnamefont {Bulanov}}, \bibinfo {author} {\bibfnamefont {J.}~\bibnamefont
  {van Tilborg}}, \bibinfo {author} {\bibfnamefont {C.~G.~R.}\ \bibnamefont
  {Geddes}}, \bibinfo {author} {\bibfnamefont {C.~B.}\ \bibnamefont
  {Schroeder}}, \bibinfo {author} {\bibfnamefont {Cs.}\ \bibnamefont {T\'oth}},
  \bibinfo {author} {\bibfnamefont {E.}~\bibnamefont {Esarey}}, \bibinfo
  {author} {\bibfnamefont {K.}~\bibnamefont {Swanson}}, \bibinfo {author}
  {\bibfnamefont {L.}~\bibnamefont {Fan-Chiang}}, \bibinfo {author}
  {\bibfnamefont {G.}~\bibnamefont {Bagdasarov}}, \bibinfo {author}
  {\bibfnamefont {N.}~\bibnamefont {Bobrova}}, \bibinfo {author} {\bibfnamefont
  {V.}~\bibnamefont {Gasilov}}, \bibinfo {author} {\bibfnamefont
  {G.}~\bibnamefont {Korn}}, \bibinfo {author} {\bibfnamefont {P.}~\bibnamefont
  {Sasorov}}, \ and\ \bibinfo {author} {\bibfnamefont {W.~P.}\ \bibnamefont
  {Leemans}},\ }\bibfield  {title} {\enquote {\bibinfo {title} {{Petawatt Laser
  Guiding and Electron Beam Acceleration to 8 GeV in a Laser-Heated Capillary
  Discharge Waveguide}},}\ }\href
  {https://link.aps.org/doi/10.1103/PhysRevLett.122.084801} {\bibfield
  {journal} {\bibinfo  {journal} {Phy. Rev. Lett.}\ }\textbf {\bibinfo {volume}
  {122}} (\bibinfo {year} {2019})}\BibitemShut {NoStop}%
\bibitem [{\citenamefont {Thomas}\ \emph {et~al.}(2020)\citenamefont {Thomas},
  \citenamefont {H\"utzen}, \citenamefont {Lehrach}, \citenamefont {Pukhov},
  \citenamefont {Ji}, \citenamefont {Wu}, \citenamefont {Geng},\ and\
  \citenamefont {B\"uscher}}]{Thomas2020}%
  \BibitemOpen
  \bibfield  {author} {\bibinfo {author} {\bibfnamefont {Johannes}\
  \bibnamefont {Thomas}}, \bibinfo {author} {\bibfnamefont {Anna}\ \bibnamefont
  {H\"utzen}}, \bibinfo {author} {\bibfnamefont {Andreas}\ \bibnamefont
  {Lehrach}}, \bibinfo {author} {\bibfnamefont {Alexander}\ \bibnamefont
  {Pukhov}}, \bibinfo {author} {\bibfnamefont {Liangliang}\ \bibnamefont {Ji}},
  \bibinfo {author} {\bibfnamefont {Yitong}\ \bibnamefont {Wu}}, \bibinfo
  {author} {\bibfnamefont {Xuesong}\ \bibnamefont {Geng}}, \ and\ \bibinfo
  {author} {\bibfnamefont {Markus}\ \bibnamefont {B\"uscher}},\ }\bibfield
  {title} {\enquote {\bibinfo {title} {Scaling laws for the depolarization time
  of relativistic particle beams in strong fields},}\ }\href {\doibase
  10.1103/PhysRevAccelBeams.23.064401} {\bibfield  {journal} {\bibinfo
  {journal} {Phys. Rev. Accel. Beams}\ }\textbf {\bibinfo {volume} {23}},\
  \bibinfo {pages} {064401} (\bibinfo {year} {2020})}\BibitemShut {NoStop}%
\bibitem [{\citenamefont {Brunel}(1987)}]{Brunel1987}%
  \BibitemOpen
  \bibfield  {author} {\bibinfo {author} {\bibfnamefont {F.}~\bibnamefont
  {Brunel}},\ }\bibfield  {title} {\enquote {\bibinfo {title} {Not-so-resonant,
  resonant absorption},}\ }\href {\doibase 10.1103/PhysRevLett.59.52}
  {\bibfield  {journal} {\bibinfo  {journal} {Phys. Rev. Lett.}\ }\textbf
  {\bibinfo {volume} {59}},\ \bibinfo {pages} {52--55} (\bibinfo {year}
  {1987})}\BibitemShut {NoStop}%
\bibitem [{\citenamefont {Kruer}\ and\ \citenamefont
  {Estabrook}(1985)}]{Kruer1985}%
  \BibitemOpen
  \bibfield  {author} {\bibinfo {author} {\bibfnamefont {W.~L.}\ \bibnamefont
  {Kruer}}\ and\ \bibinfo {author} {\bibfnamefont {Kent}\ \bibnamefont
  {Estabrook}},\ }\bibfield  {title} {\enquote {\bibinfo {title} {{J$\times$B
  heating by very intense laser light}},}\ }\href {\doibase 10.1063/1.865171}
  {\bibfield  {journal} {\bibinfo  {journal} {Phys. Fluids}\ }\textbf {\bibinfo
  {volume} {28}},\ \bibinfo {pages} {430--432} (\bibinfo {year}
  {1985})}\BibitemShut {NoStop}%
\bibitem [{\citenamefont {Zhang}\ \emph {et~al.}(2007)\citenamefont {Zhang},
  \citenamefont {Li}, \citenamefont {Sheng}, \citenamefont {Wei}, \citenamefont
  {Dong},\ and\ \citenamefont {Lu}}]{zhang2007fast}%
  \BibitemOpen
  \bibfield  {author} {\bibinfo {author} {\bibfnamefont {J.}~\bibnamefont
  {Zhang}}, \bibinfo {author} {\bibfnamefont {Y.~T.}\ \bibnamefont {Li}},
  \bibinfo {author} {\bibfnamefont {Z.~M.}\ \bibnamefont {Sheng}}, \bibinfo
  {author} {\bibfnamefont {Z.~Y.}\ \bibnamefont {Wei}}, \bibinfo {author}
  {\bibfnamefont {Q.~L.}\ \bibnamefont {Dong}}, \ and\ \bibinfo {author}
  {\bibfnamefont {X.}~\bibnamefont {Lu}},\ }\enquote {\bibinfo {title} {Fast
  electrons in high-intensity laser interactions with plasmas},}\ in\ \href
  {\doibase 10.1007/978-3-540-38156-3_16} {\emph {\bibinfo {booktitle}
  {Progress in Ultrafast Intense Laser Science II}}}\ (\bibinfo  {publisher}
  {Springer},\ \bibinfo {address} {Berlin, Heidelberg},\ \bibinfo {year}
  {2007})\ pp.\ \bibinfo {pages} {319--340}\BibitemShut {NoStop}%
\bibitem [{\citenamefont {McMaster}(1961)}]{mcmaster1961}%
  \BibitemOpen
  \bibfield  {author} {\bibinfo {author} {\bibfnamefont {William~H.}\
  \bibnamefont {McMaster}},\ }\bibfield  {title} {\enquote {\bibinfo {title}
  {Matrix representation of polarization},}\ }\href {\doibase
  10.1103/RevModPhys.33.8} {\bibfield  {journal} {\bibinfo  {journal} {Rev.
  Mod. Phys.}\ }\textbf {\bibinfo {volume} {33}},\ \bibinfo {pages} {8--28}
  (\bibinfo {year} {1961})}\BibitemShut {NoStop}%
\bibitem [{\citenamefont {Berestetskii}\ \emph {et~al.}(1982)\citenamefont
  {Berestetskii}, \citenamefont {Lifshitz},\ and\ \citenamefont
  {Pitaevskii}}]{berestetskii1982quantum}%
  \BibitemOpen
  \bibfield  {author} {\bibinfo {author} {\bibfnamefont {Vladimir~Borisovich}\
  \bibnamefont {Berestetskii}}, \bibinfo {author} {\bibfnamefont
  {Evgenii~Mikhailovich}\ \bibnamefont {Lifshitz}}, \ and\ \bibinfo {author}
  {\bibfnamefont {Lev~Petrovich}\ \bibnamefont {Pitaevskii}},\ }\href@noop {}
  {\emph {\bibinfo {title} {Quantum Electrodynamics}}}\ (\bibinfo  {publisher}
  {Pergamon},\ \bibinfo {address} {Oxford},\ \bibinfo {year}
  {1982})\BibitemShut {NoStop}%
\bibitem [{\citenamefont {Lundh}\ \emph {et~al.}(2007)\citenamefont {Lundh},
  \citenamefont {Lindau}, \citenamefont {Persson}, \citenamefont {Wahlstr\"om},
  \citenamefont {McKenna},\ and\ \citenamefont {Batani}}]{Lundh2007}%
  \BibitemOpen
  \bibfield  {author} {\bibinfo {author} {\bibfnamefont {O.}~\bibnamefont
  {Lundh}}, \bibinfo {author} {\bibfnamefont {F.}~\bibnamefont {Lindau}},
  \bibinfo {author} {\bibfnamefont {A.}~\bibnamefont {Persson}}, \bibinfo
  {author} {\bibfnamefont {C.-G.}\ \bibnamefont {Wahlstr\"om}}, \bibinfo
  {author} {\bibfnamefont {P.}~\bibnamefont {McKenna}}, \ and\ \bibinfo
  {author} {\bibfnamefont {D.}~\bibnamefont {Batani}},\ }\bibfield  {title}
  {\enquote {\bibinfo {title} {Influence of shock waves on laser-driven proton
  acceleration},}\ }\href {\doibase 10.1103/PhysRevE.76.026404} {\bibfield
  {journal} {\bibinfo  {journal} {Phys. Rev. E}\ }\textbf {\bibinfo {volume}
  {76}},\ \bibinfo {pages} {026404} (\bibinfo {year} {2007})}\BibitemShut
  {NoStop}%
\bibitem [{\citenamefont {Doumy}\ \emph {et~al.}(2004)\citenamefont {Doumy},
  \citenamefont {Qu\'er\'e}, \citenamefont {Gobert}, \citenamefont {Perdrix},
  \citenamefont {Martin}, \citenamefont {Audebert}, \citenamefont {Gauthier},
  \citenamefont {Geindre},\ and\ \citenamefont {Wittmann}}]{Doumy2004}%
  \BibitemOpen
  \bibfield  {author} {\bibinfo {author} {\bibfnamefont {G.}~\bibnamefont
  {Doumy}}, \bibinfo {author} {\bibfnamefont {F.}~\bibnamefont {Qu\'er\'e}},
  \bibinfo {author} {\bibfnamefont {O.}~\bibnamefont {Gobert}}, \bibinfo
  {author} {\bibfnamefont {M.}~\bibnamefont {Perdrix}}, \bibinfo {author}
  {\bibfnamefont {Ph.}\ \bibnamefont {Martin}}, \bibinfo {author}
  {\bibfnamefont {P.}~\bibnamefont {Audebert}}, \bibinfo {author}
  {\bibfnamefont {J.~C.}\ \bibnamefont {Gauthier}}, \bibinfo {author}
  {\bibfnamefont {J.-P.}\ \bibnamefont {Geindre}}, \ and\ \bibinfo {author}
  {\bibfnamefont {T.}~\bibnamefont {Wittmann}},\ }\bibfield  {title} {\enquote
  {\bibinfo {title} {Complete characterization of a plasma mirror for the
  production of high-contrast ultraintense laser pulses},}\ }\href {\doibase
  10.1103/PhysRevE.69.026402} {\bibfield  {journal} {\bibinfo  {journal} {Phys.
  Rev. E}\ }\textbf {\bibinfo {volume} {69}},\ \bibinfo {pages} {026402}
  (\bibinfo {year} {2004})}\BibitemShut {NoStop}%
\bibitem [{\citenamefont {L\'{e}vy}\ \emph {et~al.}(2007)\citenamefont
  {L\'{e}vy}, \citenamefont {Ceccotti}, \citenamefont {D'Oliveira},
  \citenamefont {R\'{e}au}, \citenamefont {Perdrix}, \citenamefont
  {Qu\'{e}r\'{e}}, \citenamefont {Monot}, \citenamefont {Bougeard},
  \citenamefont {Lagadec}, \citenamefont {Martin}, \citenamefont {Geindre},\
  and\ \citenamefont {Audebert}}]{Levy2007}%
  \BibitemOpen
  \bibfield  {author} {\bibinfo {author} {\bibfnamefont {Anna}\ \bibnamefont
  {L\'{e}vy}}, \bibinfo {author} {\bibfnamefont {Tiberio}\ \bibnamefont
  {Ceccotti}}, \bibinfo {author} {\bibfnamefont {Pascal}\ \bibnamefont
  {D'Oliveira}}, \bibinfo {author} {\bibfnamefont {Fabrice}\ \bibnamefont
  {R\'{e}au}}, \bibinfo {author} {\bibfnamefont {Michel}\ \bibnamefont
  {Perdrix}}, \bibinfo {author} {\bibfnamefont {Fabien}\ \bibnamefont
  {Qu\'{e}r\'{e}}}, \bibinfo {author} {\bibfnamefont {Pascal}\ \bibnamefont
  {Monot}}, \bibinfo {author} {\bibfnamefont {Michel}\ \bibnamefont
  {Bougeard}}, \bibinfo {author} {\bibfnamefont {Herv\'{e}}\ \bibnamefont
  {Lagadec}}, \bibinfo {author} {\bibfnamefont {Philippe}\ \bibnamefont
  {Martin}}, \bibinfo {author} {\bibfnamefont {Jean-Paul}\ \bibnamefont
  {Geindre}}, \ and\ \bibinfo {author} {\bibfnamefont {Patrick}\ \bibnamefont
  {Audebert}},\ }\bibfield  {title} {\enquote {\bibinfo {title} {Double plasma
  mirror for ultrahigh temporal contrast ultraintense laser pulses},}\ }\href
  {\doibase 10.1364/OL.32.000310} {\bibfield  {journal} {\bibinfo  {journal}
  {Opt. Lett.}\ }\textbf {\bibinfo {volume} {32}},\ \bibinfo {pages} {310--312}
  (\bibinfo {year} {2007})}\BibitemShut {NoStop}%
\bibitem [{\citenamefont {Reed}\ \emph {et~al.}(2009)\citenamefont {Reed},
  \citenamefont {Matsuoka}, \citenamefont {Bulanov}, \citenamefont {Tampo},
  \citenamefont {Chvykov}, \citenamefont {Kalintchenko}, \citenamefont
  {Rousseau}, \citenamefont {Yanovsky}, \citenamefont {Kodama}, \citenamefont
  {Litzenberg}, \citenamefont {Krushelnick},\ and\ \citenamefont
  {Maksimchuk}}]{Reed2009}%
  \BibitemOpen
  \bibfield  {author} {\bibinfo {author} {\bibfnamefont {Stephen~A.}\
  \bibnamefont {Reed}}, \bibinfo {author} {\bibfnamefont {Takeshi}\
  \bibnamefont {Matsuoka}}, \bibinfo {author} {\bibfnamefont {Stepan}\
  \bibnamefont {Bulanov}}, \bibinfo {author} {\bibfnamefont {Motonobu}\
  \bibnamefont {Tampo}}, \bibinfo {author} {\bibfnamefont {Vladimir}\
  \bibnamefont {Chvykov}}, \bibinfo {author} {\bibfnamefont {Galina}\
  \bibnamefont {Kalintchenko}}, \bibinfo {author} {\bibfnamefont {Pascal}\
  \bibnamefont {Rousseau}}, \bibinfo {author} {\bibfnamefont {Victor}\
  \bibnamefont {Yanovsky}}, \bibinfo {author} {\bibfnamefont {Ryousuke}\
  \bibnamefont {Kodama}}, \bibinfo {author} {\bibfnamefont {Dale~W.}\
  \bibnamefont {Litzenberg}}, \bibinfo {author} {\bibfnamefont {Karl}\
  \bibnamefont {Krushelnick}}, \ and\ \bibinfo {author} {\bibfnamefont
  {Anatoly}\ \bibnamefont {Maksimchuk}},\ }\bibfield  {title} {\enquote
  {\bibinfo {title} {{Relativistic plasma shutter for ultraintense laser
  pulses}},}\ }\href {\doibase 10.1063/1.3139860} {\bibfield  {journal}
  {\bibinfo  {journal} {Appl. Phys. Lett.}\ }\textbf {\bibinfo {volume} {94}},\
  \bibinfo {pages} {201117} (\bibinfo {year} {2009})}\BibitemShut {NoStop}%
\bibitem [{\citenamefont {Mironov}\ \emph {et~al.}(2009)\citenamefont
  {Mironov}, \citenamefont {Lozhkarev}, \citenamefont {Ginzburg},\ and\
  \citenamefont {Khazanov}}]{Mironov2009}%
  \BibitemOpen
  \bibfield  {author} {\bibinfo {author} {\bibfnamefont {Sergey}\ \bibnamefont
  {Mironov}}, \bibinfo {author} {\bibfnamefont {Vladimir}\ \bibnamefont
  {Lozhkarev}}, \bibinfo {author} {\bibfnamefont {Vladislav}\ \bibnamefont
  {Ginzburg}}, \ and\ \bibinfo {author} {\bibfnamefont {Efim}\ \bibnamefont
  {Khazanov}},\ }\bibfield  {title} {\enquote {\bibinfo {title}
  {High-efficiency second-harmonic generation of superintense ultrashort laser
  pulses},}\ }\href {\doibase 10.1364/AO.48.002051} {\bibfield  {journal}
  {\bibinfo  {journal} {Appl. Opt.}\ }\textbf {\bibinfo {volume} {48}},\
  \bibinfo {pages} {2051--2057} (\bibinfo {year} {2009})}\BibitemShut {NoStop}%
\bibitem [{\citenamefont {Aidala}\ \emph {et~al.}(2013)\citenamefont {Aidala},
  \citenamefont {Bass}, \citenamefont {Hasch},\ and\ \citenamefont
  {Mallot}}]{Aidala2013}%
  \BibitemOpen
  \bibfield  {author} {\bibinfo {author} {\bibfnamefont {Christine~A.}\
  \bibnamefont {Aidala}}, \bibinfo {author} {\bibfnamefont {Steven~D.}\
  \bibnamefont {Bass}}, \bibinfo {author} {\bibfnamefont {Delia}\ \bibnamefont
  {Hasch}}, \ and\ \bibinfo {author} {\bibfnamefont {Gerhard~K.}\ \bibnamefont
  {Mallot}},\ }\bibfield  {title} {\enquote {\bibinfo {title} {The spin
  structure of the nucleon},}\ }\href {\doibase 10.1103/RevModPhys.85.655}
  {\bibfield  {journal} {\bibinfo  {journal} {Rev. Mod. Phys.}\ }\textbf
  {\bibinfo {volume} {85}},\ \bibinfo {pages} {655--691} (\bibinfo {year}
  {2013})}\BibitemShut {NoStop}%
\bibitem [{\citenamefont {Gamaly}\ \emph {et~al.}(2002)\citenamefont {Gamaly},
  \citenamefont {Rode}, \citenamefont {Luther-Davies},\ and\ \citenamefont
  {Tikhonchuk}}]{gamaly2002}%
  \BibitemOpen
  \bibfield  {author} {\bibinfo {author} {\bibfnamefont {E.~G.}\ \bibnamefont
  {Gamaly}}, \bibinfo {author} {\bibfnamefont {A.~V.}\ \bibnamefont {Rode}},
  \bibinfo {author} {\bibfnamefont {B.}~\bibnamefont {Luther-Davies}}, \ and\
  \bibinfo {author} {\bibfnamefont {V.~T.}\ \bibnamefont {Tikhonchuk}},\
  }\bibfield  {title} {\enquote {\bibinfo {title} {{Ablation of solids by
  femtosecond lasers: Ablation mechanism and ablation thresholds for metals and
  dielectrics}},}\ }\href {\doibase 10.1063/1.1447555} {\bibfield  {journal}
  {\bibinfo  {journal} {Phys. Plasmas}\ }\textbf {\bibinfo {volume} {9}},\
  \bibinfo {pages} {949--957} (\bibinfo {year} {2002})}\BibitemShut {NoStop}%
\bibitem [{\citenamefont {Ma}\ \emph {et~al.}(2006)\citenamefont {Ma},
  \citenamefont {Sheng}, \citenamefont {Li}, \citenamefont {Zhang},
  \citenamefont {Yuan}, \citenamefont {Xu}, \citenamefont {Zheng},
  \citenamefont {Chang}, \citenamefont {Chen},\ and\ \citenamefont
  {Zheng}}]{ma2006}%
  \BibitemOpen
  \bibfield  {author} {\bibinfo {author} {\bibfnamefont {Y.~Y.}\ \bibnamefont
  {Ma}}, \bibinfo {author} {\bibfnamefont {Z.~M.}\ \bibnamefont {Sheng}},
  \bibinfo {author} {\bibfnamefont {Y.~T.}\ \bibnamefont {Li}}, \bibinfo
  {author} {\bibfnamefont {J.}~\bibnamefont {Zhang}}, \bibinfo {author}
  {\bibfnamefont {X.~H.}\ \bibnamefont {Yuan}}, \bibinfo {author}
  {\bibfnamefont {M.~H.}\ \bibnamefont {Xu}}, \bibinfo {author} {\bibfnamefont
  {Z.~Y.}\ \bibnamefont {Zheng}}, \bibinfo {author} {\bibfnamefont {W.~W.}\
  \bibnamefont {Chang}}, \bibinfo {author} {\bibfnamefont {M.}~\bibnamefont
  {Chen}}, \ and\ \bibinfo {author} {\bibfnamefont {J.}~\bibnamefont {Zheng}},\
  }\bibfield  {title} {\enquote {\bibinfo {title} {Preplasma effects on the
  emission directions of energetic electrons in relativistic laser–solid
  interactions},}\ }\href {\doibase 10.1017/S0022377806005691} {\bibfield
  {journal} {\bibinfo  {journal} {J. Plasma Phys.}\ }\textbf {\bibinfo {volume}
  {72}},\ \bibinfo {pages} {1269--1272} (\bibinfo {year} {2006})}\BibitemShut
  {NoStop}%
\bibitem [{\citenamefont {Pukhov}\ \emph {et~al.}(1999)\citenamefont {Pukhov},
  \citenamefont {Sheng},\ and\ \citenamefont {Meyer-ter Vehn}}]{Pukhov1999}%
  \BibitemOpen
  \bibfield  {author} {\bibinfo {author} {\bibfnamefont {A.}~\bibnamefont
  {Pukhov}}, \bibinfo {author} {\bibfnamefont {Z.-M.}\ \bibnamefont {Sheng}}, \
  and\ \bibinfo {author} {\bibfnamefont {J.}~\bibnamefont {Meyer-ter Vehn}},\
  }\bibfield  {title} {\enquote {\bibinfo {title} {Particle acceleration in
  relativistic laser channels},}\ }\href {\doibase 10.1063/1.873242} {\bibfield
   {journal} {\bibinfo  {journal} {Phys. Plasma}\ }\textbf {\bibinfo {volume}
  {6}},\ \bibinfo {pages} {2847--2854} (\bibinfo {year} {1999})}\BibitemShut
  {NoStop}%
\end{thebibliography}%

\end{document}